\documentclass[10pt,epsfig]{article}

\usepackage{enumerate}
\usepackage{float}
\usepackage{subfig}
\usepackage{color}
\usepackage[table]{xcolor}
\usepackage{slashed}
\usepackage{multirow}
\usepackage{cite}

\let\counterwithin\relax
\usepackage{chngcntr}
\usepackage{amssymb, amsmath,mathrsfs}
\usepackage{longtable}
\usepackage{graphics}
\usepackage{graphicx}
\usepackage{epsf}
\usepackage{epsfig}
\usepackage{float}
\usepackage{makecell}
\usepackage{multirow}
\usepackage{color}
\usepackage{xcolor}
\usepackage{simplewick}

\usepackage[utf8]{inputenc}
\usepackage{amsmath}
\usepackage{amssymb}
\usepackage{subfig}
\usepackage[normalem]{ulem}

\usepackage{mathtools}
\usepackage{amsmath}
\usepackage{diagbox}

\newcommand\undermat[2]{
	\makebox[0.5pt][l]{$\smash{\underbrace{\phantom{%
					\begin{matrix}#2\end{matrix}}}_{ \let\scriptstyle\textstyle\text{\large $#1$}}}$}#2}
\newcommand\overmat[2]{
	\makebox[-1pt][l]{$\smash{\overbrace{\phantom{%
					\begin{matrix}#2\end{matrix}}}^{ \let\scriptstyle\textstyle\text{\large $#1$}}}$}#2}    
\usepackage{tikz}
\usepackage[vcentermath]{youngtab}
\usepackage{slashed} 
\usepackage[font=small]{caption} 
\usepackage{times} 

\usepackage{bm}       
\usepackage{bbm} 

\usepackage{relsize}  

\usepackage{autobreak}  

\usepackage[colorlinks=true,
linkcolor=blue,
urlcolor=red,
citecolor=red]{hyperref}

\usepackage{makeidx} 
\usepackage{bbding} 
\usepackage{listings} 
\usepackage{ytableau}
\usepackage[utf8]{inputenc}
\usepackage[T1]{fontenc}
\usepackage{lmodern}

\ytableausetup
{boxsize=1.0em}

\long\def\rpl#1!!#2!!{\textcolor{red}{#1} \textcolor{blue}{#2}}

\usepackage[top=1in, left=0.95in, bottom=1.1in, right=0.95in]{geometry}
\def\baselinestretch{1.27}
\usepackage[toc]{appendix}

\newcommand{\beq}{\begin {equation}}
\newcommand{\eeq}{\end   {equation}}
\newcommand{\bea}{\begin {eqnarray}}
\newcommand{\eea}{\end   {eqnarray}}
\newcommand{\beqa}{\begin {eqnarray}}
\newcommand{\eeqa}{\end   {eqnarray}}
\newcommand{\baa}{\begin {array}   }
\newcommand{\eaa}{\end   {array}   }
\newcommand{\bit}{\begin {itemize} }
\newcommand{\eit}{\end   {itemize} }
\newcommand{\be }{\begin {equation}}
\newcommand{\ee }{\end   {equation}}

\newcommand{\Tr}{\textbf{Tr}}

\newcommand{\bft}{\mathbf{T}}
\newcommand{\bfu}{\mathbf{U}}
\newcommand{\bfv}{\mathbf{V}}

\newcommand{\calf}{\mathcal{F}}
\newcommand{\calt}{\mathcal{T}}
\newcommand{\calo}{\mathcal{O}}
\newcommand{\caly}{\mathcal{Y}}
\newcommand{\calm}{\mathcal{M}}
\newcommand{\calb}{\mathcal{B}}
\newcommand{\yt}[1]{\ytableaushort{#1}}
\newcommand{\lra}[1]{\langle #1 \rangle}
\newcommand{\lrs}[1]{[ #1 ]}
\newcommand{\ione}{i_1}
\newcommand{\itwo}{i_2}
\newcommand{\ithree}{i_3}
\newcommand{\jthree}{j_3}
\newcommand{\ifour}{i_4}
\newcommand{\jfour}{j_4}
\newcommand{\ifive}{i_5}
\newcommand{\jfive}{j_5}
\newcommand{\isix}{i_6}
\newcommand{\jsix}{j_6}

\begin{document}

\begin{center}

{\Large \textbf  {Complete NLO Operators in the Higgs Effective Field Theory}}\\[10mm]

Hao Sun$^{a, b}$\footnote{sunhao@itp.ac.cn}, Ming-Lei Xiao$^{c, d}$\footnote{minglei.xiao@northwestern.edu}, Jiang-Hao Yu$^{a, b, e, f, g}$\footnote{jhyu@itp.ac.cn}\\[10mm]

\noindent 
$^a${\em \small CAS Key Laboratory of Theoretical Physics, Institute of Theoretical Physics, Chinese Academy of Sciences,    \\ Beijing 100190, P. R. China}  \\
$^b${\em \small School of Physical Sciences, University of Chinese Academy of Sciences,   Beijing 100049, P.R. China}   \\
$^c${\em \small Department of Physics and Astronomy, Northwestern University, Evanston, Illinois 60208, USA}\\
$^d${\em \small High Energy Physics Division, Argonne National Laboratory, Lemont, Illinois 60439, USA}\\
$^e${\em \small Center for High Energy Physics, Peking University, Beijing 100871, China} \\
$^f${\em \small School of Fundamental Physics and Mathematical Sciences, Hangzhou Institute for Advanced Study, UCAS, Hangzhou 310024, China} \\
$^g${\em \small International Centre for Theoretical Physics Asia-Pacific, Beijing/Hangzhou, China}\\[10mm]

\date{\today}   
          
\end{center}

\begin{abstract}

We enumerate the complete and independent sets of operators at the next-to-leading order (NLO) in the Higgs effective field theory (HEFT), based on the Young tensor technique on the Lorentz, gauge and flavor structures. 
The operator-amplitude correspondence tells a type of operators forms the on-shell amplitude basis, and for operators involving in Nambu-Goldstone bosons, the amplitude basis is further reduced to the subspace satisfying the Adler zero condition in the soft momentum limit. Different from dynamical fields, the spurion should not enter into the Lorentz sector, instead it only plays a role of forming the $SU(2)$ invariant together with other dynamical fields. 
With these new treatments, for the first time we could obtain the 224 (7704) operators for one (three) generation fermions, 295 (11307) with right-handed neutrinos, and find there were 6 (9) terms of operators missing and many redundant operators can be removed in the effective theory without (with) right-handed neutrinos. 

\end{abstract}

\newpage

\setcounter{tocdepth}{3}
\setcounter{secnumdepth}{3}

\tableofcontents

\setcounter{footnote}{0}

\def\baselinestretch{1.5}
\counterwithin{equation}{section}

\newpage

\section{Introduction}

The Large Hadron Collider (LHC) has confirmed the success of the Standard Model (SM) with the discovery of the SM-like Higgs boson and the non-observation of new particles below the TeV scale. The separated scales between the SM particles and new physics suggests the effective field theory (EFT) is quite a suitable framework to parametrize new physics effects based on the SM degrees of freedom. Pioneered by Weinberg~\cite{Weinberg:1978kz}, starting from the degrees of freedom at low energy scale, one writes down the most general possible Lagrangian, including all terms consistent with presumed symmetry principle, with proper power counting rules. In this bottom-up approach, the Wilson coefficients of the effective operators are totally free and independent parameters in which all kinds of Lorentz-invariant new physics effects are encoded and parameterized.

With the assumption that new physics decouples at a high energy scale and the electroweak (EW) symmetry is in the unbroken phase, the standard model effective field theory (SMEFT) is appreciated and widely adapted in the description of the new physics in the EFT framework. In the SMEFT, starting from the fields and symmetries of the SM, one can write down all possible operators order by order according to the canonical dimension power counting. Since Weinberg wrote down the dimension 5 operator~\cite{Weinberg:1979sa}, lots of progress has been made on writing down complete and independent operators up to the mass dimension 9 in the standard model effective field theory (SMEFT)~\cite{Weinberg:1979sa, Buchmuller:1985jz, Grzadkowski:2010es, Lehman:2014jma, Liao:2016hru, Li:2020gnx, Murphy:2020rsh, Li:2020xlh, Liao:2020jmn,Li:2020zfq}, the low-energy effective field theory (LEFT)~\cite{Jenkins:2017jig,Liao:2020zyx,Li:2020tsi,Murphy:2020cly}, the standard model effective field theory with right-handed neutrinos ($\nu$SMEFT) and the low-energy effective field theory with right-handed neutrinos ($\nu$LEFT)~\cite{delAguila:2008ir,Aparici:2009fh,Bhattacharya:2015vja,Liao:2016qyd,Li:2021tsq}. A general algorithm, implemented in a Mathematica package ABC4EFT~\cite{Li:2022tec}, has been proposed to construct the independent and complete SMEFT operator bases up to any mass dimension.

However, there are still new physics scenarios that cannot be described within the SMEFT framework due to the existence of the {\it non-decoupling effects}. In these scenarios, the heavy particles obtain their physical masses predominantly from the vacuum expectation value of the Higgs boson, and thus it is not possible to integrate out the heavy particle in the unbroken phase. On the other hand, since these non-decoupling effects can only be systematically described in the EW broken phase, it is necessary to adapt the non-linear realization of the EW symmetry provided by the electroweak chiral Lagrangian with the light Higgs boson (H-EWChL)~\cite{Appelquist:1980vg,Longhitano:1980iz,Longhitano:1980tm,Feruglio:1992wf,Herrero:1993nc,Herrero:1994iu}, also known as the Higgs effective field theory (HEFT)\cite{Grinstein:2007iv,Buchalla:2012qq,Buchalla:2013rka,Buchalla:2013eza,Gavela:2014vra,Pich:2015kwa,Pich:2016lew,Krause:2018cwe,Alonso:2012px,Brivio:2013pma,Brivio:2016fzo,Pich:2018ltt,Merlo:2016prs}. The HEFT provides a more general realization of the EW symmetry breaking, which includes the SMEFT as a particular case~\cite{Falkowski:2019tft,Agrawal:2019bpm,Cohen:2020xca,Banta:2021dek,Alonso:2021rac,Gomez-Ambrosio:2022giw}. The electroweak symmetry breaking (EWSB) patterns determine the nature of the Higgs boson, such as the fundamental Higgs, Nambu-Goldstone Higgs, Coleman-Weinberg Higgs, and tadpole-induced Higgs~\cite{Agrawal:2019bpm}. Among all four scenarios, only the elementary Higgs and the Nambu-Goldstone Higgs scenarios can be described within the SMEFT framework, while all the four scenarios can be unified into the HEFT framework, in which the Higgs boson belongs to the singlet under the global symmetry.


In the SMEFT the symmetry breaking is manifestly realized under the Higgs mechanism, in which the Higgs boson makes part of an $SU(2)_L$ doublet together with the three electroweak Nambu-Goldstone fields, and the doublet Higgs is transformed linearly under the EW gauge symmetry. On the other hand, the HEFT does not make any assumption about the nature of the Higgs field and the pattern of the EWSB. The HEFT provides a general description of the EWSB sector, in terms of the three Nambu-Goldstone bosons (NGBs) transformed non-linearly under the EW symmetry, the spurion parametrizing the custodial symmetry breaking, and the observed Higgs boson as a EW singlet field. Inspired by the chiral perturbation theory (ChPT)~\cite{Weinberg:1978kz,Gasser:1983yg,Gasser:1984gg} in the quantum chromodynamics, the HEFT operators have been written order by order with the power counting~\cite{Weinberg:1978kz,Manohar:1983md,Hirn:2005fr,Buchalla:2013eza,Buchalla:2012qq,Buchalla:2016sop,Gavela:2016bzc,Krause:2018cwe,Pich:2015kwa,Pich:2016lew} of the chiral dimensions using the coset structure~\cite{Coleman:1969sm,Callan:1969sn}. The electroweak chiral Lagrangian has been constructed up to the next-to-leading-order (NLO) \cite{Grinstein:2007iv,Buchalla:2012qq,Buchalla:2013rka,Buchalla:2013eza,Krause:2018cwe,Alonso:2012px,Brivio:2013pma,Brivio:2016fzo,Gavela:2014vra,Pich:2015kwa,Pich:2016lew,Pich:2018ltt,Merlo:2016prs}, and the one-loop renormalization has been investigated systematically~\cite{Guo:2015isa,Alonso:2017tdy,Buchalla:2017jlu,Buchalla:2020kdh}. However, writing down the complete operator basis including flavor structure has not yet been finished.

In this work we revisit the HEFT operator basis and write down the complete list of the effective operators via the Young tensor method developed in Ref~\cite{Li:2020xlh,Li:2020gnx,Li:2022tec,Li:2020zfq}, with certain improvements on the NGBs and spurion. Compared with the previous works, we need the following new treatments:
\bit
\item For operators involving in the NGBs, the operator basis written by the Young tensor method needs to satisfy additional constraints from the shift symmetry of the NGBs. Based on the operator-amplitude correspondence, we impose the Adler zero condition on the contact amplitude basis and select the amplitudes satisfying the soft theorem in the soft momentum limit~\cite{Adler:1964um,Adler:1965ga, Low:2014nga, Low:2014oga, Cheung:2014dqa, Cheung:2015ota, Low:2019ynd, Dai:2020cpk,minglei-adler}.

\item For operators involving in the spurions, the spurion, as the frozen degree of freedom of dynamical field to parametrize the custodial symmetry breaking, does not enter into the Lorentz sector, but plays the role of forming the group invariant under the custodial symmetry together with other SM field degrees of freedom. We remove the spurion from the Lorentz construction but keep the spurion to be involved in the gauge and flavor structure. Different from the dynamical field treatment, we also need to remove the self contraction of the spurions in effective operators because they do not have physical effects.    

\eit
With the new treatments on the NGB and spurion, we obtain the complete and independent set of operators for both one generation and three generation fermions in the HEFT without (with) the right-handed neutrinos. Comparing with the NLO operators listed in \cite{Buchalla:2013rka,Brivio:2016fzo,Merlo:2016prs}, we find that there were {\it 6(9) terms of operators missing}, corresponding to 420(663) operators for three generation fermions, in the HEFT without (with) sterile neutrinos, and identify that there were many redundant operators that should be removed. Furthermore, the flavor structure was {\it not} considered and properly treated in literature, while we list the complete NLO operators with the complete flavor structure using the flavor tensor. Finally, we obtain the on-shell amplitude basis for the NLO operators, listed in Appendix~\ref{app:amplitude}, which would be useful for investigating the scattering amplitudes in the HEFT framework.

The paper is organized as follows. In section \ref{sec:field} we review the essential parts of electroweak chiral Lagrangian: field contents, symmetry and the power counting rules. Based on the operator-amplitude correspondence, we ultilize the Young tensor method to write down the operator basis, and impose the Adler zero condition for the operators involving in the NGBs in section \ref{sec:operator}. In section \ref{sec:boson} we list the complete bosonic operators while section \ref{sec:fermion} we list the complete NLO operators involving in three generation fermions. Finally we conclude at the section \ref{sec:conclu}. In appendix \ref{app:converting}, we summarise the conversion rules between 2-component and 4-component spinors. In appendix \ref{app:operatorlist} we list the operators comparing with the literature and Appendix \ref{app:amplitude} shows the complete on-shell amplitude basis.

\section{Electroweak Chiral Lagrangian with Light Higgs}
\label{sec:field}

In this section, we lay out the main ingredients of the H-EWChL (HEFT): the particle content (build blocks), the global and local symmetries, and the power-counting rules. We also write down the leading order (LO) Lagrangian and organize the NLO terms based on the power-counting rules.

\subsection{Building Blocks and Coset Construction}
\label{subsec:block}

First let us briefly review the Callan-Coleman-Wess-Zumino (CCWZ) construction~\cite{Coleman:1969sm,Callan:1969sn} for the symmetry breaking pattern ${\mathcal G}\to {\mathcal H}$,
which provides a systematic way to write effective Lagrangians that allow to manifest the symmetries of the theory. By the definition of the coset group,
any group element of the global symmetry $\mathfrak{g}[\alpha_A]  \in {\mathcal G} $ can be decomposed as the product
\bea
	\mathfrak{g}[\alpha_A] = e^{i \alpha_A T^A} = e^{i f_{\hat{a}}[\alpha] \hat{T}^{\hat{a}}  } \cdot  e^{i f_{a}[\alpha] {T}^{{a}}  },
\eea
where $T^a$ and $\hat{T}^{\hat{a}}$ are the unbroken and broken generators, respectively, and $f_{a, \hat{a}}[\alpha] = \alpha_{a, \hat{a}} + {\mathcal O}(\alpha^2)$. 
Since the fields along the coset directions are in a one-to-one correspondence with the NGBs, we promote the corresponding parameters to dynamical fields by defining the field matrix
\bea
	\Omega(\Pi) \equiv \exp\left[\frac{i}{2f}\Pi(x)\right],  \quad \textrm{with} \quad \Pi(x) = \Pi_{\hat{a}}  T^{\hat{a}},  
\eea
Let us determine how $\Omega(\Pi)$ and $\Pi$ transform under a element of the group ${\mathcal G}$, ${\mathcal H}$, and ${\mathcal G}/{\mathcal H}$. 
Using the decomposition relation above, under the action of a general element $\mathfrak{g}[\alpha_A] \in {\mathcal G}$, the $\Omega(\Pi)$ transforms
\bea
\mathfrak{g} \cdot \Omega(\Pi) = \Omega(\Pi^{(\mathfrak{g})}) \cdot \mathfrak{h}(\Pi; \mathfrak{g} )\, \quad \textrm{with} \quad \mathfrak{h}(\Pi; \mathfrak{g} ) = \exp(i \zeta^a[\Pi; \mathfrak{g}] T^a)
\eea
with the compensating transformation $\mathfrak{h} \in {\mathcal H}$ and equivalently
\bea
 \Omega(\Pi)  \rightarrow \Omega(\Pi^{(\mathfrak{g})}) =  \mathfrak{g} \Omega(\Pi)  \mathfrak{h}^{-1}(\Pi; \mathfrak{g}). \label{Omega_g-h}
\eea
In the particular cases of restricting the general transformation $\mathfrak{g}$ into the subgroup transformation $\mathfrak{g}_{\mathcal{H}}$ and $\mathfrak{g}_{\mathcal{G}/\mathcal{H}}$,
the above transformation on the Goldstone matrix tells the explicit form of the $\Pi^{(\mathfrak{g})}$:
\bea
	\Pi_{\hat{a}} &\rightarrow & \Pi_{\hat{a}}^{\left(\mathfrak{g}_{\mathcal{H}}\right)}=\left(e^{i \alpha_{a} T_{\pi}^{a}}\right)_{\hat{a}}^{\hat{b}} \Pi_{\hat{b}}, \quad {\textrm{with}} \quad
	\mathfrak{h}^{-1}(\Pi; \mathfrak{g}) = \mathfrak{g}_{\mathcal{H}},  \\
	\Pi_{\hat{a}} &\rightarrow & \Pi_{\hat{a}}^{\left(\mathfrak{g}_{\mathcal{G}/\mathcal{H}}\right)}=\Pi_{\hat{a}}+ 2f \alpha_{\hat{a}}+\mathcal{O}\left(\alpha \frac{\Pi^{2}}{f}+\alpha \frac{\Pi^{3}}{f^{2}}+\cdots\right)\,. \label{eq:shift}
\eea
Thus the NGBs $\Pi$ transform linearly under the subgroup ${\mathcal{H}}$ while non-linearly with the shift symmetry under the coset ${\mathcal{G}/\mathcal{H}}$.

Applying the above general construction to the electroweak symmetry breaking, the chiral symmetry breaking pattern ${\mathcal G} = SU(2)_L \times SU(2)_R$ to the custodial symmetry ${\mathcal H} = SU(2)_V$~\cite{Weinberg:1978kz,Gasser:1983yg,Gasser:1984gg} is identified and the the Goldstone matrix $\Omega$ can be written in a block-diagonal form 
\bea
	 \Omega(\Pi) \equiv \left[ \baa {cc} \xi_L(\Pi) & 0 \\ 0 & \xi_R(\Pi) \eaa\right]
	   = \left[\baa {cc} \exp\left[\frac{i}{2v}\phi_I \tau^I \right] & 0 \\ 0 & \exp\left[ -\frac{i}{2v}\phi_I \tau^I \right] \eaa\right], \quad {\textrm{with}} \quad \xi_L(\Pi) = \xi_R(\Pi)^\dagger \equiv u(\Pi)\,.
\eea
with the NGBs $\phi^I$ ($I = 1, 2, 3$). 
Identifying $\mathfrak{g} = {\textrm{diag}} (\mathfrak{g}_L(\alpha_L), \mathfrak{g}_R(\alpha_R))$ along with $\mathfrak{g}_{L,R}(\alpha_{L,R}) = \exp(\frac{i}{2}\alpha_{L,R}^I \tau_I)$, the Goldstone matrix transformation can be further reduced to 
\bea
	\xi_L(\Pi) & \rightarrow &\xi_L(\Pi^{(\mathfrak{g}_L)}) =  \mathfrak{g}_L \xi(\Pi)  \mathfrak{h}^{-1}(\Pi; \mathfrak{g}_L, \mathfrak{g}_R), \\
	\xi_R(\Pi) & \rightarrow &\xi_R(\Pi^{(\mathfrak{g}_R)}) =  \mathfrak{g}_R \xi(\Pi)  \mathfrak{h}^{-1}(\Pi; \mathfrak{g}_L, \mathfrak{g}_R).
\eea
For the symmetric coset~\footnote{The coset space ${\mathcal G}/{\mathcal H}$ is identified to be symmetric if there exists an automorphism or ``grading'' symmetry ${\mathcal R}$, under which the broken generators change sign
\bea
\mathcal{R} :
\left\{\begin{array}{l}{T_{a} \rightarrow+T_{a}} \\ 
{T_{\hat{a}} \rightarrow-T_{\hat{a}}} 
\end{array}\right.\,.  \label{R_automorphism}
\eea
For chiral symmetry breaking, the automorphism corresponds to the parity operator, whose eigenvalue are $\pm 1$ for vector and axial-vector generators, respectively. 
}, 
the compensating transformation $\mathfrak{h}^{-1}(\Pi; \mathfrak{g}_L, \mathfrak{g}_R)$ is the same for the chiral relations since they are related by the automorphism symmetry. 
Thus combining the two chiral relations would remove the $\mathfrak{h}$ and obtain a simpler form with the canonical Goldstone matrix
\bea
	\bfu (\Pi) \equiv \xi_R(\Pi) \xi_L^\dagger(\Pi) = u^2(\Pi)= \exp\left[\frac{i}{v}\phi_I \tau^I \right] \longrightarrow \mathfrak{g}_L  \bfu (\Pi) \mathfrak{g}_R^\dagger \,.
\eea
Promoting the global symmetry $G$ to the local one, we introduce the auxiliary $SU(2)_L$ and $SU(2)_R$ matrix gauge fields $\hat{W}_\mu$ and $\hat{B}_\mu$. The covariant derivative of the $\bfu$ is defined as
\beqa
D_{\mu} \bfu =\partial_{\mu}\bfu -i  \hat{W}^{\mu} \bfu + i \bfu  \hat{B}^{\mu}, \label{Dm_Sigma-Ab-AbR}
\eeqa
where $\hat{W}^{\mu}$ and $\hat{B}^{\mu}$ respects the transformation above
\bea
\hat{W}^{\mu} \longrightarrow \mathfrak{g}_{L} \hat{W}^{\mu} \mathfrak{g}_{L}^{\dagger}+i \mathfrak{g}_{L} \partial^{\mu} \mathfrak{g}_{L}^{\dagger}, \quad \hat{B}^{\mu}   \longrightarrow \mathfrak{g}_{R} \hat{B}^{\mu} \mathfrak{g}_{R}^{\dagger}+i \mathfrak{g}_{R} \partial^{\mu} \mathfrak{g}_{R}^{\dagger}\,,
\eea 
and the field strength tensors
\bea
\hat{W}_{\mu \nu} &=\partial_{\mu} \hat{W}_{\nu}-\partial_{\nu} \hat{W}_{\mu}-i\left[\hat{W}_{\mu}, \hat{W}_{\nu}\right] \quad \longrightarrow \quad \mathfrak{g}_{L} \hat{W}_{\mu \nu} \mathfrak{g}_{L}^{\dagger} \,,\\ \hat{B}_{\mu \nu} &=\partial_{\mu} \hat{B}_{\nu}-\partial_{\nu} \hat{B}_{\mu}-i\left[\hat{B}_{\mu}, \hat{B}_{\nu}\right] \quad \longrightarrow \quad \mathfrak{g}_{R} \hat{B}_{\mu \nu} \mathfrak{g}_{R}^{\dagger} \,.
\eea
The gauge fields are incorporated in the same way as the left and right sources in the ChPT, and the SM gauge fields are recovered through the identification
\bea
\hat{W}^{\mu}=-g \frac{\vec{\sigma}}{2} \vec{W}^{\mu}, \quad \hat{B}^{\mu}=-\mathcal{D}^{\prime} \frac{\sigma_{3}}{2} B^{\mu} = -\mathcal{D}^{\prime} {\mathcal T}_R B^{\mu}\,,
\eea
which explicitly breaks the $SU(2)_R$ symmetry group while preserving the $SU(2)_L \times U(1)_Y$ gauge symmetry. 
Here the right-handed spurion ${\mathcal T}_R = \frac{\sigma^3}{2}$ is introduced to represent the explicit breaking of custodial symmetry.

The SM fermion multiplets $\psi_{L, R} = P_{L, R} \psi$ can be incorporated into the $SU(2)_L$ and $SU(2)_R$ doublets respectively:~\footnote{Note that the right-handed neutrinos can be naturally included in the doublet $L_R$. Removing right-handed neutrino degree of freedom
\bea
L_R = \left(\begin{array}{c} 0 \\e_R\end{array}\right)
\eea
would recover the SM spectrum.  
}
\bea
Q_L &=& \left(\begin{array}{c}u_L\\d_L\end{array}\right) \quad \rightarrow \quad \mathfrak{g}_{L} Q_L\,, \quad \quad
Q_R = \left(\begin{array}{c}u_R\\d_R\end{array}\right) \quad \rightarrow \quad \mathfrak{g}_{R} Q_R\,, \\ 
L_L &=& \left(\begin{array}{c}\nu_L\\e_L\end{array}\right) \quad \rightarrow \quad \mathfrak{g}_{L} L_L\,, \quad \quad
L_R = \left(\begin{array}{c}\nu_R\\e_R\end{array}\right) \quad \rightarrow \quad \mathfrak{g}_{R} L_R\,,
\eea
and the fermion covariant derivative transformation takes the form $D_\mu \psi_{L, R}  \quad \rightarrow \quad \mathfrak{g}_{L, R} \psi_{L, R}$. Since the right-handed fermions are the $SU(2)_R$ doublets, the $U(1)_Y$ symmetry of fermions is promoted to the $U(1)_X$ symmetry, where $X=(B-L)/2$ is half of the baryon number $B$ minus the lepton number $L$.
%
The fermion masses are incorporated through the Yukawa terms that explicitly breaks the symmetry via the right-handed spurion field
\bea
	v \overline{\psi_L} \bfu(\Pi) {\mathcal Y}_R  \psi_R + h.c. \quad {\textrm{with}} \quad {\mathcal Y}_R  \quad \longrightarrow \quad \mathfrak{g}_{R} {\mathcal Y}_R \mathfrak{g}^\dagger_{R}\,,
\eea
In terms of the spurion field ${\mathcal T}_R = \sigma^3/2$, the Yukawa couplings take the form
\bea
{\mathcal Y}_R^Q = \frac{1 }{2}(y_u  + y_d)  + {\mathcal T}_R  (y_u  - y_d), \quad \mathcal{Y}^L_R =
\frac{1 }{2}(y_e  + y_v)  + {\mathcal T}_R  (y_e  - y_v)\,,
\eea
where $y_v = 0$ if no right-handed neutrinos.

In summary, the following building blocks are introduced to keep the Lagrangian formally invariant under the global symmetry $\mathcal{G}$ transformation:
\bea
	\bfu, \quad\psi_L, \quad\psi_R, \quad \hat{W}_{\mu \nu}, \quad\hat{B}_{\mu \nu}, \quad\hat{G}_{\mu \nu}, \quad {\mathcal T}_R, \quad{\mathcal Y}_R. 
\eea
It is also customary to take different building blocks transformed under the unbroken group $\mathcal{H}$~\cite{Pich:2015kwa,Pich:2016lew,Pich:2018ltt} by dressing the above building blocks with the $u(\Pi)$, such that
\bea
u_{\mu} = i u\left(D_{\mu} U\right)^{\dagger} u=-i u^{\dagger} D_{\mu} U u^{\dagger} \quad &\longrightarrow &  \quad \mathfrak{g}_{\mathcal H} u_{\mu} \mathfrak{g}_{\mathcal H}^\dagger \\ 
f_{\pm}^{\mu \nu} = u^{\dagger} \hat{W}^{\mu \nu} u \pm u \hat{B}^{\mu \nu} u^{\dagger} \quad &\longrightarrow &  \quad 
\mathfrak{g}_{\mathcal H} f_{\pm}^{\mu \nu} \mathfrak{g}_{\mathcal H}^\dagger\\ 
\mathcal{T} = u \mathcal{T}_{R} u^{\dagger} \quad &\longrightarrow &  \quad \mathfrak{g}_{\mathcal H} \mathcal{T} \mathfrak{g}_{\mathcal H}^\dagger\\
 u^{\dagger} \psi_{L} \quad &\longrightarrow &  \quad \mathfrak{g}_{\mathcal H} u^{\dagger} \psi_{L}\\ 
 u \psi_{R} \quad &\longrightarrow &  \quad \mathfrak{g}_{\mathcal H} u \psi_{R}\\ 
\mathcal{Y} = u \mathcal{Y}_{R} u^{\dagger}  \quad &\longrightarrow & \quad \mathfrak{g}_{\mathcal H} \mathcal{Y} \mathfrak{g}_{\mathcal H}^\dagger
\eea
very similar to the building blocks in the QCD chiral perturbation theory. 
It is also very convenient to take the form of the building blocks transformed under the group $SU(2)_L$~\cite{Grinstein:2007iv,Buchalla:2012qq,Buchalla:2013rka,Buchalla:2013eza,Gavela:2014vra,Krause:2018cwe,Alonso:2012px,Brivio:2013pma,Brivio:2016fzo,Merlo:2016prs} such that
\bea
	\bfv_\mu(x) = i\bfu(x) D_\mu \bfu(x)^\dagger, \quad  &\longrightarrow & \quad \mathfrak{g}_L \bfv_\mu \mathfrak{g}_L^\dagger \\
	\hat{W}_{\mu \nu}  \quad  &\longrightarrow & \quad \mathfrak{g}_L \hat{W}_{\mu \nu} \mathfrak{g}_L^\dagger \\
	\hat{B}_{\mu \nu} \quad  &\longrightarrow & \quad  \hat{B}_{\mu \nu} \\
	\mathbf{T} = \bfu \mathcal{T}_{R} \bfu^{\dagger} \quad &\longrightarrow & \quad \mathfrak{g}_{L} \mathbf{T} \mathfrak{g}_{L}^\dagger\\
	\psi_{L}  \quad &\longrightarrow & \quad \mathfrak{g}_{L} \psi_{L} \\ 
	 \bfu \psi_{R} \quad &\longrightarrow & \quad \mathfrak{g}_{L} \bfu \psi_{R}\\ 
	\mathbf{Y} = \bfu \mathcal{Y}_{R} \bfu^{\dagger}  \quad &\longrightarrow & \quad \mathfrak{g}_{L} \mathbf{Y} \mathfrak{g}_{L}^\dagger
\eea 
In this work, we would like to adapt the building blocks transforming under the $SU(2)_L$ symmetry, although different choices would give rise to the equivalent operator sets.

\subsection{Chiral Lagrangian and Power Counting}
\label{subsec:power}


In terms of the above building blocks, the effective chiral Lagrangian is organized as an infinite series expansion in powers of momenta in the low energy
\be
\mathcal{L} = \sum_{d_\chi=2}\mathcal{L}_{d_\chi} = \mathcal{L}_2 + \text{NLO terms} + \text{terms of higher order}\,,
\ee
where the effective operators are counted by the chiral dimension $d_\chi$ instead of the canonical dimension $d_c$.

The structure of the LO Lagrangian determines the power-counting rules of the electroweak chiral Lagrangian. Let us take
the LO Lagrangian of the HEFT as
\begin{align}
\mathcal{L}_2 =&-\frac{1}{4} {\textrm Tr}\left( G^a_{\mu\nu}G^{a\mu\nu}\right) - \frac{1}{4}\lra{W_{\mu\nu} W^{\mu\nu}} - \frac{1}{4}B_{\mu\nu}B^{\mu\nu} -\frac{g_s^2}{16\pi^2}\theta_s {\textrm Tr}\left( G^a_{\mu\nu}\tilde{G}^{a\mu\nu} \right) \notag \\
& + \frac{1}{2}\partial_\mu h \partial^\mu h -V(h) -\frac{v^2}{4}\lra{\bfv_\mu \bfv^\mu}\mathcal{F}_C(h) - \frac{v^{2}}{4} \lra{\bft \bfv_{\mu}}\lra{\bft \bfv^{\mu}} \mathcal{F}_{T}(h) \notag \\
& +i\bar{Q}_L\slashed{D} Q_L + i\bar{Q}_R\slashed{D}Q_R + i\bar{L}_L\slashed{D}L_L + i\bar{L}_R \slashed{D}L_R \notag \\
& -\frac{v}{\sqrt{2}}(\bar{Q}_L \bfu \mathcal{Y}_Q(h) Q_R) + h.c.) -\frac{v}{\sqrt{2}}(\bar{L}_L \bfu \mathcal{Y}_L(h)L_R + h.c.)\,,
\end{align}
where $\langle\dots\rangle \equiv \textrm{Tr}\left( \dots \right)$ represents the $SU(2)_L$ trace. 
The first line describes the dynamic terms of the gauge bosons and the theta term, and the second line contains the Higgs dynamic terms, the Higgs potential, the NGBs dynamic term, and the custodial symmetry breaking NGB term. The third and forth lines contain dynamic terms and the mass terms of fermions. Since the Higgs boson belongs to the singlet under the global symmetry, the Higgs potential should be polynomial function of the $h$ field  that
\bea
V(h)  \sim 1 + a_1\frac{h}{v} + a_2\frac{h^2}{v^2} + O(\frac{h^3}{v^3}).
\eea
Note that the canonical dimension is compensated by the electroweak scale $v$. 
the dimensionless function $\mathcal{F}_{C,T}(h)$ of the Higgs field $h$ appearing in the NGBs dynamic term takes the form that
\be
\mathcal{F}(h) = 1 + b_1\frac{h}{v} + b_1\frac{h^2}{v^2} + O(\frac{h^3}{v^3})\,. \label{eq:calf}
\ee
This function does not appear in the dynamic terms of fermions and the singlet Higgs $h$ because they can be removed by field redefinition or equation of motion~\cite{Brivio:2016fzo}.

As in~\cite{Pich:2016lew,Krause:2018cwe,Pich:2018ltt}, the usual custodial
breaking term $\lra{\bft \bfv_{\mu}}\lra{\bft \bfv^{\mu}}$ sometimes is classified into the NLO expansion in the chiral expansion since it is not present in the SM at the tree level. However, this would miss certain non-decoupling scenarios within the weak dynamics. It is possible that the custodial symmetry breaking could be triggered by new physics at the electroweak scale, such as new VEV from triplet scalar, the tadpole-induced Higgs, etc. Although the custodial breaking effect is limited to be small from the electroweak precision data, to incorporate non-decoupling new physics scenarios into the HEFT, we put it to be at the LO and adapt the standard chiral dimension counting on the spurion fields as shown in the following~\footnote{Adapting the spurion with no chiral dimension would give rise to the operator sets with more LO + NLO operators than the ones with the spurion with chiral dimensions, for example, in Ref.~\cite{Pich:2016lew,Krause:2018cwe,Pich:2018ltt}. }.

The power-counting scheme of the HEFT is similar to the power-counting rules in the chiral perturbation theory~\cite{Weinberg:1978kz,Gasser:1983yg,Gasser:1984gg}, with certain improvements~\cite{Weinberg:1978kz,Buchalla:2016sop,Manohar:1983md,Hirn:2005fr,Buchalla:2013eza,Buchalla:2012qq,Krause:2018cwe,Pich:2015kwa,Pich:2016lew}. To be consistent, we define the LO Lagrangian of the HEFT carries the chiral dimension 2, thus all the building block's chiral dimensions are determined. First the NGBs matrix $\bfu$ is determined to be chiral-dimensionless~\footnote{Similar to the Higgs boson, its canonical dimension is compensated by the electroweak scale. }, while the covariant derivative $D_\mu$ carries the chiral dimension 1, thus $\bfv_\mu$ is of the chiral dimension 1, which implies that the dynamic term of NGBs carries the chiral dimension 2. 

For the gauge bosons, the external gauge sources $\hat{W}^\mu$ and $\hat{B}^\mu$, appeared in the covariant derivative $D_\mu$, have the chiral dimension 1, and their corresponding field strength tensors have the chiral dimension 2. Since all the gauge boson masses $g v$ have chiral dimension 1, the gauge coupling constant $g,g'$ also carry the chiral dimension 1 while the gauge boson fields $W^\mu$ and $B^\mu$ are chiral-dimensionless.


The custodial symmetry breaking term is organized into the LO Lagrangian and thus the spurion $\mathcal T$ carries no chiral dimension. The physical Higgs $h$ always appears with the VEV in the denominator as the compensator and thus has no chiral dimension as well, while the fermions have the chiral dimension $1/2$. In particular, since the fermion masses $y v$ have chiral dimension 1, the Yukawa couplings $\caly$ would give rise to the chiral dimension 1.

The power counting rules on the chiral dimension is also consistent with the loop expansion \cite{Buchalla:2013eza,Pich:2016lew,Krause:2018cwe,Buchalla:2012qq,Hirn:2005fr}. The above power counting on the gauge and Yukawa couplings implies that 
\bea 
\frac{p^{2}}{16 \pi^{2} \mathbf{v}^{2}} \sim \frac{g^2}{(4 \pi)^{2}}, \,\frac{y^2}{(4 \pi)^{2}}, \, \frac{\lambda}{(4 \pi)^{2}} \ll 1\,.
\eea
Similar argument applies to the case that all the SM gauge bosons and fermions are weakly coupled to a strong sector. 
Thus the types of operators containing the fermion bilinear $(\bar{\psi} \psi)^n$ and/or the field strength $X^n U h$ would arise at NLO where they come with explicit factors of the couplings $y^n$ and/or $g^n$. Each fermion bilinear and each field strength tensor would shift the chiral dimension by 1.

\begin{table}
		\centering
		\begin{tabular}{c|c|c|c}
				\hline
				 Classes & $\mathcal{N}_{\text{type}}$ & $\mathcal{N}_{\text{term}}$ & $\mathcal{N}_{\text{operator}}$ \\
				\hline
				$UhD^4$ & $3+6+0+0$ & 15 & 15 \\
				$X^2Uh$ & $6+4+0+0$ & 10 & 10 \\
				$XUhD^2$ & $2+6+0+0$ & 8 & 8 \\
				$X^3$ & $4+2+0+0$ & 6 & 6 \\
				$\psi^2UhD$ & $4+8+0+0$ & 13(16) & $13{n_f}^2\quad(16{n_f}^2)$ \\
				$\psi^2UhD^2$ & $6+10+0+0$ & 60(80) & $60{n_f}^2\quad(80{n_f}^2)$ \\
				$\psi^2UhX$ & $7+7+0+0$ & 22(28) & $22{n_f}^2\quad(28{n_f}^2)$ \\
				$\psi^4$ & $12+24+4+8$ & 117(160) & $\frac{1}{4}{n_f}^2(31-6{n_f}+335{n_f}^2)\quad({n_f}^2(9-2{n_f}+125{n_f}^2))$  \\
				\hline
				\multirow{2}{*}{\text{Total}} & \multirow{2}{*}{123} & \multirow{2}{*}{261(313)} & $\frac{335 {n_f}^4}{4}-\frac{3 {n_f}^3}{2}+\frac{411 {n_f}^2}{4}+39\quad(39+133{n_f}^2-2{n_f}^2-2{n_f}^3+125{n_f}^4)$ \\
& & & $\mathcal{N}_{\text{operatrs}}(n_f=1)= 224(295),\quad \mathcal{N}_{\text{operatrs}}(n_f=3)= 7704(11307)$ \\
				\hline
		\end{tabular}
		\caption{We present the complete statistics of the NLO operators. The type of operators are separated into the four categories ($C$-$B$, $C\!\!\!\!/$-$B$, $B$-$C\!\!\!\!/$, $C\!\!\!\!/$-$B\!\!\!\!/$). The numbers of terms and operators are also listed for the SM without (with) the right-handed sterile neutrinos. }
		\label{tab:boson}
\end{table}

\begin{table}[htb]
    \centering
    \begin{center}
    \begin{tabular}{c|c|c|c|c}
\hline
$\psi^4$ & sub-types & Number & $n_f=1$ & $n_f=3$\\
\hline
\multirow{4}{*}{$(\overline{L}L)^2$} 
&
\begin{tabular}{lr}
${L_L^\dagger}^2{L_R}^2+h.c.$: & ${n_f}^2({n_f}^2+1)$
\end{tabular}
& \multirow{4}{*}{$\frac{1}{4}{n_f}^2(19{n_f}^2+6{n_f}+7)$}
& \multirow{4}{*}{8} & \multirow{4}{*}{441}\\
& 
\begin{tabular}{lr}
${L_L}^2{L_L^\dagger}^2$: & $\frac{1}{2}{n_f}^2({3{n_f}^2+2{n_f}+1})$
\end{tabular} & & \\
& 
\begin{tabular}{lr}
${L_R}^2{L_R^\dagger}^2$: & $\frac{1}{4}{n_f}^2({{n_f}^2+2{n_f}+1})$
\end{tabular} & & \\
& 
\begin{tabular}{lr}
${L_L}{L_L^\dagger}{L_R}{L_R^\dagger}$: & $2{n_f}^4$
\end{tabular} & & \\
\hline
\multirow{6}{*}{$(\overline{L}L)(\overline{Q}Q)$} 
&
\begin{tabular}{lr}
${Q_L^\dagger}{Q_R}{L^\dagger_L}{L_R} + h.c.$: & $12 {n_f}^4$
\end{tabular} 
& \multirow{6}{*}{$34{n_f}^4$}
& \multirow{6}{*}{34} & \multirow{6}{*}{2754}\\
&
\begin{tabular}{lr}
${Q_L^\dagger}{Q_L}{L_L^\dagger}{L_L}$: & $6{n_f}^4$
\end{tabular} & & \\
&
\begin{tabular}{lr}
${Q_R^\dagger}{Q_R}{L_L^\dagger}{L_L}$: & $6{n_f}^4$
\end{tabular} & & \\
&
\begin{tabular}{lr}
${Q_L^\dagger}{Q_L}{L_R^\dagger}{L_R}$: & $2{n_f}^4$
\end{tabular} & & \\
&
\begin{tabular}{lr}
${Q_R^\dagger}{Q_R}{L_R^\dagger}{L_R}$: & $2{n_f}^4$
\end{tabular} & & \\
&
\begin{tabular}{lr}
${Q_L^\dagger}{Q_R}{L_L^\dagger}{L_R} + h.c.$: & $6{n_f}^4$
\end{tabular} & & \\
\hline
\multirow{4}{*}{$(\overline{Q}Q)^2$} 
& 
\begin{tabular}{lr}
${Q_L^\dagger}^2{Q_R}^2+h.c.$: & $4{n_f}^2(3{n_f}^2+1)$
\end{tabular}
& \multirow{4}{*}{${n_f}^2(30{n_f}^2+6)$}
& \multirow{4}{*}{36} & \multirow{4}{*}{2484}\\
& 
\begin{tabular}{lr}
${Q_L^\dagger}^2{Q_L}^2$: & ${n_f}^2(3{n_f}+1)$
\end{tabular} & & \\
& 
\begin{tabular}{lr}
${Q_R^\dagger}^2{Q_R}^2$: & ${n_f}^2(3{n_f}+1)$
\end{tabular} & & \\
& 
\begin{tabular}{lr}
${Q_L^\dagger}{Q_L}{Q_R^\dagger}{Q_R}$: & $12{n_f}^4$
\end{tabular} & & \\
\hline
\multirow{4}{*}{$Q^3L$} 
& 
\begin{tabular}{lr}
${L_L}{Q_L}^3 + h.c. $: & $4{n_f}^4$
\end{tabular}
& \multirow{4}{*}{${n_f}^2(15{n_f}^2-3{n_f})$}
& \multirow{4}{*}{12} & \multirow{4}{*}{1134}\\
&
\begin{tabular}{lr}
${L_L}{Q_L}{Q_R}^2 + h.c. $: & $2{n_f}^3(3{n_f}-1)$
\end{tabular} & & \\
&
\begin{tabular}{lr}
${L_R}{Q_R}{Q_L}^2 + h.c. $: & ${n_f}^3(3{n_f}-1)$
\end{tabular} & & \\
&
\begin{tabular}{lr}
${L_R}{Q_R}^3 + h.c. $: & $2{n_f}^4$
\end{tabular} & & \\
\hline
    \end{tabular}
\end{center}
    \caption{The numbers of the independent operators in class $\psi^4$ without right-handed neutrino in detail, where $(\overline{L}L)^2$ is the pure lepton sector, $(\overline{L}L)(\overline{Q}Q)$ is the mixed quark-lepton sector, $(\overline{Q}Q)^2$ is the pure quark sector, and $Q^3L$ is the baryon-number-violating sector.}
    \label{tab:psi4level1}
\end{table}

Based on the above, a general term in the HEFT can be denoted by
\bea
{\kappa}^{k_i} \psi^{F_i} X^{V_i}_{\mu\nu} \bfu h D^{d_i}\,,
\eea
where there are a fixed number $k_i$ of the gauge or Yukawa couplings $\kappa$, $F_i$ fermion fields $\psi$, $V_i$ field-strength tensor $X_{\mu\nu}$, $d_i$ covariant derivatives $D$, and an arbitrary number of both the NGBs $\bfu$ and the Higgs boson $h$. 
The total chiral dimension of the term determines the loop order $L_i$
\bea
d_\chi = d_i + k_i + \frac{F_i}{2} + V_i = 2 L_i + 2. 
\eea
In this work, we focus on the NLO terms, which include effective operators with the chiral dimension 4. We classify them into the bosonic sector and the fermionic sector, with the operator types of each~\footnote{The class of triple gauge bosons is of the chiral dimension 6 according to the power-counting rules presented in this subsection, but it are listed here for the convenience of comparing with other literature.}
\bea
&\text{boson sector: }& UhD^4,\quad X^2 Uh,\quad XUhD^2,\quad X^3\,, \\
&\text{fermion sector: }& \psi^2 UhD,\quad \psi^2 UhD^2,\quad \psi^2Uh X,\quad \psi^4\,.
\eea
Operators of each type are presented in sec.~\ref{sec:boson} and sec.~\ref{sec:fermion}, with three generation fermions included. Based on the custodial symmetry and the baryon number symmetry, the types of operators can be further separated into the four categories:
\bit
\item custodial symmetry preserving ($C$), baryon number preserving ($B$); 
\item custodial symmetry violating ($C\!\!\!\!/$), baryon number preserving ($B$); 
\item custodial symmetry preserving ($C$), baryon number violating ($B\!\!\!\!/$); 
\item custodial symmetry violating ($C\!\!\!\!/$), baryon number violating ($B\!\!\!\!/$).
\eit
Furthermore, we also list the operators without (with) the right-handed sterile neutrinos. 
The numbers of operators in each type are listed in table \ref{tab:boson}. The specific operators are listed in sec.~\ref{sec:boson} and sec.~\ref{sec:fermion}, of which the comparison  with the operators in~\cite{Buchalla:2013rka,Brivio:2016fzo,Merlo:2016prs} is presented in Appendix~\ref{app:operatorlist}, and the on-shell amplitude form of these operators are given in Appendix~\ref{app:amplitude}. Although the one-flavor operators has been presented in literature such as~\cite{Grinstein:2007iv,Buchalla:2012qq,Buchalla:2013rka,Buchalla:2013eza,Gavela:2014vra,Krause:2018cwe,Alonso:2012px,Brivio:2013pma,Brivio:2016fzo,Pich:2015kwa,Pich:2016lew,Pich:2018ltt,Merlo:2016prs}, there are still several operators missing as discussed below. 
In particular, the numbers of the independent operators in the four-fermion sector are presented in table~\ref{tab:psi4level1}, which, together with the numbers of other classes in table~\ref{tab:boson}, are consistent with the counting result via the Hilbert series obtained in Ref.~\cite{Graf:2022rco,Sun:2022aag}.

\section{HEFT Operator Bases}
\label{sec:operator}

Based on the section \ref{sec:field}, we obtain the building blocks and their group representations under the Lorentz and internal symmetries as shown in Table~\ref{tab:buildingblocks}. 
With these building blocks, we can utilize the group theoretic techniques developed in Ref.~\cite{Li:2020xlh,Li:2020gnx,Li:2022tec,Li:2020zfq}, the so-called Young tensor method, to write down the Lorentz, gauge and flavor structures of the EFT operators. 
At the same time, if an operator involves in the NGBs, the building block $\bfv^\mu$, such operator should satisfy additional condition on the Lorentz structure: the shift symmetry in Eq.~\eqref{eq:shift}, correspondingly, the Adler zero condition on the contact on-shell amplitude. Let us expand the nonlinear field $\bfv_\mu$ as
\bea
\bfv_\mu = i \bfu D_\mu \bfu^\dagger = - \frac{1}{v} \left[ D_\mu  \Pi   + {\mathcal O}\left(D_\mu \Pi \cdot \frac{\Pi^2}{v^2}\right) \right]\,,
\eea
from which we see that the leading term  $D_\mu \Pi$ would also respect the shift symmetry. 
From the soft recursion relation for the NGBs~\cite{Cheung:2014dqa, Cheung:2015ota, Low:2019ynd, Dai:2020cpk}, operators involving in $D_\mu \Pi$ could recover all physical effects for the one involving in $\bfv_\mu$. Thus from the on-shell point of view, the two descriptions should be equivalent. 
Therefore, in this section we would like to use the $D_\mu \Pi$ as the building block and build the one-to-one correspondence between them in the operator
\beqa
\bfv_\mu\, \, \leftrightarrow\,\, D_\mu \Pi \equiv D_\mu \phi^I \tau^I \,,
\eeqa
and from now on we use the indice $I, J$ to denote the $SU(2)$ indices.
The operator involving in $D_\mu \Pi$ would also respect the Adler zero condition.
In the following, we will simply review this Young tensor method and present some examples on the construction of the HEFT operators.  

\begin{table}
\begin{center}
\begin{tabular}{|c|c|c|c|c|c|}
\hline
building blocks & spinor-helicity & Lorentz group & $SU(2)_L$ & $SU(3)_C$ & $d_\chi$ \\
\hline
$L_L$ & ${L_L}_{\alpha}$ & $(\frac{1}{2},0)$ & Fundamental & Singlet & 1 \\
\hline
$L_R$ & ${L_R}^{\dot{\alpha}}$ & $(0,\frac{1}{2})$ & Fundamental & Singlet & 1 \\
\hline
$Q_L$ & ${Q_L}_{\alpha}$ & $(\frac{1}{2},0)$ & Fundamental & Fundamental & 1 \\
\hline
$Q_R$ & ${Q_R}^{\dot{\alpha}}$ & $(0,\frac{1}{2})$ & Fundamental & Fundamental & 1 \\
\hline
$W_L$ & ${W_L}^I_{\alpha\beta}\tau^I$ & $(1,0)$ & Adjoint & Singlet & 2 \\
\hline
$W_R$ & ${W_R}^{I\dot{\alpha}\dot{\beta}}\tau^I$ & $(0,1)$ & Adjoint & Singlet & 2 \\
\hline
$G_L$ & ${G_L}_{\alpha\beta}$ & $(1,0)$ & Singlet & Adjoint & 2 \\
\hline
$G_R$ & ${G_R}^{\dot{\alpha}\dot{\beta}}$ & $(0,1)$ & Singlet & Adjoint & 2 \\
\hline
$B_L$ & ${B_L}_{\alpha\beta}$ & $(1,0)$ & Singlet & Singlet & 2 \\
\hline
$B_R$ & ${B_R}^{\dot{\alpha}\dot{\beta}}$ & $(0,1)$ & Singlet & Singlet & 2 \\
\hline
$\bfv^\mu\sim D^\mu\Pi$ & ${(D\phi^I)}_{\dot{\alpha}\beta}\tau^I$ & $(\frac{1}{2},\frac{1}{2})$ & Adjoint & Singlet & 1 \\
\hline
$D^\mu$ & $D_{\alpha\dot{\beta}}$ & $(\frac{1}{2},\frac{1}{2})$ & Singlet & Singlet & 1 \\
\hline
$\bft$ & $\bft^T\tau^I$ & $(0,0)$ & Adjoint & Singlet & 0 \\
\hline
\end{tabular}
\end{center}
\caption{The building blocks of the HEFT and their group representations under the Lorentz and internal symmetries. From the soft recursion relation of the NGBs, the building block $\bfv^\mu$ can be replaced by the linearized one $D^\mu\Pi$. }
\label{tab:buildingblocks}
\end{table}

\subsection{Operator Amplitude Correspondence}
\label{subsec:corres}

An effective operator can be decomposed into the Lorentz, gauge and flavor structures, each of which should be the singlet of the Lorentz group and the gauge groups. All the independent Lorentz (gauge) structures form a basis of a linear space, called the Lorentz (gauge) basis.

There exists a correspondence relation between the effective operators and the on-shell amplitudes. To see that, considering a general building block composed by a field of any helicity $\Phi$ and a number of derivatives $D$, it takes the form that
\beq
(D^{r-|h|}\Phi)_{\alpha^{(1)}\dots\alpha^{(r-h)}\dot{\alpha}^{(1)}\dots\dot{\alpha}^{(\alpha+h)}}\equiv \left\{\begin{array}{l}
D_{\alpha^{(1)}\dot{\alpha}^{(1)}}\dots D_{\alpha^{(r+h)}\dot{\alpha}^{(r+h)}}\Phi_{\alpha^{(r+h+1)}\dots\alpha^{(r-h)}},\quad h<0\\
D_{\alpha^{(1)}\dot{\alpha}^{(1)}}\dots D_{\alpha^{(r+h)}\dot{\alpha}^{(r+h)}}\Phi_{\dot{\alpha}^{(r-h+1)}\dots\dot{\alpha}^{(r+h)}},\quad h>0
\end{array}\right.\,, \label{eq:buildingblock}
\eeq
where $h$ is the helicity of the particle and $r$ is the half number of the spinor indices of this building block, by which the number of derivative $n_D=r-|h|$ is completely determined.
Since all operators are Lorentz scalars, all the spinor indices appearing in the effective operators should be contracted by the tensor $\epsilon_{\alpha\beta}$ or $\tilde{\epsilon}_{\dot{\alpha}\dot{\beta}}$. The numbers of needed $\epsilon,\tilde{\epsilon}$ are $n=(r-h/2)$ and $\tilde{n}=(r+h)/2$, respectively. Thus the Lorentz part of a general effective operator involving in several building blocks such as that in Eq.~\eqref{eq:buildingblock} takes the form that
\beq
\label{eq:operator}
\calo=\epsilon^{\otimes n}\tilde{\epsilon}^{\otimes\tilde{n}}\prod_i(D^{r_i-|h_i|})\Phi_i\,,
\eeq

On the other hand, a general light-like momentum can be expressed using the spinor helicity formalism as $
p^\mu_i=\lambda^\alpha_i\sigma^\mu_{\alpha\dot{\beta}}\tilde{\lambda}^{\dot{\beta}}_i$
up to a $U(1)$ little group transformation: 
\beq
\lambda_i\rightarrow e^{-i\varphi/2}\lambda_i,\quad \tilde{\lambda}_i\rightarrow e^{i\varphi/2}\tilde{\lambda}_i\,.
\eeq
where $\varphi$ denotes the $U(1)$ phase. Under this transformation, amplitudes $\calm$ transforms as $\calm\rightarrow e^{ih_i\varphi}\calm$ for the $i$-th particle. Thus a massless particle of helicity $h_i$ contributes the amplitude a factor $\lambda^{r_i-h_i}_i\tilde{\lambda}^{r_i+h_i}_i$, where $r_i$ is the half number of the spinor indices in the factor. Similarly, all spinor indices should be contracted by the $\epsilon / \tilde{\epsilon}$, and the numbers of them are the same with $n / \tilde{n}$ mentioned previously. Making the following identification,
\beq
\lambda_i^\alpha\lambda_{j\alpha}=\lra{ij}, \quad \tilde{\lambda}_{i\dot{\beta}}\tilde{\lambda}^{\dot{\beta}}_j=\lrs{ij}\,,
\eeq
we obtain that an amplitude involving in several particle $\Phi_i$ with the helicity $h_i$ takes the form that
\beq
\calm = \underbrace{\prod\lra{ij}}_{n}\underbrace{\prod\lrs{kl}}_{\tilde{n}} = \epsilon^{\otimes n}\tilde{\epsilon}^{\otimes \tilde{n}}\prod \lambda^{r_i-h_i}_i\tilde{\lambda}^{r_i+h_i}_i\,.
\eeq
Comparing with Eq.~\eqref{eq:operator}, it is natural to make the following translations,
\begin{align}
\lambda_i^{r_i\pm1}\tilde{\lambda}_i^{r_I\mp 1} &\Leftrightarrow D^{r_i-1}{X_{L/R}}_i\,, \notag\\
\lambda_i^{r_i\pm 1/2}\tilde{\lambda}_i^{r_i\mp 1/2} &\Leftrightarrow D^{r_i-1/2}\psi_i^{(\dagger)}\,, \notag\\
\lambda_i^{r_i}\tilde{\lambda}_i^{r_i} & \Leftrightarrow D^{r_i}\phi_i\,,
\end{align}
which provides the correspondence between the Lorentz part of the effective operators to contact on-shell amplitudes.

One advantage of this operator-amplitude correspondence is that the on-shell reduction techniques of the spinor helicity can be utilized to further eliminate the redundancies other than the equation of motion, such as the momentum conservation 
\beq
\sum_i\lra{li}\lrs{ik}=0
\eeq
as discussed in \cite{Li:2020gnx}. 
After the removal of all kinds of redundancies, the remaining independent Lorentz structures can be related to a set of so-called semi-standard Young tableaux (SSYTs). The shape of these Young tableaux shapes is determined by the field content of each type:

\begin{eqnarray}\label{eq:YD_shape}
\arraycolsep=0pt\def\arraystretch{1}
\rotatebox[]{90}{\text{$N-2$}}
	\left\{
	\begin{array}{cccccc}
		\yng(1,1) &\ \ldots{}&\ \yng(1,1)& \overmat{n}{\yng(1,1)&\ \ldots{}\  &\yng(1,1)} \\
		\vdotswithin{}& & \vdotswithin{}&&&\\
		\undermat{\tilde{n}}{\yng(1,1)\ &\ldots{}&\ \yng(1,1)} &&&
	\end{array}
	\right.\,,
	\\
	\nonumber 
\end{eqnarray}
where $N$ is the total number of fields. Such Young diagrams are called the primary Young diagrams and filling in numbers in such diagrams tells the independent Lorentz structures in each type of operators~\cite{Li:2020xlh,Li:2020gnx,Li:2022tec,Li:2020zfq,Henning:2019mcv,Henning:2019enq}. The filling of the primary Young diagrams follows the rule that numbers in every column increase and in every row weakly increase, and the number of each particle index is determined by
\beq
\#i = \frac{1}{2}n_D+\sum_{h_i>0}(|h_i|-2h_i),\quad i=1,2,\dots N\,,
\eeq
where $n_D$ is the number of derivatives of this type.

In practice, the above process is reversed to construct independent Lorentz structures: given a type of operators, we can directly fill in the primary Young diagram following the rule given above to get SSYTs, by which the independent and complete Lorentz basis are obtained, called the y-basis. The construction is discussed in detail in \cite{Li:2022tec}. 
The gauge structure also forms the gauge y-basis, which can be constructed by the similar Young tableau method, as discussion in \cite{Li:2022tec}. 
Combining the Lorentz and gauge basis together, general form of an effective operator is
\be
\label{op}
\mathcal{O} = \mathcal{T}\bigotimes_{n}\epsilon\bigotimes_{\tilde{n}}\tilde{\epsilon}\prod_i (D^{r_i-h_i})\Phi_i\,,
\ee
where $\mathcal{T}$ denotes the gauge factor and the rest corresponds to the on-shell amplitude as shown in Eq.~\ref{eq:operator}.
The y-basis by filling the primary Young diagrams are usually polynomials. For convenience, the m-basis is introduced by non-singular linear transformation of the y-basis~\cite{Li:2022tec}. In the m-basis, both the Lorentz and gauge structures are monomials, which constitute the monomial form of an operator. 

Besides, the correspondence between fields and spinors implies that all fermions in this work are represented by weyl spinors $\psi$, and the relation between them and Dirac spinors $\Psi$ are that
\be
\Psi_L = \left(\begin{array}{c}\psi_L\\0\end{array}\right),\quad \Psi_R = \left(\begin{array}{c}0\\\psi_R\end{array}\right),\quad \bar{\Psi}_L = \left(\begin{array}{cc}0,&\psi^\dagger_L\end{array}\right),\quad \bar{\Psi}_R = \left(\begin{array}{cc}\psi^\dagger_R,&0\end{array}\right)\,.
\ee
The transformation of bilinears between different forms of spinors can be found in Appendix \ref{app:converting} or Ref. \cite{Li:2020xlh}.

\subsection{Adler Zero Condition and Spurions}
\label{subsec:adler}

The above procedure on the operator construction is suitable for the generic Lorentz-invariant EFT with arbitrary internal symmetry up to arbitrary mass dimension. However, in the exceptional EFTs~\cite{Cheung:2016drk}, such as the theory containing the Goldstone boson and/or spurion, additional treatment on the Lorentz basis is needed. As mentioned in the beginning of this section, the building block $\bfv^\mu$ in the HEFT indicates the Goldstone nature of the longitudinal gauge bosons, satisfying the shift symmetry due to the existence of derivative in such building block. Similarly the building block $D_\mu \phi$ in effective operators respects the shift symmetry. From the operator amplitude correspondence, an operator involving the Goldstone boson corresponds to the on-shell amplitude satisfying the Adler zero condition in the soft momentum limit~\cite{Adler:1964um,Adler:1965ga, Low:2014nga, Low:2014oga, Cheung:2014dqa, Cheung:2015ota, Low:2019ynd, Dai:2020cpk}. The Adler zero condition~\cite{Adler:1964um,Adler:1965ga} states that the on-shell scattering amplitudes containing the NGBs must vanish when one external momentum $p$ of the NGBs is taken to be soft
\beqa
{\mathcal M}(p) \sim p \quad {\textrm{for}} \quad p \to 0.
\eeqa
Although this condition is trivially satisfied in the HEFT operator basis with the building block $\bfv^\mu$ and/or $D_\mu \phi$, the on-shell amplitude basis obtained in the above subsection does not automatically satisfy this condition, and thus we need to impose the Adler zero condition explicitly in the Lorentz sector for the on-shell amplitude involving in the Goldstone bosons. 
There are attempts \cite{Dai:2020cpk,minglei-adler} to enumerating independent operators in the chiral perturbation theory via imposing Adler zero condition on the corresponding amplitude basis for NGBs. In particular, the procedure in Ref.~\cite{minglei-adler} could be applied to on-shell amplitudes involving external sources such as fermions and gauge bosons in the HEFT. In the following, we present the procedure of imposing the Adler zero condition which is also shown in Ref.~\cite{minglei-adler}.




Let us consider a type of operators with $N$ particles, including at least one NGB. Based on the Young tensor method above, the Lorentz basis can be expressed as the $N$-point on-shell amplitudes  $\{\calb^{(N)}_i,i=1,2,\dots d_N\}$, where $d_N$ is the dimension of this Lorentz basis. In terms of such basis, any general amplitude takes the form that
\beq
\calm^{(N)} = \sum_{i=1}^{d_N} c_i\calb^{(N)}_i, 
\eeq
where $c_i$ are coefficients in the $\calb$ basis. If this amplitude satisfies the Adler zero condition, it vanishes when an external pion momentum $p_\pi$ becomes soft:
\beq
\calm^{(N)}( p_\pi \rightarrow 0) = 0 =\sum_{i=1}^{d_N}c_i\calb^{(N)}_i( p_\pi \rightarrow 0).
\label{eq:soft}
\eeq
Here $\calb^{(N)}_i( p_\pi \rightarrow 0)$ becomes $(N-1)$-point on-shell amplitudes $\calm^{(N-1)}_i \equiv \calb^{(N)}_i( p_\pi \rightarrow 0)$, which are generally redundant in the Lorentz basis without the pion particle. Defining the Lorentz basis of the $(N-1)$-point amplitudes $\{\calb^{(N-1)}_i,i=1,2,\dots d_{N-1}\}$, where $d_{N-1}$ is the dimension of this Lorentz basis, the amplitudes $\calm^{(N-1)}_i$can be expanded by this complete and independent basis,
\be
\calm^{(N-1)}_i \equiv \calb^{(N)}_i( p_\pi \rightarrow 0) = \sum_{j=1}^{d_{N-1}}b_{ij}\calb^{(N-1)}_j.\label{eq:soft2}
\ee
Furthermore, since any $(N-1)$-point amplitudes can be decomposed using the $N$-point basis $\{\calb^{(N)}_i,i=1,2,\dots d_N\}$, we can expand the $N-1$-point basis $\calb^{(N-1)}_j$ as
\be
\calb^{(N-1)}_j = \sum_{l=1}^{d_N}f_{jl}\calb^{(N)}_l.\label{eq:soft3}
\ee
Using the Eq.~\eqref{eq:soft2} and Eq.~\eqref{eq:soft3}, the Adler zero condition in Eq.~\eqref{eq:soft} becomes
\beq
0=\sum_{i=1}^{d_N}c_i(\sum_{j=1}^{d_{N-1}} b_{ij}(\sum_{l=1}^{d_N}f_{jl}\calb^{(N)}_l))=\sum_{l=1}^{d_N}(\sum_{l=i}^{d_N}c_i\mathcal{{K}}_{il})\calb^{(N)}_l\,,
\eeq
where we identify
\be
\mathcal{K}_{il} = \sum_{j=1}^{d_{N-1}}b_{ij}f_{jl}\,
\ee
is the product of the expansion matrice $b_{ij}$ and $f_{jl}$ in the Eq.~\eqref{eq:soft2} and Eq.~\eqref{eq:soft3}, respectively. Since the basis $\{\calb^{(N)}_i,i=1,2,\dots d_N\}$ are independent, this equation holds only if all the coefficients vanish,
\beq
0=\sum_{i=1}^{d_N}c_i\mathcal{K}_{ij},\quad (j=1,2,\dots d_N)\,. 
\eeq
This is a system of linear equations about $c_i$, whose solutions span the subspace satisfying the Adler zero condition, which constitute the amplitude basis with Goldstone bosons. 
%
In sum, for a type of operators, we first treat the NGBs as ordinary scalars and write down all the relevant on-shell amplitudes, then impose the Adler zero condition on such amplitudes to eliminate the unwanted Lorentz structures, and thus obtain the Lorentz basis.


Besides, the spurions are introduced to describe the symmetry breaking, in this case the breaking of the custodial/weak $SU(2)$ symmetry. They are treated as vacuum expectation values of a dynamical degree of freedom and thus do not enter the Lorentz sector. As mentioned previously, we are using the only spurion $\bft = \bfu\calt_R\bfu^\dagger$ in the adjoint representation $\mathbf{3}$ of the $SU(2)$ as the source of symmetry breaking. The spurions and the dynamical fields, such as the SM fields in this case, together form an $SU(2)$ singlet operator, in which one should prevent any self-contracted singlet factor that only consists of the spurions because they do not have physical effects. For example, consider an operator with two spurions, which form a $\mathbf{3}\otimes\mathbf{3}$ tensor that decomposes
\bea
\begin{array}{ccccccc}
\bft^I\bft^J &=& \bft^2\delta^{IJ} &+& \bft^{[I}\bft^{J]} &+& \bft^{(I}\bft^{J)}, \\
\mathbf{3}\otimes\mathbf{3} &=& \mathbf{1} &+& \mathbf{3} &+& \mathbf{5}.
\end{array}
\eea
The first term should be neglected since the factor $\bft^2$ is just an irrelevant constant although it is symmetric for the two spurions. The second term vanishes due to the anti-symmetry behavior of the two spurions~\footnote{This can also be seen from the identify
\be
    \epsilon^{IJK}\bft^I\bft^JA^K = \Tr(\bft\bft A) = \Tr(\bfu\frac{\sigma^3}{2}\bfu^\dagger\bfu\frac{\sigma^3}{2}\bfu^\dagger A) = \frac{1}{4}\Tr(A) = 0\,,
    \ee
    where $A$ is any building block of the adjoint representation of the $SU(2)$ group.
}, which will be carefully treated in the next section. Therefore, the only non-trivial term is the symmetric traceless tensor, indicating the dynamical fields form a 5 multiplet of the broken $SU(2)$ group. Indeed, in the decomposition of the $SU(2)$ adjoint representation product, for any number of the spurions, only the highest weight combination should be considered, which we pick out using the technique of the gauge j-basis developed in Ref.~\cite{Li:2022abx}.

\subsection{P-Basis Operator and Examples}
\label{subsec:basis}


The above discussion does not consider the operators involving in identical particles. Considering a type of operators containing $n$ identical particles $\phi$ with the flavor number $n_f$, we identify the operator as the n-rank tensor under the flavor group $\bigotimes_nSU(n_f)$. Based on the Schur-Weyl theorem, the representation space of this group can be divided into the subspaces furnishing the irreducible representations of the symmetry group $S_n$. Thus we re-organized the y-basis or m-basis to a new basis, in which operators satisfy the specific permutation symmetry, thus are called the p-basis.
The p-basis help us identify the operator satisfying the spin-statistics, for example, if the repeated particle's flavor number is 1, only operators of completely symmetrical under permutation in p-basis are physical. { If the p-basis contain operators with the mixed flavor symmetry such as ${\tiny\yng(2,1)}$, the irreducible subspace of $\bigotimes_nSU(n_f)$ marked by this Young diagram has multiplicity equal to the dimension of the irreducible representation of symmetry group $S_{n}$ presented by the same Young diagram. It can be proved that these irreducible subspaces are isomorphic to each other \cite{Li:2022tec}, and only one of them need to be reserved. After this, the remaining operators form  the so-called f-basis or p'-basis.} 

Here, we will illustrate construction of the Lorentz basis, gauge basis and the conversion among various bases using several examples. The first one is the type $\phi^2 h^2 D^4\bft^2$, which is of the class $UhD^4$.
There are 6 particles $h,h,\phi,\phi,\bft,\bft$, labeled by indices $1$-$6$ respectively, and 4 derivatives in this type, among which, the 2 Higgs, 2 NGBs and 2 spurions are identical particles. 

As mentioned in the previous subsection, we neglect the spurions $\bft$'s from the type during the construction of the Lorentz basis, and thus the particles considered in the y-basis construction are $h,h,\phi,\phi$, in which the last two must satisfy the Adler zero condition. It is straightforward to get that $n=\tilde{n}=2$, thus the primary Young diagram of this type takes form that
\be
\yng(4,4)\,.
\ee
The numbers of the 4 indices are the same, $\# i=2,i=1,2,3,4$, thus there are only 3 SSYTs that
\be
\young(1133,2244),\quad \young(1122,3344),\quad \young(1123,2344)\,,
\ee
whose corresponding amplitude in the y-basis is 
\bea
\mathcal{B}^{(y)}_1 &=& \lra{34}^2\lrs{34}^2, \notag \\
\mathcal{B}^{(y)}_2 &=& \lra{24}^2\lrs{24}^2, \notag \\
\mathcal{B}^{(y)}_3 &=& -\lra{24}\lra{34}\lrs{24}\lrs{34}\,. \label{yb1}
\eea
The Adler zero condition tells that when the momenta $p_3,p_4$ go to zero, the amplitudes become zero in the soft momentum limit. We check that all the 3 basis above satisfy that, because the Adler zero condition makes the 4-point amplitudes become 2-point amplitudes, which are all zero.

The Y-basis can be translated to the operator form via the amplitude-operator correspondences,
\bea
\mathcal{B}^{(y)}_1 &=& 4(D_\nu h)(D^\nu h)(D_\mu\phi)(D^\mu\phi), \notag\\
\mathcal{B}^{(y)}_2 &=& 4(D^\mu h)(D^\nu h)(D_\mu\phi)(D_\nu\phi), \notag\\
\mathcal{B}^{(y)}_3 &=& 4(D^\mu D^\nu h)h(D_\mu\phi)(D_\nu\phi)+4(D^\mu h)(D^\nu h)(D_\mu\phi)(D_\nu\phi), \label{eq:lmbasis1}
\eea
from which we can construct the Lorentz m-basis
\bea
\mathcal{B}^{(m)}_1 &=& (D^\mu D^\nu h)h(D_\mu\phi)(D_\nu\phi)\bft\bft, \notag \\
\mathcal{B}^{(m)}_2 &=& (D^\mu h)(D^\nu h)(D_\mu\phi)(D_\nu\phi)\bft\bft, \notag \\
\mathcal{B}^{(m)}_3 &=& (D^\mu h)(D^\nu h)(D_\mu\phi)(D_\nu\phi)\bft\bft,
\eea
with the transformation matrix
\be
{\mathcal{K}^{(my)}} = \left(\begin{array}{ccc}
0&-\frac{1}{4}&\frac{1}{4} \\\frac{1}{4}&0&0 \\0&\frac{1}{4}&0
\end{array}
\right),
\ee
where the spurions $\bft$ have been multiplied back to be consistent with the gauge basis we are going to deal with in the following, and to form the ultimate tensor product space of the Lorentz and gauge structure.

The gauge basis is simple in this type. The $SU(3)$ structure is trivial. There are only 3 SSYTs of $SU(2)$ that
\be
\young(\ithree\jthree\ifour\jfour,\ifive\jfive\isix\jsix),\quad \young(\ithree\jthree\ifive\jfive,\ifour\jfour\isix\jsix),\quad\young(\ithree\jthree\ifour\ifive,\jfour\jfive\isix\jsix),
\ee
which corresponds to the gauge y-basis
\bea
\mathcal{T}^{(y)}_{(SU(2),1} &=& \epsilon^{i_3i_5}\epsilon^{j_3j_5}\epsilon^{i_4i_6}\epsilon^{j_4j_6}{\tau^{I_3}}^{l_3}_{j_3}\epsilon_{l_3i_3}{\tau^{I_4}}^{l_4}_{j_4}\epsilon_{l_4i_4}{\tau^{I_5}}^{l_5}_{j_5}\epsilon_{l_5i_5}{\tau^{I_6}}^{l_6}_{j_6}\epsilon_{l_6i_6} \notag \\
 &=& 4\delta^{I_3I_5}\delta^{I_4I_6} \notag, \\
\mathcal{T}^{(y)}_{SU(2),2} &=& \epsilon^{i_3i_4}\epsilon^{j_3j_4}\epsilon^{i_5i_6}\epsilon^{j_5j_6}{\tau^{I_3}}^{l_3}_{j_3}\epsilon_{l_3i_3}{\tau^{I_4}}^{l_4}_{j_4}\epsilon_{l_4i_4}{\tau^{I_5}}^{l_5}_{j_5}\epsilon_{l_5i_5}{\tau^{I_6}}^{l_6}_{j_6}\epsilon_{l_6i_6} \notag \\
&=& 4\delta^{I_3I_4}\delta^{I_5I_6}, \notag \\
\mathcal{T}^{(y)}_{SU(2),3} &=& \epsilon^{i_3j_4}\epsilon^{j_3j_5}\epsilon^{i_4i_6}\epsilon^{i_5j_6}{\tau^{I_3}}^{l_3}_{j_3}\epsilon_{l_3i_3}{\tau^{I_4}}^{l_4}_{j_4}\epsilon_{l_4i_4}{\tau^{I_5}}^{l_5}_{j_5}\epsilon_{l_5i_5}{\tau^{I_6}}^{l_6}_{j_6}\epsilon_{l_6i_6} \notag \\
&=& -2\delta^{I_3I_6}\delta^{I_4I_5}+2\delta^{I_3I_5}\delta^{I_4I_6}.
\eea
The m-basis is chosen to be
\be
\mathcal{T}^{(m)}_{SU(2),1}=\delta^{I_3I_5}\delta^{I_4I_6},\quad \mathcal{T}^{(m)}_{SU(2),2}=\delta^{I_3I_4}\delta^{I_5I_6},\quad \mathcal{T}^{(m)}_{SU(2),3}=\delta^{I_3I_6}\delta^{I_4I_5}, \label{eq:gmbasis1}
\ee
with the transformation matrix
\be
\mathcal{K}^{(my)}_{SU(2)}= \left(\begin{array}{ccc}\frac{1}{4}&0&0\\0&\frac{1}{4}&0\\\frac{1}{4}&0&\frac{1}{2}\end{array}\right).
\ee
Because of identical spurions in this operator, the m-basis in Eq.~\ref{eq:gmbasis1} should be further reduced to remove the self-contraction of the two spurions. This is performed using the gauge j-basis decomposition of the two spurions: $\mathbf{3}\otimes\mathbf{3}=\mathbf{1}\oplus\mathbf{3}\oplus\mathbf{5}$. Picking up the basis with the gauge quantum number $\mathbf 5$ would select the subspace with the highest weight of the $SU(2)_L$ gauge basis, 
\be
\mathcal{T}^{(j)}_{SU(2)} = -3\mathcal{T}^{(m)}_{SU(2),1}+2\mathcal{T}^{(m)}_{SU(2),2}-3\mathcal{T}^{(m)}_{SU(2),3}=-3\delta^{I_3I_5}\delta^{I_4I_6}+2\delta^{I_3I_4}\delta^{I_5I_6}-3\delta^{I_3I_6}\delta^{I_4I_5},\label{eq:jbasis1}
\ee
which is dimension 1 and, for convenience, is denoted by the j-basis of the spurions.

The tensor product of the Lorentz and gauge m-basis \eqref{eq:lmbasis1} and \eqref{eq:gmbasis1} gives all the independent operators that
\bea
\mathcal{O}^{(m)}_1 &=&  (D^\mu D^\nu h)h(D_\mu\phi^I)(D_\nu\phi^J)\bft^I\bft^J, \notag \\
\mathcal{O}^{(m)}_2 &=&  (D^\mu h)(D^\nu h)(D_\mu\phi^I)(D_\nu\phi^J)\bft^I\bft^J, \notag \\
\mathcal{O}^{(m)}_3 &=&  (D^\mu h)(D^\nu h)(D_\mu\phi^I)(D_\nu\phi^J)\bft^I\bft^J, \notag \\
\mathcal{O}^{(m)}_4 &=&  (D^\mu D^\nu h)h(D_\mu\phi^I)(D_\nu\phi^I)\bft^J\bft^J, \notag \\
\mathcal{O}^{(m)}_5 &=&  (D^\mu h)(D^\nu h)(D_\mu\phi^I)(D_\nu\phi^I)\bft^J\bft^J, \notag \\
\mathcal{O}^{(m)}_6 &=&  (D^\mu h)(D^\nu h)(D_\mu\phi^I)(D_\nu\phi^I)\bft^J\bft^J, \notag \\
\mathcal{O}^{(m)}_7 &=&  (D^\mu D^\nu h)h(D_\mu\phi^I)(D_\nu\phi^J)\bft^J\bft^I, \notag \\
\mathcal{O}^{(m)}_8 &=&  (D^\mu h)(D^\nu h)(D_\mu\phi^I)(D_\nu\phi^J)\bft^J\bft^I, \notag \\
\mathcal{O}^{(m)}_9 &=&  (D^\mu h)(D^\nu h)(D_\mu\phi^I)(D_\nu\phi^J)\bft^J\bft^I, \label{mbasis1}
\eea
which is the m-basis of the operators. Due to the repeated fields, we need to convert them to the physical basis, the p-basis. Furthermore, it is obvious that not all operators above are independent, for example, the first 3 and the last 3 operators are actually the same, since the permutation of spurions does not change anything, and the operators $\calo^{(m)}_{4-6}$ contain spurion self-contractions and should be eliminated. 
Thus, we need to take the tensor product of Lorentz m-basis \eqref{eq:lmbasis1} and gauge j-basis \eqref{eq:jbasis1} to get the operator j-basis with the spurion self-contraction removed,
{\small
\bea
\calo^{(j)}_1 &=& -3(D^\mu D^\nu h)h(D_\mu\phi^I)(D_\nu\phi^J)\bft^I\bft^J+2(D^\mu D^\nu h)h(D_\mu\phi^I)(D_\nu\phi^I)\bft^J\bft^J-3(D^\mu D^\nu h)h(D_\mu\phi^I)(D_\nu\phi^J)\bft^J\bft^I, \notag \\
\calo^{(j)}_2 &=& -3(D^\mu h)(D^\nu h)(D_\mu\phi^I)(D_\nu\phi^J)\bft^I\bft^J+2(D^\mu h)(D^\nu h)(D_\mu\phi^I)(D_\nu\phi^I)\bft^J\bft^J-3(D^\mu h)(D^\nu h)(D_\mu\phi^I)(D_\nu\phi^J)\bft^J\bft^I,\notag \\
\calo^{(j)}_3 &=& -3(D^\mu h)(D^\nu h)(D_\mu\phi^I)(D_\nu\phi^J)\bft^I\bft^J + 2(D^\mu h)(D^\nu h)(D_\mu\phi^I)(D_\nu\phi^I)\bft^J\bft^J -3(D^\mu h)(D^\nu h)(D_\mu\phi^I)(D_\nu\phi^J)\bft^J\bft^I.\notag \\ \label{m'basis1}
\eea
}

To obtain the p-basis, we need the generators of the $S_2$ group, which characterises the permutation properties of all repeated fields $h,\phi,\bft$. By the way, the two generators of $S_2$ are identical, we present one of them in the rest of this paper.
In the Lorentz y-basis, the generators of the $S_2(h)$ and $S_2(\phi)$, denoted by  $\mathcal{D}^{(y)}_{\mathcal{B}}(h)$ and $\mathcal{D}^{(y)}_{\mathcal{B}}(\phi)$, respectively, can be obtained by permuting the y-basis \eqref{yb1}, which takes the form that
\be
\mathcal{D}^{(y)}_{\mathcal{B}}(h) = \mathcal{D}^{(y)}_{\mathcal{B}}[(1,2)]=\left(
\begin{array}{ccc}
 1 & 0 & 0 \\
 1 & 1 & -2 \\
 1 & 0 & -1 \\
\end{array}
\right),\quad \mathcal{D}^{(y)}_{\mathcal{B}}(\phi)= \mathcal{D}^{(y)}_{\mathcal{B}}[(3,4)]=\left(
\begin{array}{ccc}
 1 & 0 & 0 \\
 1 & 1 & -2 \\
 1 & 0 & -1 \\
\end{array}
\right), 
\ee
while these generators after transforming to the Lorentz m-basis are
\be
\mathcal{D}^{(m)}_{\mathcal{B}}(h)={\mathcal{K}^{(my)}}\mathcal{D}^{(y)}_{\mathcal{B}}(h){{\mathcal{K}^{(my)}}}^{-1} = \left(
\begin{array}{ccc}
 1 & 0 & 0 \\
 0 & 1 & 0 \\
 -2 & 1 & -1 \\
\end{array}
\right),\quad \mathcal{D}^{(m)}_{\mathcal{B}}(\phi)={\mathcal{K}^{(my)}}\mathcal{D}^{(y)}_{\mathcal{B}}(\phi){{\mathcal{K}^{(my)}}}^{-1} = \left(
\begin{array}{ccc}
 1 & 0 & 0 \\
 0 & 1 & 0 \\
 -2 & 1 & -1 \\
\end{array}
\right).
\ee
and since the spurion does not carry Lorentz structure, the generator of the $S_2(\bft)$ takes
\be
\mathcal{D}^{(y)}_{\mathcal{B}}(\bft) = \mathcal{D}^{(m)}_{\mathcal{B}}(\bft) = \mathbf{I}_{3\times 3},
\ee

As for the $SU(2)$ structure, the generators $\mathcal{D}^{(m)}_{SU(2)}(h)$ is the identity matrix $\mathbf{I}_{3\times 3}$, since $h$ is the $SU(2)$ singlet, while the others $\mathcal{D}^{(m)}_{SU(2)}(\phi),\mathcal{D}^{(m)}_{SU(2)}(\bft)$ can be obtained by permuting the gauge m-basis Eq.~\ref{eq:gmbasis1} directly,
\be
\mathcal{D}^{(m)}_{SU(2)}(\phi) = \mathcal{D}^{(m)}_{SU(2)}[(3,4)]=\left(\begin{array}{ccc}0&0&1\\0&1&0\\1&0&0\end{array}\right),\quad \mathcal{D}^{(m)}_{SU(2)}(\bft) = \mathcal{D}^{(m)}_{SU(2)}[(5,6)]=\left(\begin{array}{ccc}0&0&1\\0&1&0\\1&0&0\end{array}\right).
\ee
At the same time, the generators in the 1-dimension j-basis in Eq.~\eqref{eq:jbasis1} are just identities,
\be
\mathcal{D}^{(j)}(h) = \mathcal{D}^{(j)}(\phi) = \mathcal{D}^{(j)}(\bft) = 1.
\ee

The complete generators of the operator m-basis in Eq.~\eqref{mbasis1} are
\bea
\mathcal{D}^{(m)}(h)&=& \mathcal{D}^{(m)}_{SU(2)}(h)\otimes \mathcal{D}^{(m)}_{\mathcal{B}}(h)=\left(
\begin{array}{ccccccccc}
 1 & 0 & 0 & 0 & 0 & 0 & 0 & 0 & 0 \\
 0 & 1 & 0 & 0 & 0 & 0 & 0 & 0 & 0 \\
 -2 & 1 & -1 & 0 & 0 & 0 & 0 & 0 & 0 \\
 0 & 0 & 0 & 1 & 0 & 0 & 0 & 0 & 0 \\
 0 & 0 & 0 & 0 & 1 & 0 & 0 & 0 & 0 \\
 0 & 0 & 0 & -2 & 1 & -1 & 0 & 0 & 0 \\
 0 & 0 & 0 & 0 & 0 & 0 & 1 & 0 & 0 \\
 0 & 0 & 0 & 0 & 0 & 0 & 0 & 1 & 0 \\
 0 & 0 & 0 & 0 & 0 & 0 & -2 & 1 & -1 \\
\end{array}
\right),
\eea
\bea
\mathcal{D}^{(m)}(\phi)&=& \mathcal{D}^{(m)}_{SU(2)}(\phi)\otimes \mathcal{D}^{(m)}_{\mathcal{B}}(\phi)=\left(
\begin{array}{ccccccccc}
 0 & 0 & 0 & 0 & 0 & 0 & 1 & 0 & 0 \\
 0 & 0 & 0 & 0 & 0 & 0 & 0 & 1 & 0 \\
 0 & 0 & 0 & 0 & 0 & 0 & -2 & 1 & -1 \\
 0 & 0 & 0 & 1 & 0 & 0 & 0 & 0 & 0 \\
 0 & 0 & 0 & 0 & 1 & 0 & 0 & 0 & 0 \\
 0 & 0 & 0 & -2 & 1 & -1 & 0 & 0 & 0 \\
 1 & 0 & 0 & 0 & 0 & 0 & 0 & 0 & 0 \\
 0 & 1 & 0 & 0 & 0 & 0 & 0 & 0 & 0 \\
 -2 & 1 & -1 & 0 & 0 & 0 & 0 & 0 & 0 \\
\end{array}
\right),
\eea
\bea
\mathcal{D}^{(m)}(\bft)&=& \mathcal{D}^{(m)}_{SU(2)}(\bft)\otimes \mathcal{D}^{(m)}_{\mathcal{B}}(\bft)=\left(
\begin{array}{ccccccccc}
 0 & 0 & 0 & 0 & 0 & 0 & 1 & 0 & 0 \\
 0 & 0 & 0 & 0 & 0 & 0 & 0 & 1 & 0 \\
 0 & 0 & 0 & 0 & 0 & 0 & 0 & 0 & 1 \\
 0 & 0 & 0 & 1 & 0 & 0 & 0 & 0 & 0 \\
 0 & 0 & 0 & 0 & 1 & 0 & 0 & 0 & 0 \\
 0 & 0 & 0 & 0 & 0 & 1 & 0 & 0 & 0 \\
 1 & 0 & 0 & 0 & 0 & 0 & 0 & 0 & 0 \\
 0 & 1 & 0 & 0 & 0 & 0 & 0 & 0 & 0 \\
 0 & 0 & 1 & 0 & 0 & 0 & 0 & 0 & 0 \\
\end{array}
\right),
\eea
while those of the j-basis in Eq.~\eqref{m'basis1}are just the generators of Lorentz m-basis,
\be
\mathcal{D}^{(j)}(h) =\mathcal{D}^{(y)}_{\mathcal{B}}(h),\quad \mathcal{D}^{(j)}(\phi) =\mathcal{D}^{(y)}_{\mathcal{B}}(\phi), \quad\mathcal{D}^{(j)}(\bft) = \mathcal{D}^{(y)}_{\mathcal{B}}(\bft).
\label{eq:generatorjbasis}
\ee

In this type all particles carry no flavor number, thus only the symmetrical representations {\tiny\yng(2)} in the flavor structure are physical. The idempotent element $\caly[{\tiny\yt{12},\yt{34},\yt{56}}]$ of this operators m-basis in Eq.~\eqref{mbasis1} takes the form that
\bea
\caly[{\tiny\yt{12},\yt{34},\yt{56}}] &=& \caly[{\tiny\yt{12}}]\caly[{\tiny\yt{34}}]\caly[{\tiny\yt{56}}]\notag \\
&=& (\mathbf{I}_{3\times 3}+\mathcal{D}^{(m)}(h))(\mathbf{I}_{3\times 3}+\mathcal{D}^{(m)}(\phi))(\mathbf{I}_{3\times 3}+\mathcal{D}^{(m)}(\bft)) \notag \\
&=& \left(
\begin{array}{ccccccccc}
 4 & 0 & 0 & 0 & 0 & 0 & 4 & 0 & 0 \\
 0 & 4 & 0 & 0 & 0 & 0 & 0 & 4 & 0 \\
 -4 & 2 & 0 & 0 & 0 & 0 & -4 & 2 & 0 \\
 0 & 0 & 0 & 8 & 0 & 0 & 0 & 0 & 0 \\
 0 & 0 & 0 & 0 & 8 & 0 & 0 & 0 & 0 \\
 0 & 0 & 0 & -8 & 4 & 0 & 0 & 0 & 0 \\
 4 & 0 & 0 & 0 & 0 & 0 & 4 & 0 & 0 \\
 0 & 4 & 0 & 0 & 0 & 0 & 0 & 4 & 0 \\
 -4 & 2 & 0 & 0 & 0 & 0 & -4 & 2 & 0 \\
\end{array}
\right).\label{eq:mgen1}
\eea
This idempotent element has rank 4, thus this subspace has dimension 4. Its is actually arbitrary to choose the 4 independent operators, in this example we prefer choosing independent operators projected by the $1$st, $2$nd, $4$th and $5$th rows in the idempotent element, because they are monomials by coincidence,
\bea
\calo^{(p)}_1 &=& 4\calo^{(m)}_1 + 4 \calo^{(m)}_7 = 8(D^\mu D^\nu h)h(D_\mu\phi^I)(D_\nu\phi^J)\bft^I\bft^J, \notag \\
\calo^{(p)}_2 &=& 4\calo^{(m)}_2 + 4 \calo^{(m)}_8 = 8(D^\mu h)(D^\nu h)(D_\mu\phi^I)(D_\nu\phi^J)\bft^I\bft^J, \notag \\
\calo^{(p)}_3 &=& 8\calo^{(m)}_4 = 8(D^\mu D^\nu h)h(D_\mu\phi^I)(D_\nu\phi^I)\bft^J\bft^J, \notag \\
\calo^{(p)}_4 &=& 8\calo^{(m)}_5 = 8(D^\mu h)(D^\nu h)(D_\mu\phi^I)(D_\nu\phi^I)\bft^J\bft^J.\label{pbasis1}
\eea
However, the last 2 contain the spurion self-contractions, thus should be eliminated. 
This elimination can be seen if we directly use  the generator of the operator j-basis in Eq.~\ref{eq:generatorjbasis} to calculate the idempotent element of them
\be
\caly'[{\tiny\young(12),\young(34),\young(56)}]=\caly'[{\tiny\young(12)}]\caly'[{\tiny\young(34)}]\caly'[{\tiny\young(56)}]=
 (\mathbf{I}_{3\times 3}+\mathcal{D}^{(j)}(h))(\mathbf{I}_{3\times 3}+\mathcal{D}^{(j)}(\phi))(\mathbf{I}_{3\times 3}+\mathcal{D}^{(j)}(\bft)) =
\left(
\begin{array}{ccc}
 4 & 0 & 0 \\
 0 & 4 & 0 \\
 -4 & 2 & 0 \\
\end{array}
\right),
\ee
which is of rank 2, thus there are only 2 independent operators. Though it is arbitrary to choose these 2 operators from j-basis, we choose $\calo^{(j)}_{1,2}$ here, and also denote them as the final p-basis
\be
\calo^{(p)}_1 = \caly[{\tiny\young(12),\young(34),\young(56)}](D^\mu D^\nu h)h(D_\mu\phi^I)(D_\nu\phi^J)\bft^I\bft^J,\quad \calo^{(p)}_2 = \caly[{\tiny\young(12),\young(34),\young(56)}](D^\mu h)(D^\nu h)(D_\mu\phi^I)(D_\nu\phi^J)\bft^I\bft^J,
\ee
which correspond to the operators $\calo^{UhD^4}_{12,13}$ in next section \ref{sec:boson} and their explicit form are given by $\calo^{(j)}_{1,2}$.
Several comments are in order. 
First, this is consistent with the observation of the original result \eqref{pbasis1}, although the 2 operators $\calo^{(j)}_{1,2}$ are not the same with the $\calo^{(p)}_{1,2}$. Second, usually operators in the p-basis are polynomials, thus in this work, we always write the p-basis with action of the idempotent elements to avoid the complication of writing the complete polynomials. 
In the rest of this paper, we will omit the idempotent elements in the type that the identical particles carry no flavor number, as in this example. 

The second example is taken to be the type ${Q_L^\dagger}{Q_L}{Q_R^\dagger}{Q_R}\bft$, which was missing in previous works Ref.~\cite{Brivio:2016fzo}. 
The shape of the primary Young diagram is 
\be
\yng(2,2),
\ee
and there is only one SSYT
\be
\young(11,22).
\ee
The corresponding operator form is
\be
\mathcal{B}^{(y)}_1 = {\psi_1}^\alpha {\psi_2}_\alpha {\bar{\psi_3}}_{\dot{\beta}}{\bar{\psi}_4}^{\dot{\beta}} = \epsilon^{\alpha\gamma}\epsilon_{\dot{\beta}\dot{\delta}}{\psi_1}_\gamma{\psi_2}_\alpha{\bar{\psi}_3}^{\dot{\delta}}{\bar{\psi}_4}^{\dot{\beta}} = -\frac{1}{2}(\psi_1\sigma^\mu\bar{\psi}_3)(\psi_2\sigma_\mu \bar{\psi}_4).
\ee
Thus the Lorentz m-basis is just the y-basis, $\mathcal{B}^{(m)}_1=\mathcal{B}^{(y)}_1$. As for the SU(3) gauge y-basis, the primary Young diagram is  
\be
\yng(2,2,2),
\ee
and there are 2 SSYTs
\be
\young(\ione\itwo,\ithree\ifour,\jthree\jfour),\quad \young(\ione\ithree,\itwo\ifour,\jthree\jfour).
\ee
They correspond to the y-basis
\bea
& \mathcal{T}^{(y)}_{SU(3),1} &=\epsilon_{i_1i_3j_3}\epsilon_{i_2i_4j_4}\epsilon^{i_3j_3k_3}\epsilon^{i_4j_4k_4} = 36 \delta_{i_1}^{k_3}\delta_{i_2}^{k_4}, \notag \\
& \mathcal{T}^{(y)}_{SU(3),1} &=\epsilon_{i_1i_2i_3}\epsilon_{j_3i_4j_4}\epsilon^{i_3j_3k_3}\epsilon^{i_4j_4k_4} = 18 (\delta_{i_1}^{k_4}\delta_{i_2}^{k_3}-\delta_{i_1}^{k_3}\delta_{i_2}^{k_4}) ,
\eea
and the m-basis can be taken as
\be
\mathcal{T}^{(m)}_{SU(3),1} = \delta_{i_1}^{k_3}\delta_{i_2}^{k_4},\quad \mathcal{T}^{(m)}_{SU(3),1} = \delta_{i_1}^{k_4}\delta_{i_2}^{k_3},
\ee
with the transformation matrix 
\be
\mathcal{K}_{SU(3)}^{(my)} = \left(\begin{array}{cc}
\frac{1}{36} & 0 \\ \frac{1}{36} & \frac{1}{18}
\end{array}\right).
\ee
The primary Young diagram of the  $SU(2)$ group is 
\be
\yng(3,3),
\ee
and there are 3 SSYTs that
\be
\young(\ione\itwo\ithree,\ifour\ifive\jfive),\quad \young(\ione\itwo\ifour,\ithree\ifive\jfive),\quad \young(\ione\ithree\ifour,\itwo\ifive\jfive).
\ee
The gauge y-basis reads 
\bea
&\mathcal{T}^{(y)}_{SU(2),1} &= \epsilon_{i_1i_4}\epsilon_{i_3j_3}\epsilon_{i_2i_5}{\tau^{I_5}}^{i_5}_{l_5}\epsilon^{l_5j_5}\epsilon^{i_2j_2}\epsilon^{i_4j_4} = \delta^{j_4}_{i_1}{\tau^{I_5}}^{j_2}_{i_3}, \notag \\
&\mathcal{T}^{(y)}_{SU(2),2} &= \epsilon_{i_1i_3}\epsilon_{i_2i_5}\epsilon_{i_4j_5}{\tau^{I_5}}^{i_5}_{l_5}\epsilon^{l_5j_5}\epsilon^{i_2j_2}\epsilon^{i_4j_4} = -\delta^{j_4}_{i_3}{\tau^{I_5}}^{j_2}_{i_1} + \delta^{j_4}_{i_1}{\tau^{I_5}}^{j_2}_{i_3}, \notag \\
&\mathcal{T}^{(y)}_{SU(2),3} &= \epsilon_{i_1i_2}\epsilon_{i_3i_5}\epsilon_{i_4j_5}{\tau^{I_5}}^{i_5}_{l_5}\epsilon^{l_5j_5}\epsilon^{i_2j_2}\epsilon^{i_4j_4} = -\delta^{j_2}_{i_1}{\tau^{I_5}}^{j_4}_{i_3} .
\eea
Thus we choose the gauge m-basis as
\be
\mathcal{T}^{(m)}_{SU(2),1}= \delta^{j_4}_{i_1}{\tau^{I_5}}^{j_2}_{i_3},\quad \mathcal{T}^{(m)}_{SU(2),2} = \delta^{j_4}_{i_3}{\tau^{I_5}}^{j_2}_{i_1},\quad \mathcal{T}^{(m)}_{SU(2),3} = \delta^{j_2}_{i_1}{\tau^{I_5}}^{j_4}_{i_3},
\ee
with the transformation matrix
\be
\mathcal{K}_{SU(2)}^{(my)} = \left(\begin{array}{ccc}
1&0&0\\1&-1&0\\0&0&-1
\end{array}\right).
\ee
The tensor product of the 2 space is of six-dimension, and all the 6 operators are physical since there is no repeated fields in this type, which are presented as $\mathcal{O}^{Uh\psi^4}_{29-34}$,
\bea
\calo^{Uh\psi^4}_{29} &=& ({\overline{q}_L}_s\gamma_\mu\bft{q_L}_p)({\overline{q}_R}_r\gamma^\mu{q_R}_t)\calf^{Uh\psi^4}_{29}(h),\\
\calo^{Uh\psi^4}_{30} &=& ({\overline{q}_L}_s\gamma_\mu\lambda^A\bft{q_L}_p)({\overline{q}_R}_r\gamma^\mu\lambda^A{q_R}_t)\calf^{Uh\psi^4}_{30}(h),\\
\calo^{Uh\psi^4}_{31} &=& ({\overline{q}_L}_s\gamma_\mu{q_L}_p)({\overline{q}_R}_r\gamma^\mu\bfu^\dagger\bft\bfu{q_R}_t)\calf^{Uh\psi^4}_{31}(h),\\
\calo^{Uh\psi^4}_{32} &=& ({\overline{q}_L}_s\gamma\mu\lambda^A{q_L}_p)({\overline{q}_R}_r\gamma^\mu\lambda^A\bfu^\dagger\bft\bfu{q_R}_t)\calf^{Uh\psi^4}_{32}(h),\\
{\color{red}\calo^{Uh\psi^4}_{33}} &=&({\overline{q}_L}_s\gamma_\mu\tau^I\bft{q_L}_p)({\overline{q}_R}_r\gamma^\mu\bfu^\dagger\tau^I\bfu{q_R}_t)\calf^{Uh\psi^4}_{33}(h),\\
{\color{red}\calo^{Uh\psi^4}_{34}} &=& ({\overline{q}_L}_s\gamma_\mu\lambda^A\tau^I\bft{q_L}_p)({\overline{q}_R}_r\gamma^\mu\lambda^A\bfu^\dagger\tau^I\bfu{q_R}_t)\calf^{Uh\psi^4}_{34}(h), 
\eea


The last example is the type ${Q_L}^3{L_L}\bft^2$, which is neglected in the previous literature. Neglecting the spurions, there are 4 fermions in this type, ${L_L},Q_L,Q_L,Q_L$, and the primary Young diagrams of the Lorentz basis, the $SU(2)$ gauge basis and the $SU(3)$ gauge basis are
\be
\yng(2,2),\quad \yng(4,4),\quad \yng(1,1,1).
\ee
For the Lorentz basis, there are 2 SSYTs,
\be
\young(13,24),\quad \young(12,34),
\ee
which correspond to the y-basis that
\bea
\mathcal{B}^{(y)}_1 &=& ({L_L}{Q_L})({Q_L}{Q_L}), \notag \\
\mathcal{B}^{(y)}_2 &=& \frac{1}{2}({L_L}{Q_L})({Q_L}{Q_L}) -\frac{1}{8} ({L_L}\sigma^{\mu\nu}{Q_L})({Q_L}\sigma_{\mu\nu}{Q_L}).
\eea
We choose the m-basis as
\be
\mathcal{B}^{(m)}_1 = ({L_L}{Q_L})({Q_L}{Q_L})\bft\bft,\quad \mathcal{B}^{(m)}_1 = ({L_L}\sigma^{\mu\nu}{Q_L})({Q_L}\sigma_{\mu\nu}{Q_L})\bft\bft.
\ee
with the transformation matrix
\be
{\mathcal{K}^{(my)}} = \left(
\begin{array}{cc}
 1 & 0 \\
 4 & -8 \\
\end{array}
\right),
\ee
where spurions has been multiplied back.

At the same time, we present the generators of $S_3$ and $S_2$, which are of the symmetric groups of the ${Q_L}$ and $\bft$, respectively,
\be
\mathcal{D}^{(m)}_{\mathcal{B}}(Q_L)_1 =\mathcal{D}^{(y)}_{\mathcal{B}}[(2,3)]=\left(
\begin{array}{cc}
 -\frac{1}{2} & \frac{1}{8} \\
 6 & \frac{1}{2} \\
\end{array}
\right),\quad \mathcal{D}^{(m)}_{\mathcal{B}}({Q_L})_2 = \mathcal{D}^{(y)}_{\mathcal{B}}[(2,3,4)] = \left(
\begin{array}{cc}
 -\frac{1}{2} & \frac{1}{8} \\
 -6 & -\frac{1}{2} \\
\end{array}
\right),\quad \mathcal{D}^{(m)}_{\mathcal{B}}(\bft) = \left(
\begin{array}{cc}
 1 & 0 \\
 0& 1 \\
\end{array}
\right),\label{eq:mgen2}
\ee
where the $S_3$ group has 2 different generators, and the generator of the $\bft$ is the identity matrix. 

The $SU(3)$ group structure is simple in this type, there is only 1 SSYT,
\be
\young(2,3,4),
\ee
and the corresponding y-basis is $\mathcal{B}^{(y)}_1=\epsilon^{a_2a_3a_4}$, which is also the m-basis. The generator of $S_2$ in this $SU(3)$ m-basis is trivial because the spurion does not carry $SU(3)$ quantum number, while the generators of $S_3$ takes the form that
\be
\mathcal{D}^{(m)}_{SU(3)}(Q_L)_1 = -1,\quad \mathcal{D}^{(m)}_{SU(3)}(Q_L)_2 = 1.
\ee

As for the $SU(2)$ group, there are 6 gauge SSYTs, thus 6 m-basis, and we present them directly,
\bea
\mathcal{T}^{(m)}_{SU(2),1} &=& \epsilon^{i_2i_3}\epsilon^{i_4m}\epsilon^{I_1I_2K}{\tau^K}_m^{i_1},\notag \\
\mathcal{T}^{(m)}_{SU(2),2} &=& \epsilon^{i_3m}\epsilon^{i_4l}{\tau^{I_1}}_m^{i_1}{\tau^{I_2}}_n^{i_2}, \notag \\
\mathcal{T}^{(m)}_{SU(2),3} &=& \delta^{I_1I_2}\epsilon^{i_1i_4}\epsilon^{i_2i_3}, \notag \\
\mathcal{T}^{(m)}_{SU(2),4} &=& \epsilon^{i_1 m}\epsilon^{i_3n}{\tau^{I_1}}_n^{i_4}{\tau^{I_2}}_m^{i_2}, \notag \\
\mathcal{T}^{(m)}_{SU(2),5} &=& \epsilon^{i_1 m}\epsilon^{i_2i_4}\epsilon^{I_1I_2K}{\tau^K}_m^{i_3}, \notag \\
\mathcal{T}^{(m)}_{SU(2),6} &=& \epsilon^{i_3m}\epsilon^{i_4n}{\tau^{I_1}}_n^{i_1}{\tau^{I_2}}_m^{i_2}.\label{mbais2}
\eea
Similar to the discussion near Eq.~\ref{eq:jbasis1}, taking the gauge j-basis of the spurions, the highest-weight subspace of the $SU(2)$ m-basis is of dimension-1, and we take the gauge j-basis with the gauge quantum number $\mathbf{5}$ 
\be
\mathcal{D}^{(j)}_{SU(2)} = i\mathcal{T}^{(m)}_{SU(2),1}-\mathcal{T}^{(m)}_{SU(2),2}-\mathcal{T}^{(m)}_{SU(2),3}-3\mathcal{T}^{(m)}_{SU(2),4}+2i\mathcal{T}^{(m)}_{SU(2),5}-2\mathcal{T}^{(m)}_{SU(2),6}.\label{m'basis2}
\ee
The generators in the m-basis \eqref{mbais2} are
\be
\mathcal{D}^{(m)}_{SU(2)}(Q_L)_1 = \left(
\begin{array}{cccccc}
 -1 & 0 & 0 & 0 & 0 & 0 \\
 -i & 1 & -1 & 0 & 0 & 0 \\
 0 & 0 & -1 & 0 & 0 & 0 \\
 -i & 0 & 1 & 1 & 0 & 0 \\
 -1 & 0 & 0 & 0 & 1 & 0 \\
 0 & 0 & 0 & 0 & 0 & 1 \\
\end{array}
\right),
\ee
\be
\mathcal{D}^{(m)}_{SU(2)}(Q_L)_1 = \left(
\begin{array}{cccccc}
 -1 & 0 & 0 & 0 & 1 & 0 \\
 0 & 0 & 0 & 0 & 0 & 1 \\
 i & -1 & 0 & 0 & -i & 1 \\
 -i & 0 & 1 & 1 & 0 & 0 \\
 -1 & 0 & 0 & 0 & 0 & 0 \\
 -i & 1 & -1 & 0 & 0 & 0 \\
\end{array}
\right),
\ee
\be
\mathcal{D}^{(m)}_{SU(2)}(\bft) = \left(
\begin{array}{cccccc}
 -1 & 0 & 0 & 0 & 0 & 0 \\
 -i & 0 & 1 & 1 & 0 & 0 \\
 0 & 0 & 1 & 0 & 0 & 0 \\
 -i & 1 & -1 & 0 & 0 & 0 \\
 0 & 0 & 0 & 0 & -1 & 0 \\
 i & -1 & 1 & 1 & -2 i & 1 \\
\end{array}
\right),
\ee
while those in the j-basis \eqref{m'basis2} are just identities,
\be
\mathcal{D}^{(j)}_{SU(2)}(Q_L)_1 = \mathcal{D}^{(j)}_{SU(2)}(Q_L)_2 = \mathcal{D}^{(j)}_{SU(2)}(\bft) = 1.\label{eq:mgen3}
\ee

The tensor product of the $SU(3)$ m-basis, the $SU(2)$ m-basis and the Lorentz m-basis gives the operator m-basis, in which there are 12 operators,
\bea
\calo^{(m)}_1 &=& {{\tau^{K}}_{m}^{i}}{\epsilon^{abc}}{\epsilon^{IJK}}{\epsilon^{jk}}{\epsilon^{lm}}{{\mathbf{T}}^{I}}{{\mathbf{T}}^{J}}({{{L_L}_{p}}_{i}}{{{Q_L}_{r}}_{aj}})({{{Q_L}_{s}}_{bk}}{{{Q_L}_{t}}_{cl}}),\notag \\
\calo^{(m)}_2 &=& {{\tau^{K}}_{m}^{i}}{\epsilon^{abc}}{\epsilon^{IJK}}{\epsilon^{jk}}{\epsilon^{lm}}{{\mathbf{T}}^{I}}{{\mathbf{T}}^{J}}({{{L_L}_{p}}_{i}}{\sigma_{\mu\nu}}{{{Q_L}_{r}}_{aj}})({{{Q_L}_{s}}_{bk}}{\sigma^{\mu\nu}}{{{Q_L}_{t}}_{cl}}),\notag \\
\calo^{(m)}_3 &=& {{\tau^{I}}_{m}^{i}}{{\tau^{J}}_{n}^{j}}{\epsilon^{abc}}{\epsilon^{km}}{\epsilon^{ln}}{{\mathbf{T}}^{I}}{{\mathbf{T}}^{J}}({{{L_L}_{p}}_{i}}{{{Q_L}_{r}}_{aj}})({{{Q_L}_{s}}_{bk}}{{{Q_L}_{t}}_{cl}}),\notag \\
\calo^{(m)}_4 &=& {{\tau^{I}}_{m}^{i}}{{\tau^{J}}_{n}^{j}}{\epsilon^{abc}}{\epsilon^{km}}{\epsilon^{ln}}{{\mathbf{T}}^{I}}{{\mathbf{T}}^{J}}({{{L_L}_{p}}_{i}}{\sigma_{\mu\nu}}{{{Q_L}_{r}}_{aj}})({{{Q_L}_{s}}_{bk}}{\sigma^{\mu\nu}}{{{Q_L}_{t}}_{cl}}),\notag \\
\calo^{(m)}_5 &=& { i{\mathbf{T}^I}{\mathbf{T}^I}}{\epsilon^{abc}}{\epsilon^{il}}{\epsilon^{jk}}({{{L_L}_{p}}_{i}}{{{Q_L}_{r}}_{aj}})({{{Q_L}_{s}}_{bk}}{{{Q_L}_{t}}_{cl}}),\notag \\
\calo^{(m)}_6 &=& { i{\mathbf{T}^I}{\mathbf{T}^I}}{\epsilon^{abc}}{\epsilon^{il}}{\epsilon^{jk}}({{{L_L}_{p}}_{i}}{\sigma_{\mu\nu}}{{{Q_L}_{r}}_{aj}})({{{Q_L}_{s}}_{bk}}{\sigma^{\mu\nu}}{{{Q_L}_{t}}_{cl}}),\notag \\
\calo^{(m)}_7 &=& {{\tau^{I}}_{n}^{l}}{{\tau^{J}}_{m}^{j}}{\epsilon^{abc}}{\epsilon^{im}}{\epsilon^{kn}}{{\mathbf{T}}^{I}}{{\mathbf{T}}^{J}}({{{L_L}_{p}}_{i}}{{{Q_L}_{r}}_{aj}})({{{Q_L}_{s}}_{bk}}{{{Q_L}_{t}}_{cl}}),\notag \\
\calo^{(m)}_8 &=& {{\tau^{I}}_{n}^{l}}{{\tau^{J}}_{m}^{j}}{\epsilon^{abc}}{\epsilon^{im}}{\epsilon^{kn}}{{\mathbf{T}}^{I}}{{\mathbf{T}}^{J}}({{{L_L}_{p}}_{i}}{\sigma_{\mu\nu}}{{{Q_L}_{r}}_{aj}})({{{Q_L}_{s}}_{bk}}{\sigma^{\mu\nu}}{{{Q_L}_{t}}_{cl}}),\notag \\
\calo^{(m)}_9 &=& {{\tau^{K}}_{m}^{k}}{\epsilon^{abc}}{\epsilon^{IJK}}{\epsilon^{im}}{\epsilon^{jl}}{{\mathbf{T}}^{I}}{{\mathbf{T}}^{J}}({{{L_L}_{p}}_{i}}{{{Q_L}_{r}}_{aj}})({{{Q_L}_{s}}_{bk}}{{{Q_L}_{t}}_{cl}}),\notag \\
\calo^{(m)}_{10} &=& {{\tau^{K}}_{m}^{k}}{\epsilon^{abc}}{\epsilon^{IJK}}{\epsilon^{im}}{\epsilon^{jl}}{{\mathbf{T}}^{I}}{{\mathbf{T}}^{J}}({{{L_L}_{p}}_{i}}{\sigma_{\mu\nu}}{{{Q_L}_{r}}_{aj}})({{{Q_L}_{s}}_{bk}}{\sigma^{\mu\nu}}{{{Q_L}_{t}}_{cl}}),\notag \\
\calo^{(m)}_{11} &=& {{\tau^{I}}_{n}^{i}}{{\tau^{J}}_{m}^{j}}{\epsilon^{abc}}{\epsilon^{km}}{\epsilon^{ln}}{{\mathbf{T}}^{I}}{{\mathbf{T}}^{J}}({{{L_L}_{p}}_{i}}{{{Q_L}_{r}}_{aj}})({{{Q_L}_{s}}_{bk}}{{{Q_L}_{t}}_{cl}}),\notag \\
\calo^{(m)}_{12} &=& {{\tau^{I}}_{n}^{i}}{{\tau^{J}}_{m}^{j}}{\epsilon^{abc}}{\epsilon^{km}}{\epsilon^{ln}}{{\mathbf{T}}^{I}}{{\mathbf{T}}^{J}}({{{L_L}_{p}}_{i}}{\sigma_{\mu\nu}}{{{Q_L}_{r}}_{aj}})({{{Q_L}_{s}}_{bk}}{\sigma^{\mu\nu}}{{{Q_L}_{t}}_{cl}}). \label{eq:ombasis1}
\eea
To remove the spurion self-contractions in above, we take the $SU(2)$ j-basis in Eq.~\eqref{m'basis2}, the operator j-basis is obtained after the tensor product 
\bea
\calo^{(j)}_1 &=& i\calo^{(m)}_1-\calo^{(m)}_3-\calo^{(m)}_5-3\calo^{(m)}_7+2i\calo^{(m)}_9-2\calo^{(m)}_{11},\notag \\
\calo^{(j)}_2 &=& i\calo^{(m)}_2-\calo^{(m)}_4-\calo^{(m)}_6-3\calo^{(m)}_8+2i\calo^{(m)}_{10}-2\calo^{(m)}_{12}.\label{eq:om'basis1}
\eea

    Following the procedure of converting the m-basis to the f-basis~\cite{Li:2020xlh,Li:2020gnx,Li:2022tec,Li:2020zfq}, we present the resulting 4 operators in the f-basis projected from the operator m-basis in Eq.~\eqref{eq:ombasis1}
\bea
\calo^{(p)}_1 &=& \caly[{\tiny\yt{234},\yt{56}}]{{\tau^{I}}_{m}^{i}}{{\tau^{J}}_{n}^{j}}{\epsilon^{abc}}{\epsilon^{km}}{\epsilon^{ln}}{{\mathbf{T}}^{I}}{{\mathbf{T}}^{J}}({{{L_L}_{p}}_{i}}{\sigma_{\mu\nu}}{{{Q_L}_{r}}_{aj}})({{{Q_L}_{s}}_{bk}}{\sigma^{\mu\nu}}{{{Q_L}_{t}}_{cl}}),\notag \\
\calo^{(p)}_2 &=& \caly[{\tiny\yt{23,4},\yt{56}}]{{\tau^{I}}_{m}^{i}}{{\tau^{J}}_{n}^{j}}{\epsilon^{abc}}{\epsilon^{km}}{\epsilon^{ln}}{{\mathbf{T}}^{I}}{{\mathbf{T}}^{J}}({{{L_L}_{p}}_{i}}{\sigma_{\mu\nu}}{{{Q_L}_{r}}_{aj}})({{{Q_L}_{s}}_{bk}}{\sigma^{\mu\nu}}{{{Q_L}_{t}}_{cl}}),\notag \\
\calo^{(p)}_3 &=& \caly[{\tiny\yt{23,4},\yt{56}}] { i{\mathbf{T}^I}{\mathbf{T}^I}}{\epsilon^{abc}}{\epsilon^{il}}{\epsilon^{jk}}({{{L_L}_{p}}_{i}}{\sigma_{\mu\nu}}{{{Q_L}_{r}}_{aj}})({{{Q_L}_{s}}_{bk}}{\sigma^{\mu\nu}}{{{Q_L}_{t}}_{cl}}),\notag \\
\calo^{(p)}_4 &=& \caly[{\tiny\yt{2,3,4},\yt{56}}]{{\tau^{I}}_{m}^{i}}{{\tau^{J}}_{n}^{j}}{\epsilon^{abc}}{\epsilon^{km}}{\epsilon^{ln}}{{\mathbf{T}}^{I}}{{\mathbf{T}}^{J}}({{{L_L}_{p}}_{i}}{\sigma_{\mu\nu}}{{{Q_L}_{r}}_{aj}})({{{Q_L}_{s}}_{bk}}{\sigma^{\mu\nu}}{{{Q_L}_{t}}_{cl}}).
\eea
To further eliminate the spurion self-contraction, we take the idempotent elements in the j-basis in Eq.~\eqref{eq:om'basis1}. Since the generators in Eq.~\ref{eq:mgen3} are identities in the $SU(2)$ j-basis in Eq.~\eqref{m'basis2}, the generators in the operator j-basis in Eq.~\eqref{eq:om'basis1} are
\bea
\mathcal{D}^{(j)}(Q_L)_1 &=& \mathcal{D}^{(j)}_{SU(3)}(Q_L)_1\otimes \mathcal{D}^{(m)}_{\mathcal{B}}(Q_L)_1 = \left(
\begin{array}{cc}
 -\frac{1}{2} & \frac{1}{8} \\
 6 & \frac{1}{2} \\
\end{array}
\right) ,\\
\mathcal{D}^{(j)}(Q_L)_2 &=& \mathcal{D}^{(j)}_{SU(3)}(Q_L)_1\otimes \mathcal{D}^{(m)}_{\mathcal{B}}(Q_L)_1 = \left(
\begin{array}{cc}
 -\frac{1}{2} & \frac{1}{8} \\
 -6 & -\frac{1}{2} \\
\end{array}
\right) ,\\
\mathcal{D}^{(j)}(\bft) &=& \mathcal{D}^{(m)}_{\mathcal{B}}(\bft) = \left(
\begin{array}{cc}
 1 & 0 \\
 0 & 1 \\
\end{array}
\right).
\eea
The 3 idempotent elements are 
\be
\caly[{\tiny\yt{234},\yt{56}}] = \caly[{\tiny\yt{234}}]\caly[{\tiny\yt{56}}],\quad \caly[{\tiny\yt{23,4},\yt{56}}] = \caly[{\tiny\yt{23,4}}]\caly[{\tiny\yt{56}}],\quad \caly[{\tiny\yt{2,3,4},\yt{56}}] = \caly[{\tiny\yt{23,4}}]\caly[{\tiny\yt{56}}],
\ee
where the idempotent elements of the $S_2$ is calculated in the same way as the first example in Eq.~\eqref{eq:mgen1}, 
\be
\mathcal{Y}[\mathcal{{\tiny\yt{56}}}] = \mathbf{I}_{3\times 3} + \mathcal{D}^{(j)}(\bft),
\ee
but those of the $S_3$ are a little bit complicated, and we give their expressions in terms of generators,
\bea
\caly[\tiny\yt{234}] &=& \mathbf{I}_{2\times 2} + \mathcal{D}^{(j)}(Q_L)_1 + {\mathcal{D}^{(j)}(Q_L)_1}{\mathcal{D}^{(j)}(Q_L)_2}{\mathcal{D}^{(j)}(Q_L)_1}^{-1} \notag \\
&+& {\mathcal{D}^{(j)}(Q_L)_1}{\mathcal{D}^{(j)}(Q_L)_2}{\mathcal{D}^{(j)}(Q_L)_1}{\mathcal{D}^{(j)}(Q_L)_2}^{-1}{\mathcal{D}^{(j)}(Q_L)_1} \notag\\
&+& {\mathcal{D}^{(j)}(Q_L)_1}{\mathcal{D}^{(j)}(Q_L)_2}{\mathcal{D}^{(j)}(Q_L)_1}{\mathcal{D}^{(j)}(Q_L)_2}^{-1}{\mathcal{D}^{(j)}(Q_L)_1}{\mathcal{D}^{(j)}(Q_L)_2}{\mathcal{D}^{(j)}(Q_L)_1}{\mathcal{D}^{(j)}(Q_L)_2}^{-1}\notag \\
&+& {\mathcal{D}^{(j)}(Q_L)_2},\\
\caly[\tiny\yt{23,4}] &=& \mathbf{I}_{2\times 2} - \mathcal{D}^{(j)}(Q_L)_1 - {\mathcal{D}^{(j)}(Q_L)_1}{\mathcal{D}^{(j)}(Q_L)_2}{\mathcal{D}^{(j)}(Q_L)_1}^{-1} \notag \\
&+& {\mathcal{D}^{(j)}(Q_L)_1}{\mathcal{D}^{(j)}(Q_L)_1}{\mathcal{D}^{(j)}(Q_L)_2}{\mathcal{D}^{(j)}(Q_L)_1}^{-1},\\
\caly[\tiny\yt{2,3,4}] &=& \mathbf{I}_{2\times 2} - \mathcal{D}^{(j)}(Q_L)_1 - {\mathcal{D}^{(j)}(Q_L)_1}{\mathcal{D}^{(j)}(Q_L)_2}{\mathcal{D}^{(j)}(Q_L)_1}^{-1} \notag \\
&+& {\mathcal{D}^{(j)}(Q_L)_1}{\mathcal{D}^{(j)}(Q_L)_2}{\mathcal{D}^{(j)}(Q_L)_1}{\mathcal{D}^{(j)}(Q_L)_2}^{-1}{\mathcal{D}^{(j)}(Q_L)_1}\notag \\
&+& {\mathcal{D}^{(j)}(Q_L)_1}{\mathcal{D}^{(j)}(Q_L)_2}{\mathcal{D}^{(j)}(Q_L)_1}{\mathcal{D}^{(j)}(Q_L)_2}^{-1}{\mathcal{D}^{(j)}(Q_L)_1}{\mathcal{D}^{(j)}(Q_L)_2}{\mathcal{D}^{(j)}(Q_L)_1}{\mathcal{D}^{(j)}(Q_L)_2}^{-1} \notag\\
&+& {\mathcal{D}^{(j)}(Q_L)_2}\,.
\eea
Thus we obtain the idempotent elements that
\be
    \caly[{\tiny\yt{234},\yt{56}}] = \caly[{\tiny\yt{2,3,4},\yt{56}}] = \left(
\begin{array}{cc}
 0 & 0 \\
 0 & 0 \\
\end{array}
\right),\quad \caly[{\tiny\yt{23,4},\yt{56}}] = \left(
\begin{array}{cc}
 \frac{1}{2} & \frac{1}{8} \\
 2 & \frac{1}{2} \\
\end{array}
\right), 
\ee
in which the second one is of rank 1. It implies there is only one operator in the f-basis, with the flavor symmetry {\tiny\yng(2,1)},
\be
{\color{red}\calo^{Uh\psi^4}_{140}} = \caly[{\tiny\yt{rs,t}}]\epsilon^{abc}\epsilon^{ln}\epsilon^{km}((\bft{{l_L}^T})_{pm}C(\bft{q_L})_{ran})({{q_L}^T}_{rak}C{q_L}_{tcl})\calf^{Uh\psi^4}_{140}(h)\,, 
\ee
which is missing in the previous literature such as the Ref.~\cite{Merlo:2016prs}.
Several comments are in order. 
First, the particles indices $2,3,4$ in the idempotent element of the f-basis operators, have been replaced to the flavor indices $r,s,t$, which is the convention of writing the operators with repeated fields carrying flavor numbers in this paper.
Second, unlike the $Q^3 L$ operators, the operators projected by $\caly[{\tiny\yt{234},\yt{56}}]$ and $\caly[{\tiny\yt{2,3,4},\yt{56}}]$ are all eliminated in this type. Thus only certain flavor symmetry is allowed for this type of operators.

In summary, the Young tensor method first constructs effective operators in the on-shell y-basis, which is obtained by filling primary Young diagram. If such operators contain the NGBs, the Adler zero condition needs to be imposed to the on-shell amplitude. After that, transforming the y-basis to the m-basis obtain monomial operators for both Lorentz and gauge structures, with the spurion included in the operators. Finally, repeated fields should be considered for both the dynamical fields and the spurion in the operators. In this stage, all physical operators are classified by permutations of repeated fields, this basis is the p-basis. For operators involving the flavor structure, the p-basis should be reduced to the f-basis respecting the flavor symmetry. In this work, we always write operators in the p/f-basis with the idempotent elements $\caly$ of the symmetric group in front of them to indicate their permutation property, but in the case that repeated field have flavor number 1, the idempotent elements are omitted.   


\section{Complete Bosonic Operator List at NLO}
\label{sec:boson}

The NLO bosonic operators are divided into the following types: $UhD^4, X^2Uh, XUhD^2, X^3$. 
Although the Higgs boson $h$ in the HEFT is gauge singlet and thus can appear in the effective operators freely by the means of dimensionless function $\calf$ in Eq.~\eqref{eq:calf}, there are cases that the number of $h$ can not be arbitrary. In this section the full list of NLO bosonic operators will be presented, and we will explain the explicit form of function $\calf$ in every type.

\paragraph{Type: $UhD^4$:}

This type includes the operators with 4 derivatives applied on the NGBs and the Higgs, thus the building blocks contain $\bfv_\mu$, $h$ and the spurion $\bft$. The contractions of $SU(2)$ group among $\bft,\bfv_\mu$ are represented by the matrix trace $\lra{\dots}$. In particular, the operator $\calo^{UhD^4}_{14}$ in this type is missing in Ref.~\cite{Brivio:2016fzo}.
\begin{align}
&\calo^{UhD^4}_1=\lra{\bfv_\mu\bfv^\mu}^2\calf^{UhD^4}_1(h), & \calo^{UhD^4}_{2} &=\lra{\bfv_\mu\bfv_\nu}\lra{\bfv^\mu\bfv^\nu}\calf^{UhD^4}_{2}(h),\notag\\
&\calo^{UhD^4}_{3} = \lra{\bft\bfv_\mu}\lra{\bft\bfv_\nu}\lra{\bfv^\mu\bfv^\nu}\calf^{UhD^4}_{3}(h), & \calo^{UhD^4}_{4} &= \lra{\bft\bfv_\mu}\lra{\bft\bfv^\mu}\lra{\bfv_\nu\bfv^\nu}\calf^{UhD^4}_{4}(h), \notag\\
&\calo^{UhD^4}_{5} = (\lra{\bft\bfv_\mu}\lra{\bft\bfv^\mu})^2\calf^{UhD^4}_{5}(h), &  \calo^{UhD^4}_6 &= \lra{\bfv_\mu\bfv^\mu}\lra{\bft\bfv_\nu}\frac{D^\nu h}{v}\calf^{UhD^4}_6(h), \notag\\
&\calo^{UhD^4}_7 = \lra{\bfv_\mu\bfv_\nu}\lra{\bft\bfv^\mu}\frac{D^\nu h}{v}\calf^{UhD^4}_7(h), & \calo^{UhD^4}_{8} &= \lra{\bft\bfv_\mu\bfv_\nu}\lra{\bft\bfv^\mu}\frac{D^\nu h}{v}\calf^{UhD^4}_{8}(h), \notag\\
&\calo^{UhD^4}_{9}=\lra{\bft\bfv_\mu}\lra{\bft\bfv^\mu}\lra{\bft\bfv_\nu}\frac{D^\nu h}{v}\calf^{UhD^4}_{9}(h), & \calo^{UhD^4}_{10} &=\lra{\bfv_\mu\bfv_\nu}\frac{h D^\mu D^\nu h}{v^2}\calf^{UhD^4}_{10}(h), \notag\\
&\calo^{UhD^4}_{11} =\lra{\bfv_\mu\bfv^\mu}\frac{D_\nu h D^\nu h}{v^2}\calf^{UhD^4}_{11}(h), & \calo^{UhD^4}_{12} &= \lra{\bft\bfv_\mu}\lra{\bft\bfv_\nu}\frac{hD^\mu D^\nu}{v^2}\calf^{UhD^4}_{12}(h), \notag\\
&\calo^{UhD^4}_{13}=\lra{\bft\bfv_\mu}\lra{\bft\bfv^\mu}\frac{D_\nu h D^\nu h}{v^2}\calf^{UhD^4}_{13}(h), & {\color{red}\calo^{UhD^4}_{14}}&=\lra{\bft\bfv_\mu}\frac{hD^\nu hD_\nu D^\mu h}{v^3}\calf^{UhD^4}_{14}(h), \notag\\
&\calo^{UhD^4}_{15}=\frac{h^2(D_\mu D_\nu h)(D^\mu D^\nu h)}{v^4}\calf^{UhD^4}_{15}(h).
\end{align}
The function $\calf^{UhD^4}_i$ in this type takes the form that
\beq
\calf^{UhD^4}_i(h) = 1+\sum_{n=1}c^{UhD^4}_n(\frac{h}{v})^n,\quad(i=1,2,\dots,15).
\eeq

\paragraph{Type $X^2Uh$:}
The operators in this type involves in two gauge field strength tensors with two gauge coupling constants to be consistent with the power-counting. We recombine the building blocks $X_{\mu\nu}$ and $\tilde{X}_{\mu\nu}$ as
\be
X_L = \frac{1}{2}(X-i\tilde{X}),\quad X_R = \frac{1}{2}(X+i\tilde{X}),\quad X=B,W,G,
\ee
by which, we obtain the 10 operators in this type:
\begin{align}
	&\calo^{X^2Uh}_1={g'}^2{B_L}^{\mu\nu} {B_L}_{\mu\nu}\calf^{X^2Uh}_1(h), & \calo^{X^2Uh}_2 &= {g'}^2{B_R}^{\mu\nu} {B_R}_{\mu\nu}\calf^{X^2Uh}_2(h), \notag \\
	&\calo^{X^2Uh}_3=g^2_s\lra{{G_L}_{\mu\nu}{G_L}^{\mu\nu}}\calf^{X^2Uh}_3(h), &\calo^{X^2Uh}_4 &=g^2_s\lra{{G_R}_{\mu\nu}{G_R}^{\mu\nu}}\calf^{X^2Uh}_4(h),\notag\\
	&\calo^{X^2Uh}_5=g^2\lra{{W_L}_{\mu\nu}{W_L}^{\mu\nu}}\calf^{X^2Uh}_5(h), & \calo^{X^2Uh}_6 &=g^2\lra{{W_R}_{\mu\nu}{W_R}^{\mu\nu}}\calf^{X^2Uh}_6(h), \notag\\
	&\calo^{X^2Uh}_7=gg'{B_L}_{\mu\nu}\lra{{W_L}^{\mu\nu}\bft}\calf^{X^2Uh}_7(h), &\calo^{X^2Uh}_8 &=gg'{B_R}_{\mu\nu}\lra{{W_R}^{\mu\nu}\bft}\calf^{X^2Uh}_8(h), \notag\\
	&\calo^{X^2Uh}_9=g^2\lra{{W_L}_{\mu\nu} \bft}^2\calf^{X^2Uh}_9(h), &\calo^{X^2Uh}_{10} &=g^2\lra{{W_R}_{\mu\nu} \bft}^2\calf^{X^2Uh}_{10}(h).
\end{align}
The $\calf$ in this type takes the form that
\beq
\calf^{X^2Uh}_i(h) = 1+\sum_{n=1}c^{X^2Uh}_n(\frac{h}{v})^n,\quad (i=1,2,\dots,10).
\eeq

\paragraph{Type $XUhD^2$:}
The operators in this type involves in one gauge field strength tensor and two derivatives. Operators $\calo^{XUhD^2}_{2,4,8}$ are missing in Ref.~\cite{Brivio:2016fzo}. It should be noted that the frist term in the expansion of the $\calf$ function in this type is $h/v$ instead of 1. The full list of the operators in this type is:
\begin{align}
		\calo^{XUhD^2}_1 &= g\lra{{W_L}_{\mu\nu}[\bfv^\mu,\bfv^\nu]}\calf^{XUhD^2}_1(h),	& {\color{red}\calo^{XUhD^2}_2} &= g\lra{{W_R}_{\mu\nu}[\bfv^\mu,\bfv^\nu]}\calf^{XUhD^2}_2(h), \notag\\
		\calo^{XUhD^2}_3 &= g'{B_L}_{\mu\nu}\lra{\bft[\bfv^\mu,\bfv^\nu]}\calf^{XUhD^2}_3(h), & {\color{red}\calo^{XUhD^2}_4} &= g'{B_R}_{\mu\nu}\lra{\bft[\bfv^\mu,\bfv^\nu]}\calf^{XUhD^2}_4(h),\notag\\
\calo^{XUhD^2}_5 &= g\lra{{W_L}_{\mu\nu}\bfv^\mu}\lra{\bft\bfv^\nu}\calf^{XUhD^2}_5(h), &\calo^{XUhD^2}_6 &= g\lra{{W_R}_{\mu\nu}\bfv^\mu}\lra{\bft\bfv^\nu}\calf^{XUhD^2}_6(h), \notag\\
\calo^{XUhD^2}_7 &= g\lra{{W_L}_{\mu\nu}\bft}\lra{\bft[\bfv^\mu,\bfv^\nu]}\calf^{XUhD^2}_7(h), & {\color{red}\calo^{XUhD^2}_8} &= g\lra{\bft\bfv^\mu}\lra{\bft[{W_R}_{\mu\nu},\bfv^\nu]}\calf^{XUhD^2}_8(h).
\end{align}
The $\calf$ function takes the form that
\beq
\calf^{XUhD^2}_i(h)=\frac{h}{v}+\sum_{n=2}c^{XUhD^2}_n(\frac{h}{v})^n,\quad(i=1,2,\dots,8).
\eeq
Note that the derivative can not only apply on the NGBs, generating operators above, but also on the Higgs boson $h$, which, however gives no more operators. The argument is as follows: If a $\bfv_\mu$ in the operators above is replaced by $D_\mu h$, the operators would have at least two $h$'s, since there is at least 1 $h$ in $\calf$. Thus the operators would contain the repeated field $h$ and spin-statistics needs to be applied. Because the $h$ carries no flavor number, only the operators containing the symmetric $h$'s are physical in the f-basis, while the Young tensor method utilized in this paper implies that such operators do not exist in this type.
Considering this, several extra operators appeared in Ref.~\cite{Brivio:2016fzo} of this type are actually redundant.

\paragraph{Type $X^3$:} 
In this type, the operators are built of three gauge field strengths, and there are only 4 operators:
\begin{align}
		\calo^{X^3}_1 &=f^{ABC}{{G_L}^A}_{\mu\nu}{{G_L}^C}^{\mu\lambda}{{{G_L}^B}^\nu}_\lambda\calf^{X^3}_1(h), & \calo^{X^3}_2 &=f^{ABC}{G_R^A}_{\mu\nu}{G_R^C}^{\mu\lambda}{{G_R^B}^\nu}_\lambda\calf^{X^3}_2(h), \notag\\
		\calo^{X^3}_3 &= \epsilon^{IJK}{W_L^I}_{\mu\nu}{W_L^K}^{\nu\lambda}{{W_L^J}^\nu}_\lambda\calf^{x^3}_3(h), &\calo^{X^3}_4 &= \epsilon^{IJK}{W_R^I}_{\mu\nu}{W_R^K}^{\nu\lambda}{{W_R^J}^\nu}_\lambda\calf^{X^3}_4(h)\,,\notag \\
		\calo^{X^3}_5 &= \epsilon^{IJK}\bft^K{B_L}_{\mu\nu}{W^J_L}^{\lambda\mu}{{W^I_L}_\lambda}^\nu\calf^{X^3}_5(h), & 	\calo^{X^3}_6 &= \epsilon^{IJK}\bft^K{B_R}_{\mu\nu}{W^J_R}^{\lambda\mu}{{W^I_R}_\lambda}^\nu\calf^{X^3}_6(h).
\end{align}
The $\calf$ function in this type takes the form that
\beq
\calf^{X^3}_i(h)=1+\sum_{n=1}c^{X^3}_n(\frac{h}{v})^n\,,\quad (i=1,2,\dots,6).
\eeq


\section{Complete NLO Operators involving in Fermion}
\label{sec:fermion}

The fermion sector contains the operators composed by a two-fermion current $\bar{\psi}\Gamma\psi$ and several bosonic fields, called the two-fermion operators, and the operators composed by four fermions, called the 4-fermion operators. Besides, the 4-fermion operators violating the baryon numbers are also presented in this section. Compared with previous works Ref.~\cite{Brivio:2016fzo,Merlo:2016prs}, there are 9(6) terms of operators in this sector missed for the HEFT with (without) the sterile neutrino, and we present them here,
\begin{align}
     {\color{red}\calo^{Uh\psi^4}_{33}} &= ({\overline{q}_L}_s\gamma_\mu\tau^I\bft{q_L}_p)({\overline{q}_R}_r\gamma^\mu\bfu^\dagger\tau^I\bfu{q_R}_t)\calf^{Uh\psi^4}_{33}(h), \notag\\
     {\color{red}\calo^{Uh\psi^4}_{34}} &= ({\overline{q}_L}_s\gamma_\mu\lambda^A\tau^I\bft{q_L}_p)({\overline{q}_R}_r\gamma^\mu\lambda^A\bfu^\dagger\tau^I\bfu{q_R}_t)\calf^{Uh\psi^4}_{34}(h), \notag \\
	 {\color{red}\calo^{Uh\psi^4}_{89}} &= {\color{gray}({\overline{l}_L}_s\gamma_\mu\tau^I{l_L}_p)({\overline{l}_R}_t\sigma^\mu\tau^I\bfu^\dagger\bft\bfu{l_R}_s)\calf^{Uh\psi^4}_{89}(h),}\notag \\
	 {\color{red}\calo^{Uh\psi^4}_{107}} &= ({\overline{l}_L}_s\gamma_\mu\tau^I\bft{l_L}_p)({\overline{q}_L}_t\gamma^\mu\tau^I{q_L}_r)\calf^{Uh\psi^4}_{107}(h), \notag\\
	 {\color{red}\calo^{Uh\psi^4}_{113}} &= {\color{gray}({\overline{l}_R}_s\gamma_\mu\tau^I\bft{l_R}_p)({\overline{q}_R}_t\gamma^\mu\tau^I{q_R}_r)\calf^{Uh\psi^4}_{113}(h),}\notag \\
	 {\color{red}\calo^{Uh\psi^4}_{119}} &= {\color{gray}({\overline{l}_R}_s\gamma_\mu\bfu^\dagger\tau^I\bft\bfu{l_R}_p)({\overline{q}_L}_t\gamma^\mu\tau^I{q_L}_r)\calf^{Uh\psi^4}_{119}(h),} \notag \\
	 {\color{red}\calo^{Uh\psi^4}_{125}} &= ({\overline{l}_L}_s\gamma_\mu\tau^I\bft{l_L}_p)({\overline{q}_R}_t\gamma^\mu\bfu^\dagger\tau^I\bfu{q_R}_r)\calf^{Uh\psi^4}_{125}(h), & \notag\\
	 {\color{red}\calo^{Uh\psi^4}_{140}} &= \caly[{\tiny\yt{rs,t}}]\epsilon^{abc}\epsilon^{ln}\epsilon^{km}((\bft{l_L}^T)_{pm}C(\bft{q_L})_{ran})({q_L}^T_{rak}C{q_L}_{tcl}) \calf^{Uh\psi^4}_{159}(h),  \notag \\
	 {\color{red}\calo^{Uh\psi^4}_{160}} &= \caly[{\tiny\yt{rs,t}}]\epsilon^{abc}\epsilon^{km}\epsilon^{ln}((\bft{l_R}^T)_{pm}C(\bft{q_R})_{ran})({q_R}^T_{sbk}C{q_R}_{tcl}) \calf^{Uh\psi^4}_{160}(h). 
\end{align}

The missing operators are marked by the red color and the operators involving right-handed neutrino are presented by the gray color.  
Besides, previous works only consider the fermion operators with flavor number 1, while here we consider the general flavor structures.
Note that the last two terms of operators only appear when there are three generation fermions, because only the mixed flavor symmetry structure is allowed. 

\subsection{Fermion Current Operators}

\subsubsection{Quark Current Operators}

\paragraph{Type $\psi^2 UhD$: }
In this type, the operators are composed by a fermion-current and a single derivative. The full list is
\begin{align}
\calo^{\psi^2 UhD}_1 &= ({\overline{q}_L}_p\gamma^\mu\bfv_\mu{q_L}_r)\calf^{\psi^2 UhD}_1(h), & \calo^{\psi^2 UhD}_2 &= ({\overline{q}_L}_p\gamma^\mu {q_L}_{r})\lra{\bft\bfv_\mu}\calf^{\psi^2 UhD}_2(h), \notag\\
\calo^{\psi^2 UhD}_3 &= ({\overline{q}_L}_p \gamma^\mu[\bfv_\mu,\bft]{q_L}_r)\calf^{\psi^2 UhD}_3(h), & \calo^{\psi^2 UhD}_4 &= ({\overline{q}_L}_p\gamma^\mu\bft{q_L}_r)\lra{\bft\bfv_\mu}\calf^{\psi^2 UhD}_4(h), \notag\\
\calo^{\psi^2 UhD}_5 &= ({\overline{q}_R}_p\gamma^\mu\bfu^\dagger \bfv_\mu\bfu {q_R}_r)\calf^{\psi^2 UhD}_5(h), & \calo^{\psi^2 UhD}_6 &= ({\overline{q}_R}_p\gamma^\mu {q_R}_{r})\lra{\bft\bfv_\mu}\calf^{\psi^2 UhD}_6(h), \notag\\
\calo^{\psi^2 UhD}_7 &= ({\overline{q}_R}_p \gamma^\mu\bfu^\dagger[\bfv_\mu,\bft]\bfu{q_R}_r)\calf^{\psi^2 UhD}_7(h), & \calo^{\psi^2 UhD}_8 &=({\overline{q}_R}_p\gamma^\mu\bfu^\dagger\bft\bfu{q_R}_r)\lra{\bft\bfv_\mu}\calf^{\psi^2 UhD}_8(h).
\end{align}
The $\calf$ functions in this type takes the form that
\beq
\calf^{\psi^2UhD}_i(h)=1+\sum_{n=1}c_n^{\psi^2UhD}(\frac{h}{v})^n,\quad (i=1,2,\dots,8).
\eeq

\paragraph{Type $\psi^2 UhD^2$: }

\begin{align}
\calo^{\psi^2 UhD^2}_1 &= ({\overline{q}_L}_p\bfu{q_R}_r)\lra{\bfv_\mu\bfv^\mu}\calf^{\psi^2 UhD^2}_1(h), & \calo^{\psi^2 UhD^2}_2 &= ({\overline{q}_L}_p\sigma^{\mu\nu}[\bfv_\mu,\bfv_\nu]\bfu{q_R}_r)\calf^{\psi^2 UhD^2}_2(h), \notag\\
\calo^{\psi^2 UhD^2}_3 &= ({\overline{q}_L}_p\sigma^{\mu\nu}\bfu{q_R}_r)\lra{\bft[\bfv_\mu,\bfv_\nu]}\calf^{\psi^2 UhD^2}_3(h), & \calo^{\psi^2 UhD^2}_4 &= ({\overline{q}_L}_p\bfv_\mu\bfu{q_R}_r)\lra{\bft\bfv^\mu} \calf^{\psi^2 UhD^2}_4(h),\notag\\
\calo^{\psi^2 UhD^2}_5 &= ({\overline{q}_L}_p\sigma^{\mu\nu}\bfv_\mu\bfu{q_R}_r)\lra{\bft\bfv_\nu} \calf^{\psi^2 UhD^2}_5(h), & \calo^{\psi^2 UhD^2}_6 &= ({\overline{q}_L}_p\bft\bfu{q_R}_r)\lra{\bfv_\mu\bfv^\mu}\calf^{\psi^2 UhD^2}_6(h),
\end{align}
\begin{align}
\calo^{\psi^2 UhD^2}_7 &= ({\overline{q}_L}\sigma^{\mu\nu}[\bft,\bfv_\mu]\bfu{q_R}_r)\lra{\bft\bfv_\nu}\calf^{\psi^2 UhD^2}_7(h), & \calo^{\psi^2 UhD^2}_8 &= ({\overline{q}_L}_p[\bft,\bfv_\mu]\bfu{q_R}_r)\lra{\bft\bfv^\mu}\calf^{\psi^2 UhD^2}_8(h), \notag\\
\calo^{\psi^2 UhD^2}_9 &= ({\overline{q}_L}_p\sigma^{\mu\nu}\bft\bfu{q_R}_r)\lra{\bft[\bfv_\mu,\bfv_\nu]}\calf^{\psi^2 UhD^2}_9(h), & \calo^{\psi^2 UhD^2}_{10} &= ({\overline{q}_L}_p\bft\bfu{q_R}_r)\lra{\bft\bfv_\mu}\lra{\bft\bfv^\mu}\calf^{\psi^2 UhD^2}_{10}(h), \notag\\
\calo^{\psi^2 UhD^2}_{11} &= ({\overline{q}_L}_p\bfv_\mu\bfu{q_R}_r)\frac{D^\mu h}{v}\calf^{\psi^2 UhD^2}_{11}(h), & \calo^{\psi^2 UhD^2}_{12} &= (D^\mu{\overline{q}_L}_p\bfv_\mu\bfu{q_R}_r)\calf^{\psi^2 UhD^2}_{12}(h), 
\end{align}
\begin{align}
\calo^{\psi^2 UhD^2}_{13} &= ({\overline{q}_L}_p[\bfv_\mu,\bft]\bfu{q_R}_r)\frac{D^\mu h}{v}\calf^{\psi^2 UhD^2}_{13}(h), & \calo^{\psi^2 UhD^2}_{14} &= (D^\mu {\overline{q}_L}_p[\bfv_\mu,\bft]\bfu{q_R}_r)\calf^{\psi^2 UhD^2}_{14}(h), \notag\\
\calo^{\psi^2 UhD^2}_{15} &= ({\overline{q}_L}_p\bfu{q_R}_r)\lra{\bft\bfv_\mu}\frac{D^\mu h}{v}\calf^{\psi^2 UhD^2}_{15}(h), & \calo^{\psi^2 UhD^2}_{16} &= (D^\mu{\overline{q}_L}_p\bfu{q_R}_r)\lra{\bft\bfv_\mu}\calf^{\psi^2 UhD^2}_{16}(h), \notag\\
\calo^{\psi^2 UhD^2}_{17} &= ({\overline{q}_L}_p\bft\bfu{q_R}_r)\lra{\bft\bfv_\mu}\frac{D^\mu h}{v}\calf^{\psi^2 UhD}_{17}(h), & \calo^{\psi^2 UhD^2}_{18} &= (D^\mu{\overline{q}_L}_p \bft\bfu{q_R}_r)\lra{\bft\bfv_\mu}\calf^{\psi^2 UhD^2}_{18}(h), \notag\\
\calo^{\psi^2 UhD^2}_{19} &= ({\overline{q}_L}_p\bfu {q_R}_r)\frac{D_\mu h D^\mu h}{v^2}\calf^{\psi^2 UhD^2}_{19}(h), & \calo^{\psi^2 UhD^2}_{20} &= ({\overline{q}_L}_p\bft\bfu{q_R}_r)\frac{D_\mu h D^\mu h}{v^2}\calf^{\psi^2 UhD^2}_{20}(h).
\end{align}
$\calf$ functions in this type takes the form that
\beq
\calf^{\psi^2 UhD^2}_i=1+\sum_{n=1}c_n^{\psi^2UhD^2}(\frac{h}{v})^n,\quad (i=1,2,\dots,20).
\eeq

\paragraph{Type $\psi^2 UhX$: }\
This type is similar with the type $\psi^2 UhD^2$, operators which are listed as following:
\begin{align}
\calo^{\psi^2 UhX}_1 &= g'({\overline{q}_L}\sigma^{\mu\nu}\bfu{q_R}_r){B_R}_{\mu\nu}\calf^{\psi^2 UhX}_1(h), & \calo^{\psi^2 UhX}_2 &= g'({\overline{q}_L}_p\sigma^{\mu\nu} \bft\bfu{q_R}_r){B_R}_{\mu\nu} \calf^{\psi^2 UhX}_2(h), \notag\\
\calo^{\psi^2 UhX}_3 &= g_s({\overline{q}_L}_p\sigma^{\mu\nu}{G_R}_{\mu\nu}\bfu{q_R}_r) \calf^{\psi^2 UhX}_3(h), & \calo^{\psi^2 UhX}_4 &= g_s({\overline{q}_L}_p\sigma^{\mu\nu}{G_R}_{\mu\nu}\bft\bfu{q_R}_r) \calf^{\psi^2 UhX}_4(h), \notag\\
\calo^{\psi^2 UhX}_5 &= g({\overline{q}_L}_p\sigma^{\mu\nu}{W_R}\bfu{q_R}_r)\calf^{\psi^2 UhX}_5(h), & \calo^{\psi^2 UhX}_6 &= g({\overline{q}_L}_p\sigma^{\mu\nu}[\bft,{W_R}]\bfu{q_R}_r)\calf^{\psi^2 UhX}_6(h), \notag\\
\calo^{\psi^2 UhX}_7 &= g({\overline{q}_L}_p\sigma^{\mu\nu}\bfu{q_R}_r)\lra{\bft W_R} \calf^{\psi^2 UhX}_7(h), & \calo^{\psi^2 UhX}_8 &=g ({\overline{q}_L}_p\sigma^{\mu\nu}\bft\bfu{q_R}_r)\lra{\bft {W_R}}\calf^{\psi^2 UhX}_8(h).
\end{align}
The $\calf$ functions in this type is that
\beq
\calf^{\psi^2 UhX}_i=1+\sum_n c^{\psi^2 UhX}_n(\frac{h}{v})^n,\quad (i=1,2,\dots,8).
\eeq

\subsubsection{Lepton-Current Operators}

The lepton-current operators are similar with the quark-current operators with the quark fields replaced by the lepton fields. With this analogy, all previous operators can be transformed to lepton-current operators. For example,
\begin{align}
    &\calo^{\psi^2UhD}_2\Rightarrow ({\overline{l}_L}_p\gamma^\mu{l_L}_r)\lra{\bft\bfv_\mu}\calf^{\psi^2UhD}_{10}(h), \notag\\
    &\calo^{\psi^2UhD^2}_{19}\Rightarrow ({\overline{l}_L}_p\bfu{l_R}_r)\frac{D_\mu hD^\mu h}{v^2}\calf^{{\psi^2 UhD^2}}_{39}(h).
\end{align}
However, it should be emphasized that the quarks have $SU(3)$ quantum numbers while leptons do not have. Thus $\calo^{\psi^2UhX}_{7,8}$ have no correspondences in lepton-current operators. Here we present them classified by classes.
\paragraph{$\psi^2UhD$:}
\begin{align}
		\calo^{\psi^2 UhD}_9 &= ({\overline{l}_L}_p\gamma^\mu\bfv_\mu{l_L}_r)\calf^{\psi^2 UhD}_9(h), & \calo^{\psi^2 UhD}_{10} &= ({\overline{l}_L}_p\gamma^\mu {l_L}_{r})\lra{\bft\bfv_\mu}\calf^{\psi^2 UhD}_{10}(h), \notag\\
		\calo^{\psi^2 UhD}_{11} &= ({\overline{l}_L}_p \gamma^\mu[\bfv_\mu,\bft]{l_L}_r)\calf^{\psi^2 UhD}_{11}(h), & \calo^{\psi^2 UhD}_{12} &= ({\overline{l}_L}_p\gamma^\mu\bft{l_L}_r)\lra{\bft\bfv_\mu}\calf^{\psi^2 UhD}_{12}(h), \notag\\
		\calo^{\psi^2 UhD}_{13} &= ({\overline{l}_R}_p\gamma^\mu\bfu^\dagger \bfv_\mu\bfu {l_R}_r)\calf^{\psi^2 UhD}_{13}(h), & \calo^{\psi^2 UhD}_{14} &= {\color{gray}({\overline{l}_R}_p\gamma^\mu {l_R}_{r})\lra{\bft\bfv_\mu}\calf^{\psi^2 UhD}_{14}(h),} \notag\\
		\calo^{\psi^2 UhD}_{15} &= {\color{gray}({\overline{l}_R}_p \gamma^\mu\bfu^\dagger[\bfv_\mu,\bft]\bfu{l_R}_r)\calf^{\psi^2 UhD}_{15}(h),} & \calo^{\psi^2 UhD}_{16} &={\color{gray}({\overline{l}_R}_p\gamma^\mu\bfu^\dagger\bft\bfu{l_R}_r)\lra{\bft\bfv_\mu}\calf^{\psi^2 UhD}_{16}(h).}
\end{align}

\paragraph{$\psi^2UhD^2$:}
\begin{align}
		\calo^{\psi^2 UhD^2}_{21} &= ({\overline{l}_L}_p\bfu{l_R}_r)\lra{\bfv_\mu\bfv^\mu}\calf^{\psi^2 UhD^2}_{21}(h), & \calo^{\psi^2 UhD^2}_{22} &= ({\overline{l}_L}_p\sigma^{\mu\nu}[\bfv_\mu,\bfv_\nu]\bfu{l_R}_r)\calf^{\psi^2 UhD^2}_{22}(h), \notag\\
		\calo^{\psi^2 UhD^2}_{23} &= ({\overline{l}_L}_p\sigma^{\mu\nu}\bfu{l_R}_r)\lra{\bft[\bfv_\mu,\bfv_\nu]}\calf^{\psi^2 UhD^2}_{23}(h), & \calo^{\psi^2 UhD^2}_{24} &= ({\overline{l}_L}_p\bfv_\mu\bfu{l_R}_r)\lra{\bft\bfv^\mu} \calf^{\psi^2 UhD^2}_{24}(h),\notag\\
		\calo^{\psi^2 UhD^2}_{25} &= ({\overline{l}_L}_p\sigma^{\mu\nu}\bfv_\mu\bfu{l_R}_r)\lra{\bft\bfv_\nu} \calf^{\psi^2 UhD^2}_{25}(h), & \calo^{\psi^2 UhD^2}_{26} &= {\color{gray}({\overline{l}_L}_p\bft\bfu{l_R}_r)\lra{\bfv_\mu\bfv^\mu}\calf^{\psi^2 UhD^2}_{26}(h),} \notag\\
		\calo^{\psi^2 UhD^2}_{27} &= {\color{gray}({\overline{l}_L}\sigma^{\mu\nu}[\bft,\bfv_\mu]\bfu{l_R}_r)\lra{\bft\bfv_\nu}\calf^{\psi^2 UhD^2}_{27}(h),} & \calo^{\psi^2 UhD^2}_{28} &= {\color{gray}({\overline{l}_L}_p[\bft,\bfv_\mu]\bfu{l_R}_r)\lra{\bft\bfv^\mu}\calf^{\psi^2 UhD^2}_{28}(h),} \notag\\
		\calo^{\psi^2 UhD^2}_{29} &= {\color{gray}({\overline{l}_L}_p\sigma^{\mu\nu}\bft\bfu{l_R}_r)\lra{\bft[\bfv_\mu,\bfv_\nu]}\calf^{\psi^2 UhD^2}_{29}(h),} & \calo^{\psi^2 UhD^2}_{30} &= {\color{gray}({\overline{l}_L}_p\bft\bfu{l_R}_r)\lra{\bft\bfv_\mu}\lra{\bft\bfv^\mu}\calf^{\psi^2 UhD^2}_{30}(h),} \notag\\
\calo^{\psi^2 UhD^2}_{31} &= ({\overline{l}_L}_p\bfv_\mu\bfu{l_R}_r)\frac{D^\mu h}{v}\calf^{\psi^2 UhD^2}_{31}(h), & \calo^{\psi^2 UhD^2}_{32} &= (D^\mu{\overline{l}_L}_p\bfv_\mu\bfu{l_R}_r)\calf^{\psi^2 UhD^2}_{32}(h), \notag\\
\calo^{\psi^2 UhD^2}_{33} &= {\color{gray}({\overline{l}_L}_p[\bfv_\mu,\bft]\bfu{l_R}_r)\frac{D^\mu h}{v}\calf^{\psi^2 UhD^2}_{33}(h),} & \calo^{\psi^2 UhD^2}_{34} &= {\color{gray}(D^\mu {\overline{l}_L}_p[\bfv_\mu,\bft]\bfu{l_R}_r)\calf^{\psi^2 UhD^2}_{34}(h),} \notag\\
\calo^{\psi^2 UhD^2}_{35} &= ({\overline{l}_L}_p\bfu{l_R}_r)\lra{\bft\bfv_\mu}\frac{D^\mu h}{v}\calf^{\psi^2 UhD^2}_{35}(h), & \calo^{\psi^2 UhD^2}_{36} &= (D^\mu{\overline{l}_L}_p\bfu{l_R}_r)\lra{\bft\bfv_\mu}\calf^{\psi^2 UhD^2}_{36}(h), \notag\\
\calo^{\psi^2 UhD^2}_{37} &= {\color{gray}({\overline{l}_L}_p\bft\bfu{l_R}_r)\lra{\bft\bfv_\mu}\frac{D^\mu h}{v}\calf^{\psi^2 UhD}_{37}(h),} & \calo^{\psi^2 UhD^2}_{38} &= {\color{gray}(D^\mu{\overline{l}_L}_p \bft\bfu{l_R}_r)\lra{\bft\bfv_\mu}\calf^{\psi^2 UhD^2}_{38}(h),} \notag\\
\calo^{\psi^2 UhD^2}_{39} &= ({\overline{l}_L}_p\bfu {l_R}_r)\frac{D_\mu h D^\mu h}{v^2}\calf^{\psi^2 UhD^2}_{39}(h), & \calo^{\psi^2 UhD^2}_{40} &= {\color{gray}({\overline{l}_L}_p\bft\bfu{l_R}_r)\frac{D_\mu h D^\mu h}{v^2}\calf^{\psi^2 UhD^2}_{40}(h).}
\end{align}

\paragraph{$\psi^2UhX$:}
\begin{align}
		\calo^{\psi^2 UhX}_9 &= g'({\overline{l}_L}\sigma^{\mu\nu}\bfu{l_R}_r){B_R}_{\mu\nu}\calf^{\psi^2 UhX}_9(h), & \calo^{\psi^2 UhX}_{10} &= {\color{gray}g'({\overline{l}_L}_p\sigma^{\mu\nu} \bft\bfu{l_R}_r){B_R}_{\mu\nu} \calf^{\psi^2 UhX}_{10}(h),} \notag\\
		\calo^{\psi^2 UhX}_{11} &= g({\overline{l}_L}_p\sigma^{\mu\nu}{W_R}\bfu{l_R}_r)\calf^{\psi^2 UhX}_{11}(h), & \calo^{\psi^2 UhX}_{12} &= {\color{gray}g({\overline{l}_L}_p\sigma^{\mu\nu}[\bft,{W_R}]\bfu{l_R}_r)\calf^{\psi^2 UhX}_{12}(h),} \notag\\
		\calo^{\psi^2 UhX}_{13} &= g({\overline{l}_L}_p\sigma^{\mu\nu}\bfu{l_R}_r)\lra{\bft W_R} \calf^{\psi^2 UhX}_{13}(h), & \calo^{\psi^2 UhX}_{14} &= {\color{gray}({\overline{l}_L}_p\sigma^{\mu\nu}\bft\bfu{l_R}_r)\lra{\bft {W_R}}\calf^{\psi^2 UhX}_{14}(h).}
\end{align}

\subsection{Four-Fermion operators}

The four-fermion operators are divided into four classes: pure quark operators, pure lepton operators, mixed quark-lepton operators and baryon-number-violating operators. In this sector, the operators presented in \cite{Buchalla:2013rka,Brivio:2016fzo,Merlo:2016prs} are not complete, and in the following we will list all the four-fermion operators with the flavor structures. In Appendix~\ref{app:operatorlist}, the detailed comparisons among literature and our result are tabulated.  The missing operators are marked by the red color and the operators involving right-handed neutrino are presented by the gray color.  

\subsubsection{Pure Quark Operators}
\paragraph{Type ${Q_L^\dagger}^2{Q_R}^2$: }

{\small
\begin{align}
\calo^{Uh\psi^4}_1 &= \caly[{\tiny\yt{pr},\yt{st}}]({\overline{q}_L}_p\bfu{q_R}_s)({\overline{q}_L}_r\bfu{q_R}_t) \calf^{Uh\psi^4}_1(h), & \calo^{Uh\psi^4}_2 &= \caly[{\tiny\yt{pr},\yt{st}}]({\overline{q}_L}_p\lambda^A\bfu{q_R}_s)({\overline{q}_L}_r\lambda^A\bfu{q_R}_t)\calf^{Uh\psi^4}_2(h), \notag\\
\calo^{Uh\psi^4}_3 &= \caly[{\tiny\yt{pr},\yt{st}}]({\overline{q}_L}_p\tau^I\bfu{q_R}_s)({\overline{q}_L}_r\tau^I\bfu{q_R}_t)\calf^{Uh\psi^4}_3(h), & \calo^{Uh\psi^4}_4 &= \caly[{\tiny\yt{pr},\yt{st}}]({\overline{q}_L}_p\lambda^A\tau^I\bfu{q_R}_s)({\overline{q}_L}_r\lambda^A\tau^I\bfu{q_R}_t)\calf^{Uh\psi^4}_4(h), \notag\\
{\color{red}\calo^{Uh\psi^4}_5} &=\caly[{\tiny\yt{p,r},\yt{s,t}}]({\overline{q}_L}_p\bfu{q_R}_s)({\overline{q}_L}_r\bfu{q_R}_t)\calf^{Uh\psi^4}_5(h), & {\color{red}\calo^{Uh\psi^4}_6} &=\caly[{\tiny\yt{p,r},\yt{s,t}}]({\overline{q}_L}_p\tau^I\bfu{q_R}_s)({\overline{q}_L}_r\tau^I\bfu{q_R}_t)\calf^{Uh\psi^4}_6(h), 
\end{align}
\begin{align}
{\color{red}\calo^{Uh\psi^4}_7} &= \caly[{\tiny\yt{p,r},\yt{s,t}}]({\overline{q}_L}_p\lambda^A\bfu{q_R}_s)({\overline{q}_L}_r\lambda^A\bfu{q_R}_t)\calf^{Uh\psi^4}_7(h), & {\color{red}\calo^{Uh\psi^4}_8} &= \caly[{\tiny\yt{p,r},\yt{s,t}}]({\overline{q}_L}_p\lambda^A\sigma^I\bfu{q_R}_s)({\overline{q}_L}_r\lambda^A\sigma^I\bfu{q_R}_t)\calf^{Uh\psi^4}_8(h), \notag\\
\calo^{Uh\psi^4}_9 &= \caly[{\tiny\yt{pr},\yt{st}}]({\overline{q}_L}_p\lambda^A\bfu{q_R}_s)({\overline{q}_L}_r\lambda^A\bft\bfu{q_R}_t)\calf^{Uh\psi^4}_9(h), & \calo^{Uh\psi^4}_{10} &= \caly[{\tiny\yt{pr},\yt{st}}]({\overline{q}_L}_p\bfu{q_R}_s)({\overline{q}_L}_r\bft\bfu{q_R}_t)\calf^{Uh\psi^4}_{10}(h), \notag\\
{\color{red}\calo^{Uh\psi^4}_{11}} &= \caly[{\tiny\yt{pr},\yt{s,t}}]({\overline{q}_L}_p\lambda^A\bfu{q_R}_s)({\overline{q}_L}_r\lambda^A\bft\bfu{q_R}_t)\calf^{Uh\psi^4}_{11}(h), & {\color{red}\calo^{Uh\psi^4}_{12}} &= \caly[{\tiny\yt{pr},\yt{s,t}}]({\overline{q}_L}_p\bfu{q_R}_s)({\overline{q}_L}_r\bft\bfu{q_R}_t)\calf^{Uh\psi^4}_{12}(h),
\end{align}
\begin{align}
{\color{red}\calo^{Uh\psi^4}_{13}} &= \caly[{\tiny\yt{pr},\yt{s,t}}]({\overline{q}_L}_p\lambda^A\tau^I\bfu{q_R}_s)({\overline{q}_L}_r\lambda^A\tau^I\bft\bfu{q_R}_t)\calf^{Uh\psi^4}_{13}(h), & {\color{red}\calo^{Uh\psi^4}_{14}} &= \caly[{\tiny\yt{pr},\yt{s,t}}]({\overline{q}_L}_p\tau^I\bfu{q_R}_s)({\overline{q}_L}_r\tau^I\bft\bfu{q_R}_t)\calf^{Uh\psi^4}_{14}(h), \notag\\
{\color{red}\calo^{Uh\psi^4}_{15}} &= \caly[{\tiny\yt{p,r},\yt{st}}]({\overline{q}_L}_p\lambda^A\bfu{q_R}_s)({\overline{q}_L}_r\lambda^A\bft\bfu{q_R}_t)\calf^{Uh\psi^4}_{15}(h), & {\color{red}\calo^{Uh\psi^4}_{16}} &= \caly[{\tiny\yt{p,r},\yt{st}}]({\overline{q}_L}_p\bfu{q_R}_s)({\overline{q}_L}_r\bft\bfu{q_R}_t)\calf^{Uh\psi^4}_{16}(h), \notag\\
{\color{red}\calo^{Uh\psi^4}_{17}} &= \caly[{\tiny\yt{p,r},\yt{st}}]({\overline{q}_L}_p\lambda^A\tau^I\bfu{q_R}_s)({\overline{q}_L}_r\lambda^A\tau^I\bft\bfu{q_R}_t)\calf^{Uh\psi^4}_{17}(h), & {\color{red}\calo^{Uh\psi^4}_{18}} &= \caly[{\tiny\yt{p,r},\yt{st}}]({\overline{q}_L}_p\tau^I\bfu{q_R}_s)({\overline{q}_L}_r\tau^I\bft\bfu{q_R}_t)\calf^{Uh\psi^4}_{18}(h),
\end{align}
\begin{align}
{\color{red}\calo^{Uh\psi^4}_{19}} &= \caly[{\tiny\yt{p,r},\yt{s,t}}]({\overline{q}_L}_p\lambda^A\bfu{q_R}_s)({\overline{q}_L}_r\lambda^A\bft\bfu{q_R}_t)\calf^{Uh\psi^4}_{19}(h), & {\color{red}\calo^{Uh\psi^4}_{20}} &= \caly[{\tiny\yt{p,r},\yt{s,t}}]({\overline{q}_L}_p\bfu{q_R}_s)({\overline{q}_L}_r\bft\bfu{q_R}_t)\calf^{Uh\psi^4}_{20}(h),\notag\\
\calo^{Uh\psi^4}_{21} &= \caly[{\tiny\yt{pr},\yt{st}}]({\overline{q}_L}_p\bft\bfu{q_R}_s)({\overline{q}_L}_r\bft\bfu{q_R}_t)\calf^{Uh\psi^4}_{21}(h), & \calo^{Uh\psi^4}_{22} &= \caly[{\tiny\yt{pr},\yt{st}}]({\overline{q}_L}_p\lambda^A\bft\bfu{q_R}_s)({\overline{q}_L}_r\lambda^A\bft\bfu{q_R}_t)\calf^{Uh\psi^4}_{22}(h), \notag\\
{\color{red}\calo^{Uh\psi^4}_{23}} &= \caly[{\tiny\yt{p,r},\yt{s,t}}]({\overline{q}_L}_p\bft\bfu{q_R}_s)({\overline{q}_L}_r\bft\bfu{q_R}_t)\calf^{Uh\psi^4}_{23}(h), & {\color{red}\calo^{Uh\psi^4}_{24}} &= \caly[{\tiny\yt{p,r},\yt{s,t}}]({\overline{q}_L}_p\lambda^A\bft\bfu{q_R}_s)({\overline{q}_L}_r\lambda^A\bft\bfu{q_R}_t)\calf^{Uh\psi^4}_{24}(h),
\end{align}
}
where
\beq
\calf^{Uh\psi^4}_i=1+\sum_{n=1}c^{Uh\psi^4}_n(\frac{h}{v})^n,\quad(i=1,2,\dots,24).
\eeq

\paragraph{Type $Q^\dagger_L Q_L Q^\dagger_R Q_R$: }
{\small
\begin{align}
	\calo^{Uh\psi^4}_{25} &= ({\overline{q}_L}_s\gamma_\mu{q_L}_p)({\overline{q}_R}_r\gamma^\mu {q_R}_t)\calf^{Uh\psi^4}_{25}(h), & \calo^{Uh\psi^4}_{26} &= ({\overline{q}_L}_s\gamma_\mu\tau^I{q_L}_p)({\overline{q}_R}_r\gamma^\mu\bfu^\dagger\tau^I\bfu{q_R}_t)\calf^{Uh\psi^4}_{26}(h), \notag\\
	\calo^{Uh\psi^4}_{27} &= ({\overline{q}_L}_s\gamma_\mu\lambda^A{q_L}_p)({\overline{q}_R}_r\gamma^\mu\lambda^A {q_R}_t)\calf^{Uh\psi^4}_{27}(h), & \calo^{Uh\psi^4}_{28} &= ({\overline{q}_L}_s\gamma_\mu\lambda^A\tau^I{q_L}_p)({\overline{q}_R}_r\gamma^\mu\lambda^A\bfu^\dagger\tau^I\bfu{q_R}_t)\calf^{Uh\psi^4}_{26}(h), \notag \\
	\calo^{Uh\psi^4}_{29} &= ({\overline{q}_L}_s\gamma_\mu\bft{q_L}_p)({\overline{q}_R}_r\gamma^\mu{q_R}_t)\calf^{Uh\psi^4}_{29}(h), & \calo^{Uh\psi^4}_{30} &= ({\overline{q}_L}_s\gamma_\mu\lambda^A\bft{q_L}_p)({\overline{q}_R}_r\gamma^\mu\lambda^A{q_R}_t)\calf^{Uh\psi^4}_{30}(h),
\end{align}
\begin{align}
	\calo^{Uh\psi^4}_{31} &= ({\overline{q}_L}_s\gamma_\mu{q_L}_p)({\overline{q}_R}_r\gamma^\mu\bfu^\dagger\bft\bfu{q_R}_t)\calf^{Uh\psi^4}_{31}(h), & \calo^{Uh\psi^4}_{32} &= ({\overline{q}_L}_s\gamma\mu\lambda^A{q_L}_p)({\overline{q}_R}_r\gamma^\mu\lambda^A\bfu^\dagger\bft\bfu{q_R}_t)\calf^{Uh\psi^4}_{32}(h),\notag\\
	{\color{red}\calo^{Uh\psi^4}_{33}} &= ({\overline{q}_L}_s\gamma_\mu\tau^I\bft{q_L}_p)({\overline{q}_R}_r\gamma^\mu\bfu^\dagger\tau^I\bfu{q_R}_t)\calf^{Uh\psi^4}_{33}(h), & {\color{red}\calo^{Uh\psi^4}_{34}} &= ({\overline{q}_L}_s\gamma_\mu\lambda^A\tau^I\bft{q_L}_p)({\overline{q}_R}_r\gamma^\mu\lambda^A\bfu^\dagger\tau^I\bfu{q_R}_t)\calf^{Uh\psi^4}_{34}(h), \notag \\
	\calo^{Uh\psi^4}_{35} &= ({\overline{q}_L}_s\gamma_\mu\lambda^A\bft{q_L}_p)({\overline{q}_R}_r\gamma^\mu\lambda^A\bfu^\dagger\bft\bfu{q_R}_t)\calf^{Uh\psi^4}_{35}(h), & \calo^{Uh\psi^4}_{36} &= ({\overline{q}_L}_s\gamma_\mu\bft{q_L}_p)({\overline{q}_R}_r\gamma^\mu\bfu^\dagger\bft\bfu{q_R}_t) \calf^{Uh\psi^4}_{36}(h),
\end{align}
}
where
\beq
\calf^{Uh\psi^4}_i=1+\sum_{n=1}c^{Uh\psi^4}_n(\frac{h}{v})^n,\quad(i=25,26,\dots,36).
\eeq

\paragraph{Type ${Q^\dagger_L}^2 {Q_L}^2\& {Q^\dagger_R}^2{Q_R}^2$: }

{\small
\begin{align}
\calo^{Uh\psi^4}_{37} &= \caly[{\tiny\yt{pr},\yt{st}}]({\overline{q}_L}_s\gamma_\mu{q_L}_p)({\overline{q}_L}_t\gamma^\mu{q_L}_r)\calf^{Uh\psi^4}_{37}(h), & \calo^{Uh\psi^4}_{38} &= \caly[{\tiny\yt{pr},\yt{st}}]({\overline{q}_L}_s\gamma_\mu\tau^I{q_L}_p)({\overline{q}_L}_t\gamma^\mu\tau^I{q_L}_r)\calf^{Uh\psi^4}_{38}(h), \notag \\
{\color{red}\calo^{Uh\psi^4}_{39}} &= \caly[{\tiny\yt{p,r},\yt{s,t}}]({\overline{q}_L}_s\gamma_\mu{q_L}_p)({\overline{q}_L}_t\gamma^\mu{q_L}_r)\calf^{Uh\psi^4}_{39}(h), & {\color{red}\calo^{Uh\psi^4}_{40}} &= \caly[{\tiny\yt{p,r},\yt{s,t}}]({\overline{q}_L}_s\gamma_\mu\tau^I{q_L}_p)({\overline{q}_L}_t\gamma^\mu\tau^I{q_L}_r)\calf^{Uh\psi^4}_{40}(h), \notag \\
\calo^{Uh\psi^4}_{41} &= \caly[{\tiny\yt{pr},\yt{st}}]({\overline{q}_L}_s\gamma^\mu{q_L}_p)({\overline{q}_L}_t\gamma_\mu\bft{q_L}_r)\calf^{Uh\psi^4}_{41}(h), & {\color{red}\calo^{Uh\psi^4}_{42}} &= \caly[{\tiny\yt{p,r},\yt{st}}]({\overline{q}_L}_s\gamma^\mu{q_L}_p)({\overline{q}_L}_t\gamma_\mu\bft{q_L}_r)\calf^{Uh\psi^4}_{42}(h), 
\end{align}
\begin{align}
{\color{red}\calo^{Uh\psi^4}_{43}} &= \caly[{\tiny\yt{p,r},\yt{st}}]({\overline{q}_L}_s\gamma^\mu\bft{q_L}_p)({\overline{q}_L}_t\gamma_\mu{q_L}_r)\calf^{Uh\psi^4}_{43}(h), & {\color{red}\calo^{Uh\psi^4}_{44}} &= \caly[{\tiny\yt{pr},\yt{s,t}}]({\overline{q}_L}_s\gamma^\mu{q_L}_p)({\overline{q}_L}_t\gamma_\mu\bft{q_L}_r)\calf^{Uh\psi^4}_{44}(h),\notag \\
{\color{red}\calo^{Uh\psi^4}_{45}} &= \caly[{\tiny\yt{pr},\yt{s,t}}]({\overline{q}_L}_s\gamma^\mu\bft{q_L}_p)({\overline{q}_L}_t\gamma_\mu{q_L}_r)\calf^{Uh\psi^4}_{45}(h), &  {\color{red}\calo^{Uh\psi^4}_{46}} &= \caly[{\tiny\yt{p,r},\yt{s,t}}]({\overline{q}_L}_s\gamma^\mu{q_L}_p)({\overline{q}_L}_t\gamma_\mu\bft{q_L}_r)\calf^{Uh\psi^4}_{46}(h),\notag\\
\calo^{Uh\psi^4}_{47} &= \caly[{\tiny\yt{pr},\yt{st}}]({\overline{q}_L}_s\gamma_\mu\bft{q_L}_p)({\overline{q}_L}_t\gamma^\mu\bft{q_L}_s)\calf^{Uh\psi^4}_{47}(h), & {\color{red}\calo^{Uh\psi^4}_{48}} &= \caly[{\tiny\yt{p,r},\yt{s,t}}]({\overline{q}_L}_s\gamma_\mu\bft{q_L}_p)({\overline{q}_L}_t\gamma^\mu\bft{q_L}_s)\calf^{Uh\psi^4}_{48}(h), 
\end{align}

\begin{align}
\calo^{Uh\psi^4}_{49} &= \caly[{\tiny\yt{pr},\yt{st}}]({\overline{q}_R}_s\gamma_\mu{q_R}_p)({\overline{q}_R}_t\gamma^\mu{q_R}_r)\calf^{Uh\psi^4}_{49}(h), \notag \\
\calo^{Uh\psi^4}_{50} &= \caly[{\tiny\yt{pr},\yt{st}}]({\overline{q}_R}_s\gamma_\mu\bfu^\dagger\tau^I\bfu{q_R}_p)({\overline{q}_R}_t\gamma^\mu\bfu^\dagger\tau^I\bfu{q_R}_r)\calf^{Uh\psi^4}_{50}(h), \notag \\
{\color{red}\calo^{Uh\psi^4}_{51}} &= \caly[{\tiny\yt{p,r},\yt{s,t}}]({\overline{q}_R}_s\gamma_\mu{q_R}_p)({\overline{q}_R}_t\gamma^\mu{q_R}_r)\calf^{Uh\psi^4}_{51}(h),\notag \\
{\color{red}\calo^{Uh\psi^4}_{52}} &= \caly[{\tiny\yt{p,r},\yt{s,t}}]({\overline{q}_R}_s\gamma_\mu\bfu^\dagger\tau^I\bfu{q_R}_p)({\overline{q}_R}_t\gamma^\mu\bfu^\dagger\tau^I\bfu{q_R}_r)\calf^{Uh\psi^4}_{52}(h), \notag \\
\calo^{Uh\psi^4}_{53} &= \caly[{\tiny\yt{pr},\yt{st}}]({\overline{q}_R}_s\gamma^\mu{q_R}_p)({\overline{q}_R}_t\gamma_\mu\bfu^\dagger\bft\bfu{q_R}_r)\calf^{Uh\psi^4}_{53}(h), \notag \\
{\color{red}\calo^{Uh\psi^4}_{54}} &= \caly[{\tiny\yt{p,r},\yt{st}}]({\overline{q}_R}_s\gamma^\mu{q_R}_p)({\overline{q}_R}_t\gamma_\mu\bfu^\dagger\bft\bfu{q_R}_r)\calf^{Uh\psi^4}_{54}(h),
\end{align}
\begin{align}
{\color{red}\calo^{Uh\psi^4}_{55}} &= \caly[{\tiny\yt{p,r},\yt{st}}]({\overline{q}_R}_s\gamma^\mu\bfu^\dagger\bft\bfu{q_R}_p)({\overline{q}_R}_t\gamma_\mu{q_R}_r)\calf^{Uh\psi^4}_{55}(h),\notag \\
{\color{red}\calo^{Uh\psi^4}_{56}} &= \caly[{\tiny\yt{pr},\yt{s,t}}]({\overline{q}_R}_s\gamma^\mu{q_R}_p)({\overline{q}_R}_t\gamma_\mu\bfu^\dagger\bft\bfu{q_R}_r)\calf^{Uh\psi^4}_{56}(h), \notag\\
{\color{red}\calo^{Uh\psi^4}_{57}} &= \caly[{\tiny\yt{pr},\yt{s,t}}]({\overline{q}_R}_s\gamma^\mu\bfu^\dagger\bft\bfu{q_R}_p)({\overline{q}_R}_t\gamma_\mu{q_R}_r)\calf^{Uh\psi^4}_{57}(h), \notag \\
{\color{red}\calo^{Uh\psi^4}_{58}} &= \caly[{\tiny\yt{p,r},\yt{s,t}}]({\overline{q}_R}_s\gamma^\mu{q_R}_p)({\overline{q}_R}_t\gamma_\mu\bfu^\dagger\bft\bfu{q_R}_r)\calf^{Uh\psi^4}_{58}(h),\notag\\
\calo^{Uh\psi^4}_{59} &= \caly[{\tiny\yt{pr},\yt{st}}]({\overline{q}_R}_s\gamma_\mu\bfu^\dagger\bft\bfu{q_R}_p)({\overline{q}_R}_t\gamma^\mu\bfu^\dagger\bft\bfu{q_R}_s)\calf^{Uh\psi^4}_{59}(h),\notag \\
{\color{red}\calo^{Uh\psi^4}_{60}} &= \caly[{\tiny\yt{p,r},\yt{s,t}}]({\overline{q}_R}_s\gamma_\mu\bfu^\dagger\bft\bfu{q_R}_p)({\overline{q}_R}_t\gamma^\mu\bfu^\dagger\bft\bfu{q_R}_s)\calf^{Uh\psi^4}_{60}(h),
\end{align}
}
where
\beq
\calf^{Uh\psi^4}_i=1+\sum_{n=1}c^{Uh\psi^4}_n(\frac{h}{v})^n,\quad(i=36,37,\dots,60).
\eeq

\subsubsection{Pure Lepton Operators}

\paragraph{Types ${L_L^\dagger}^2{L_R}^2,{L_L}^2{L_L^\dagger}^2,{L_R}^2{L_R^\dagger}^2,{L_L}{L_R^\dagger}{L_R}{L_L^\dagger}$: }
{\small
\begin{align}
\calo^{Uh\psi^4}_{61} &= \caly[{\tiny\yt{pr},\yt{st}}]({\overline{l}_L}_p\bfu{l_R}_s)({\overline{l}_L}_r\bfu{l_R}_t)\calf^{Uh\psi^4}_{61}(h), & \calo^{Uh\psi^4}_{62} &={\color{gray} \caly[{\tiny\yt{pr},\yt{st}}]({\overline{l}_L}_p\tau^I\bfu{l_R}_s)({\overline{l}_L}_r\tau^I\bfu{l_R}_t)\calf^{Uh\psi^4}_{62}(h)}, \notag\\
{\color{red}\calo^{Uh\psi^4}_{63}} &= \caly[{\tiny\yt{p,r},\yt{s,t}}]({\overline{l}_L}_p\bfu{l_R}_s)({\overline{l}_L}_r\bfu{l_R}_t)\calf^{Uh\psi^4}_{63}(h), & {\color{red}\calo^{Uh\psi^4}_{64}} &= {\color{gray}\caly[{\tiny\yt{p,r},\yt{s,t}}]({\overline{l}_L}_p\tau^I\bfu{l_R}_s)({\overline{l}_L}_r\tau^I\bfu{l_R}_t)\calf^{Uh\psi^4}_{64}(h)},\notag\\
\calo^{Uh\psi^4}_{65} &= {\color{gray}\caly[{\tiny\yt{pr},\yt{st}}]({\overline{l}_L}_p\bfu{l_R}_t)({\overline{l}_L}_r\bft\bfu{l_R}_s)\calf^{Uh\psi^4}_{65}(h),} & {\color{red}\calo^{Uh\psi^4}_{66}} &= {\color{gray}\caly[{\tiny\yt{pr},\yt{s,t}}]({\overline{l}_L}_p\bfu{l_R}_s)({\overline{l}_L}_r\bft\bfu{l_R}_t)\calf^{Uh\psi^4}_{66}(h),} 
\end{align}
\begin{align}
{\color{red}\calo^{Uh\psi^4}_{67}} &= {\color{gray}\caly[{\tiny\yt{pr},\yt{s,t}}]({\overline{l}_L}_p\tau^I\bfu{l_R}_s)({\overline{l}_L}_r\tau^I\bft\bfu{l_R}_t)\calf^{Uh\psi^4}_{67}(h),} & {\color{red}\calo^{Uh\psi^4}_{68}} &= {\color{gray}\caly[{\tiny\yt{p,r},\yt{st}}]({\overline{l}_L}_p\bfu{l_R}_s)({\overline{l}_L}_r\bft\bfu{l_R}_t)\calf^{Uh\psi^4}_{68}(h),} \notag\\
{\color{red}\calo^{Uh\psi^4}_{69}} &= {\color{gray}\caly[{\tiny\yt{p,r},\yt{st}}]({\overline{l}_L}_p\tau^I\bfu{l_R}_s)({\overline{l}_L}_r\tau^I\bft\bfu{l_R}_t)\calf^{Uh\psi^4}_{69}(h),} &  {\color{red}\calo^{Uh\psi^4}_{70}} &= {\color{gray}\caly[{\tiny\yt{p,r},\yt{s,t}}]({\overline{l}_L}_p\bfu{l_R}_t)({\overline{l}_L}_r\bft\bfu{l_R}_s)\calf^{Uh\psi^4}_{70}(h),} \notag\\
\calo^{Uh\psi^4}_{71} &= {\color{gray}\caly[{\tiny\yt{pr},\yt{st}}]({\overline{l}_L}_p\bft\bfu{l_R}_s)({\overline{l}_L}_r\bft\bfu{l_R}_t)\calf^{Uh\psi^4}_{71}(h),} & {\color{red}\calo^{Uh\psi^4}_{72}} &= {\color{gray}\caly[{\tiny\yt{p,r},\yt{s,t}}]({\overline{l}_L}_p\bft\bfu{l_R}_s)({\overline{l}_L}_r\bft\bfu{l_R}_t)\calf^{Uh\psi^4}_{72}(h),}
\end{align}
\begin{align}
\calo^{Uh\psi^4}_{73} &= \caly[{\tiny\yt{pr},\yt{st}}]({\overline{l}_L}_s\gamma_\mu{l_L}_p)({\overline{l}_L}_t\gamma^\mu{l_L}_s)\calf^{Uh\psi^4}_{73}(h), & {\color{red}\calo^{Uh\psi^4}_{74}} &= \caly[{\tiny\yt{p,r},\yt{s,t}}]({\overline{l}_L}_s\gamma_\mu{l_L}_p)({\overline{l}_L}_t\gamma^\mu{l_L}_s)\calf^{Uh\psi^4}_{74}(h),\notag\\
\calo^{Uh\psi^4}_{75} &= \caly[{\tiny\yt{pr},\yt{st}}]({\overline{l}_L}_s\gamma_\mu{l_L}_p)({\overline{l}_L}_t\gamma^\mu\bft{l_L}_s)\calf^{Uh\psi^4}_{75}(h), & {\color{red}\calo^{Uh\psi^4}_{76}} &= \caly[{\tiny\yt{pr},\yt{s,t}}]({\overline{l}_L}_s\gamma_\mu{l_L}_p)({\overline{l}_L}_t\gamma^\mu\bft{l_L}_s)\calf^{Uh\psi^4}_{76}(h), \notag\\
{\color{red}\calo^{Uh\psi^4}_{77}} &= \caly[{\tiny\yt{p,r},\yt{st}}]({\overline{l}_L}_s\gamma_\mu{l_L}_p)({\overline{l}_L}_t\gamma^\mu\bft{l_L}_s)\calf^{Uh\psi^4}_{77}(h), & \calo^{Uh\psi^4}_{78} &= \caly[{\tiny\yt{pr},\yt{st}}]({\overline{l}_L}_s\gamma_\mu\bft{l_L}_p)({\overline{l}_L}_t\gamma^\mu\bft{l_L}_s)\calf^{Uh\psi^4}_{78}(h), 
\end{align}
\begin{align}
\calo^{Uh\psi^4}_{79} &= \caly[{\tiny\yt{pr},\yt{st}}]({\overline{l}_R}_s\gamma_\mu{l_R}_p)({\overline{l}_R}_t\gamma^\mu{l_R}_s)\calf^{Uh\psi^4}_{79}(h), \notag \\
{\color{red}\calo^{Uh\psi^4}_{80}} &= {\color{gray}\caly[{\tiny\yt{p,r},\yt{s,t}}]({\overline{l}_R}_s\gamma_\mu{l_R}_p)({\overline{l}_R}_t\gamma^\mu{l_R}_s)\calf^{Uh\psi^4}_{80}(h)},\notag\\
\calo^{Uh\psi^4}_{81} &= {\color{gray}\caly[{\tiny\yt{pr},\yt{st}}]({\overline{l}_R}_s\gamma_\mu{l_R}_p)({\overline{l}_R}_t\gamma^\mu\bfu^\dagger\bft\bfu{l_R}_s)\calf^{Uh\psi^4}_{81}(h),}\notag \\
{\color{red}\calo^{Uh\psi^4}_{82}} &= {\color{gray}\caly[{\tiny\yt{pr},\yt{s,t}}]({\overline{l}_R}_s\gamma_\mu{l_R}_p)({\overline{l}_R}_t\gamma^\mu\bfu^\dagger\bft\bfu{l_R}_s)\calf^{Uh\psi^4}_{82}(h),} \notag\\
{\color{red}\calo^{Uh\psi^4}_{83}} &= {\color{gray}\caly[{\tiny\yt{p,r},\yt{st}}]({\overline{l}_R}_s\gamma_\mu{l_R}_p)({\overline{l}_R}_t\gamma^\mu\bfu^\dagger\bft\bfu{l_R}_s)\calf^{Uh\psi^4}_{83}(h),}\notag \\
\calo^{Uh\psi^4}_{84} &= {\color{gray}\caly[{\tiny\yt{pr},\yt{st}}]({\overline{l}_R}_s\gamma_\mu\bfu^\dagger\bft\bfu{l_R}_p)({\overline{l}_R}_t\gamma^\mu\bfu^\dagger\bft\bfu{l_R}_s)\calf^{Uh\psi^4}_{84}(h),} 
\end{align}
\begin{align}
\calo^{Uh\psi^4}_{85} &= ({\overline{l}_L}_s\gamma_\mu{l_L}_p)({\overline{l}_R}_t\gamma^\mu{l_R}_s)\calf^{Uh\psi^4}_{85}(h), & \calo^{Uh\psi^4}_{86} &= ({\overline{l}_L}_s\gamma_\mu\tau^I{l_L}_p)({\overline{l}_R}_t\gamma^\mu\bfu^\dagger\tau^I\bfu{l_R}_s)\calf^{Uh\psi^4}_{86}(h), \notag\\
\calo^{Uh\psi^4}_{87} &= {\color{gray}({\overline{l}_L}_s\gamma_\mu\bft{l_L}_p)({\overline{l}_R}_t\gamma^\mu{l_R}_s)\calf^{Uh\psi^4}_{87}(h)}, & \calo^{Uh\psi^4}_{88} &= {\color{gray}({\overline{l}_L}_s\gamma_\mu{l_L}_p)({\overline{l}_R}_t\gamma^\mu\bfu^\dagger\bft\bfu{l_R}_s)\calf^{Uh\psi^4}_{88}(h),} \notag \\
{\color{red}\calo^{Uh\psi^4}_{89}} &= {\color{gray}({\overline{l}_L}_s\gamma_\mu\tau^I{l_L}_p)({\overline{l}_R}_t\gamma^\mu\tau^I\bfu^\dagger\bft\bfu{l_R}_s)\calf^{Uh\psi^4}_{89}(h),} & \calo^{Uh\psi^4}_{90} &= {\color{gray}({\overline{l}_L}_s\gamma_\mu\bft{l_L}_p)({\overline{l}_R}_t\gamma^\mu\bfu^\dagger\bft\bfu{l_R}_s)\calf^{Uh\psi^4}_{90}(h),}
\end{align}
}
where
\beq
\calf^{Uh\psi^4}_i=1+\sum_{n=1}c^{Uh\psi^4}_n(\frac{h}{v})^n,\quad(i=61,62,\dots,90).
\eeq

\subsubsection{Mixed Quark-Lepton Operators}
\paragraph{Type: $Q^\dagger_L Q_R L^\dagger_L L_R$: }
{\small
\begin{align}
    \calo^{Uh\psi^4}_{91} &= ({\overline{l}_L}_p\bfu{l_R}_r)({\overline{q}_L}_s\bfu{q_R}_t)\calf^{Uh\psi^4}_{91}(h), & \calo^{Uh\psi^4}_{92} &= ({\overline{l}_L}_p\sigma_{\mu\nu}\bfu{l_R}_r)({\overline{q}_L}_s\sigma^{\mu\nu}\bfu{q_R}_t)\calf^{Uh\psi^4}_{92}(h), \notag\\
    \calo^{Uh\psi^4}_{93} &= ({\overline{l}_L}_p\tau^I\bfu{l_R}_r)({\overline{q}_L}_s\tau^I\bfu{q_R}_t)\calf^{Uh\psi^4}_{93}(h), & \calo^{Uh\psi^4}_{94} &= ({\overline{l}_L}_p\sigma_{\mu\nu}\tau^I\bfu{l_R}_r)({\overline{q}_L}_s\sigma^{\mu\nu}\tau^I\bfu{q_R}_t)\calf^{Uh\psi^4}_{94}(h), \notag\\
    \calo^{Uh\psi^4}_{95} &= ({\overline{l}_L}_p\bfu{l_R}_r)({\overline{q}_L}_s\bft\bfu{q_R}_t)\calf^{Uh\psi^4}_{95}(h), & \calo^{Uh\psi^4}_{96} &= ({\overline{l}_L}_p\sigma_{\mu\nu}\bfu{l_R}_r)({\overline{q}_L}_s\sigma^{\mu\nu}\bft\bfu{q_R}_t)\calf^{Uh\psi^4}_{96}(h), \notag\\
\end{align}
\begin{align}
    \calo^{Uh\psi^4}_{97} &= {\color{gray}({\overline{l}_L}_p\bft\bfu{l_R}_r)({\overline{q}_L}_s\bfu{q_R}_t)\calf^{Uh\psi^4}_{97}(h),} & \calo^{Uh\psi^4}_{98} &= {\color{gray}({\overline{l}_L}_p\sigma_{\mu\nu}\bft\bfu{l_R}_r)({\overline{q}_L}_s\sigma^{\mu\nu}\bfu{q_R}_t)\calf^{Uh\psi^4}_{98}(h),} \notag\\
    \calo^{Uh\psi^4}_{99} &= {\color{gray}({\overline{l}_L}_p\tau^I\bfu{l_R}_r)({\overline{q}_L}_s\tau^I\bft\bfu{q_R}_t)\calf^{Uh\psi^4}_{99}(h)}, & \calo^{Uh\psi^4}_{100} &= {\color{gray}({\overline{l}_L}_p\sigma_{\mu\nu}\tau^I\bfu{l_R}_r)({\overline{q}_L}_s\sigma^{\mu\nu}\tau^I\bft\bfu{q_R}_t)\calf^{Uh\psi^4}_{100}(h)}, \notag\\
    \calo^{Uh\psi^4}_{101} &= {\color{gray}({\overline{l}_L}_p\bft\bfu{l_R}_r)({\overline{q}_L}_s\bft\bfu{q_R}_t)\calf^{Uh\psi^4}_{101}(h),} & \calo^{Uh\psi^4}_{102} &= {\color{gray}({\overline{l}_L}_p\sigma_{\mu\nu}\bft\bfu{l_R}_r)({\overline{q}_L}_s\sigma^{\mu\nu}\bft\bfu{q_R}_t)\calf^{Uh\psi^4}_{102}(h).}
\end{align}
}

\paragraph{Type $Q^\dagger_LQ_LL^\dagger_LL_L\& Q^\dagger_RQ_RL^\dagger_R L_R$: }
{\small
\begin{align}
    \calo^{Uh\psi^4}_{103} &= ({\overline{l}_L}_s\gamma_\mu{l_L}_p)({\overline{q}_L}_t\gamma^\mu{q_L}_r)\calo^{Uh\psi^4}_{103}(h), & \calo^{Uh\psi^4}_{104} &= ({\overline{l}_L}_s\gamma_\mu\tau^I{l_L}_p)({\overline{q}_L}_t\gamma^\mu\tau^I{q_L}_r)\calo^{Uh\psi^4}_{104}(h), \notag\\
    \calo^{Uh\psi^4}_{105} &= ({\overline{l}_L}_s\gamma_\mu\bft{l_L}_p)({\overline{q}_L}_t\gamma^\mu{q_L}_r)\calf^{Uh\psi^4}_{105}(h), & \calo^{Uh\psi^4}_{106} &= ({\overline{l}_L}_s\gamma_\mu{l_L}_p)({\overline{q}_L}_t\gamma^\mu\bft{q_L}_r)\calf^{Uh\psi^4}_{106}(h), \notag\\
	{\color{red}\calo^{Uh\psi^4}_{107}} &= ({\overline{l}_L}_s\gamma_\mu\tau^I\bft{l_L}_p)({\overline{q}_L}_t\gamma^\mu\tau^I{q_L}_r)\calf^{Uh\psi^4}_{107}(h), & \calo^{Uh\psi^4}_{108} &= ({\overline{l}_L}_s\gamma_\mu\bft{l_L}_p)({\overline{q}_L}_t\gamma^\mu\bft{q_L}_r)\calf^{Uh\psi^4}_{108}(h), \notag\\
    \calo^{Uh\psi^4}_{109} &= ({\overline{l}_R}_s\gamma_\mu{l_R}_p)({\overline{q}_R}_t\gamma^\mu{q_R}_r)\calo^{Uh\psi^4}_{109}(h), & \calo^{Uh\psi^4}_{110} &= {({\overline{l}_R}_s\gamma_\mu\bfu^\dagger\tau^I\bfu{l_R}_p)({\overline{q}_R}_t\gamma^\mu\bfu^\dagger\tau^I\bfu{q_R}_r)\calo^{Uh\psi^4}_{110}(h),} \notag\\
    \calo^{Uh\psi^4}_{111} &= {\color{gray}({\overline{l}_R}_s\gamma_\mu\bfu^\dagger\bft\bfu{l_R}_p)({\overline{q}_R}_t\gamma^\mu{q_R}_r)\calf^{Uh\psi^4}_{111}(h),} & \calo^{Uh\psi^4}_{112} &={\color{gray} ({\overline{l}_R}_s\gamma_\mu{l_R}_p)({\overline{q}_R}_t\gamma^\mu\bfu^\dagger\bft\bfu{q_R}_r)\calf^{Uh\psi^4}_{112}(h)}, \notag\\
	{\color{red}\calo^{Uh\psi^4}_{113}} &= {\color{gray}({\overline{l}_R}_s\gamma_\mu\bfu^\dagger\tau^I\bft\bfu{l_R}_p)({\overline{q}_R}_t\gamma^\mu\bfu^\dagger\tau^I\bfu{Q_R}_r)\calf^{Uh\psi^4}_{113}(h),} & \calo^{Uh\psi^4}_{114} &= {\color{gray}({\overline{l}_R}_s\gamma_\mu\bfu^\dagger\bft\bfu{l_R}_p)({\overline{q}_R}_t\gamma^\mu\bfu^\dagger\bft\bfu{Q_R}_r)\calf^{Uh\psi^4}_{114}(h).}
\end{align}
}

\paragraph{Type $L^\dagger_L L_L Q^\dagger_R Q_R\& L^\dagger_R L_R Q^\dagger_L Q_L$: }
{\small
\begin{align}
    \calo^{Uh\psi^4}_{115} &= ({\overline{l}_R}_s\gamma_\mu{l_R}_p)({\overline{q}_L}_s\gamma^\mu{q_L}_r)\calf^{Uh\psi^4}_{115}(h), & \calo^{Uh\psi^4}_{116} &=  ({\overline{l}_R}_s\gamma_\mu\bfu^\dagger\tau^I\bfu{l_R}_p)({\overline{q}_L}_s\gamma^\mu\tau^I{q_L}_r)\calf^{Uh\psi^4}_{116}(h),\notag \\
    \calo^{Uh\psi^4}_{117} &=  {\color{gray}({\overline{l}_R}_s\gamma_\mu{l_R}_p)({\overline{q}_L}_t\gamma^\mu\bft{q_L}_r)\calf^{Uh\psi^4}_{117}(h)}, & \calo^{Uh\psi^4}_{118} &= {\color{gray}({\overline{l}_R}_s\gamma_\mu\bfu^\dagger\bft\bfu{l_R}_p)({\overline{q}_L}_t\gamma^\mu{q_L}_r)\calf^{Uh\psi^4}_{118}(h),} \notag\\
	{\color{red}\calo^{Uh\psi^4}_{119}} &= {\color{gray}({\overline{l}_R}_s\gamma_\mu\bfu^\dagger\tau^I\bft\bfu{l_R}_p)({\overline{q}_L}_t\gamma^\mu\tau^I{q_L}_r)\calf^{Uh\psi^4}_{119}(h),} & \calo^{Uh\psi^4}_{120} &= {\color{gray}({\overline{l}_R}_s\gamma_\mu\bfu^\dagger\bft\bfu{l_R}_p)({\overline{q}_L}_t\gamma^\mu\bft{q_L}_r)\calf^{Uh\psi^4}_{120}(h),} \notag\\
    \calo^{Uh\psi^4}_{121} &= ({\overline{l}_L}_s\gamma_\mu{L_L}_p)({\overline{q}_R}_s\gamma^\mu{q_R}_r)\calf^{Uh\psi^4}_{121}(h), & \calo^{Uh\psi^4}_{122} &=  ({\overline{l}_L}_s\gamma_\mu\tau^I{L_L}_p)({\overline{q}_R}_s\gamma^\mu\bfu^\dagger\tau^I\bfu{q_R}_r)\calf^{Uh\psi^4}_{122}(h),\notag \\
    \calo^{Uh\psi^4}_{123} &=  ({\overline{l}_L}_s\gamma_\mu{L_L}_p)({\overline{q}_R}_t\gamma^\mu\bfu^\dagger\bft\bfu{q_R}_r)\calf^{Uh\psi^4}_{123}(h), & \calo^{Uh\psi^4}_{124} &= ({\overline{l}_L}_s\gamma_\mu\bft{L_L}_p)({\overline{q}_R}_t\gamma^\mu{q_R}_r)\calf^{Uh\psi^4}_{124}(h), \notag\\
	{\color{red}\calo^{Uh\psi^4}_{125}} &= ({\overline{l}_L}_s\gamma_\mu\tau^I\bft{L_L}_p)({\overline{q}_R}_t\gamma^\mu\bfu^\dagger\tau^I\bfu{q_R}_r)\calf^{Uh\psi^4}_{125}(h), & \calo^{Uh\psi^4}_{126} &= ({\overline{l}_L}_s\gamma_\mu\bft{L_L}_p)({\overline{q}_R}_t\gamma^\mu\bfu^\dagger\bft\bfu{q_R}_r)\calf^{Uh\psi^4}_{126}(h).
\end{align}
}

\paragraph{Type $Q^\dagger_LQ_RL^\dagger_RL_L$: }
{\small
\begin{align}
    \calo^{Uh\psi^4}_{127} &= ({\overline{q}_L}_s\gamma_\mu{l_L}_p)({\overline{l}_R}_r\gamma^\mu{q_R}_t)\calf^{Uh\psi^4}_{127}(h), & \calo^{Uh\psi^4}_{128} &= ({\overline{q}_L}_s\gamma_\mu\tau^I{l_L}_p)({\overline{l}_R}_r\gamma^\mu\bfu^\dagger\tau^I\bfu{q_R}_t)\calf^{Uh\psi^4}_{128}(h), \notag\\
    \calo^{Uh\psi^4}_{129} &= ({\overline{q}_L}_s\gamma_\mu\bft{l_L}_p)({\overline{l}_R}_r\gamma^\mu{q_R}_t)\calf^{Uh\psi^4}_{129}(h), & \calo^{Uh\psi^4}_{130} &= {\color{gray}({\overline{q}_L}_s\gamma_\mu{l_L}_p)({\overline{l}_R}_r\gamma^\mu\bfu^\dagger\bft\bfu{q_R}_t)\calf^{Uh\psi^4}_{130}(h),} \notag\\
    \calo^{Uh\psi^4}_{131} &= {\color{gray}({\overline{q}_L}_s\gamma_\mu\tau^I\bft{l_L}_p)({\overline{l}_R}_r\gamma^\mu\bfu^\dagger\tau^I\bfu{q_R}_t)\calf^{Uh\psi^4}_{131}(h)}, & \calo^{Uh\psi^4}_{132} &= {\color{gray}({\overline{q}_L}_s\gamma_\mu\bft{l_L}_p)({\overline{l}_R}_r\gamma^\mu\bfu^\dagger\bft\bfu{q_R}_t)\calf^{Uh\psi^4}_{132}(h).}
\end{align}
}
$\calf$ functions of all these types are of the same form:
\beq
\calf^{Uh\psi^4}_i=1+\sum_{n=1}c^{Uh\psi^4}_n(\frac{h}{v})^n,\quad(i=91,92,\dots,132).
\eeq

\subsubsection{Baryon-number-violating Operators}

\paragraph{Types ${L_L}{Q_L}^3, {L_L}{Q_L}{Q_R}^2, {L_R}{Q_R}{Q_L}^2, {L_R}{Q_R}^3$: }

{
\small
\begin{align}
\calo^{Uh\psi^4}_{133} &=\caly[{\tiny\yt{rst}}]\epsilon^{abc}\epsilon^{ik}\epsilon^{jl}({{l_L}^T}_{pi}C{q_L}_{raj})({{q_L}^T}_{sbk}C{q_L}_{tcl})\calf^{Uh\psi^4}_{133}(h), \notag\\
{\color{red}\calo^{Uh\psi^4}_{134}} &=\caly[{\tiny\yt{rs,t}}]\epsilon^{abc}\epsilon^{ik}\epsilon^{jl}({{l_L}^T}_{pi}C{q_L}_{raj})({{q_L}^T}_{sbk}C{q_L}_{tcl})\calf^{Uh\psi^4}_{134}(h), \notag\\
{\color{red}\calo^{Uh\psi^4}_{135}} &=\caly[{\tiny\yt{r,s,t}}]\epsilon^{abc}\epsilon^{ik}\epsilon^{jl}({{l_L}^T}_{pi}C{q_L}_{raj})({{q_L}^T}_{sbk}C{q_L}_{tcl})\calf^{Uh\psi^4}_{135}(h), 
\end{align}
\begin{align}
\calo^{Uh\psi^4}_{136} &= \caly[{\tiny\yt{rst}}]\epsilon^{abc}\epsilon^{ij}\epsilon^{km}({{l_L}^T}_{pi}C\sigma_{\mu\nu}{q_L}_{raj})({{q_L}^T}_{sbk}C\sigma^{\mu\nu}(\bft{q_L})_{tcm})\calf^{Uh\psi^4}_{136}(h), \notag\\
{\color{red}\calo^{Uh\psi^4}_{137}} &= \caly[{\tiny\yt{rs,t}}]\epsilon^{abc}\epsilon^{ij}\epsilon^{km}({{l_L}^T}_{pi}C{q_L}_{raj})({{q_L}^T}_{sbk}(\bft{q_L})_{tcm})\calf^{Uh\psi^4}_{137}(h), \notag\\
{\color{red}\calo^{Uh\psi^4}_{138}} &= \caly[{\tiny\yt{rs,t}}]\epsilon^{abc}\epsilon^{ij}\epsilon^{km}({{l_L}^T}_{pi}C\sigma_{\mu\nu}{q_L}_{raj})({{q_L}^T}_{sbk}C\sigma^{\mu\nu}(\bft{q_L})_{tcm})\calf^{Uh\psi^4}_{138}(h), 
\end{align}
\begin{align}
{\color{red}\calo^{Uh\psi^4}_{139}} &= \caly[{\tiny\yt{r,s,t}}]\epsilon^{abc}\epsilon^{ij}\epsilon^{km}({{l_L}^T}_{pi}C{q_L}_{raj})({{q_L}^T}_{sbk}C(\bft{q_L})_{tcm})\calf^{Uh\psi^4}_{139}(h), \notag\\
{\color{red}\calo^{Uh\psi^4}_{140}} &= \caly[{\tiny\yt{rs,t}}]\epsilon^{abc}\epsilon^{ln}\epsilon^{km}((\bft{{l_L}^T})_{pm}C(\bft{q_L})_{ran})({{q_L}^T}_{rak}C{q_L}_{tcl})\calf^{Uh\psi^4}_{140}(h), \notag\\
\calo^{Uh\psi^4}_{141} &= \caly[{\tiny\yt{st}}]\epsilon^{abc}\epsilon^{ik}\epsilon^{jl}({{l_L}^T}_{pi}C\gamma_\mu C{q_R}_{sbk})({{q_L}^T}_{raj}C\gamma^\mu C{q_R}_{tcl})\calf^{Uh\psi^4}_{141}(h), 
\end{align}
\begin{align}
{\color{red}\calo^{Uh\psi^4}_{142}} &= \caly[{\tiny\yt{s,t}}]\epsilon^{abc}\epsilon^{ik}\epsilon^{jl}({{l_L}^T}_{pi}C\gamma_\mu C{q_R}_{sbk})({{q_L}^T}_{raj}C\gamma^\mu C{q_R}_{tcl})\calf^{Uh\psi^4}_{142}(h), \notag\\
\calo^{Uh\psi^4}_{143} &= \caly[{\tiny\yt{st}}]\epsilon^{abc}\epsilon^{jl}\epsilon^{km}((\bft{l_L}^T)_{pm}C\gamma_\mu C(\bfu{q_R}_{sb})_k)({{q_L}^T}_{raj}C\gamma^\mu C(\bfu{q_R}_{tc})_l)\calf^{Uh\psi^4}_{143}(h), \notag\\
{\color{red}\calo^{Uh\psi^4}_{144}} &= \caly[{\tiny\yt{s,t}}]\epsilon^{abc}\epsilon^{jl}\epsilon^{km}((\bft{l_L}^T)_{pm}C\gamma_\mu C(\bfu{q_R}_{sb})_k)({{Q_L}^T}_{raj}C\gamma^\mu C(\bfu{q_R}_{tc})_l)\calf^{Uh\psi^4}_{144}(h), 
\end{align}
\begin{align}
\calo^{Uh\psi^4}_{145} &= \caly[{\tiny\yt{s,t}}]\epsilon^{abc}\epsilon^{ij}\epsilon^{km}({{l_L}^T}_{pi}C \gamma_\mu C(\bfu{q_R}_{sb})_k)({{q_L}^T}_{raj}C\gamma^\mu C(\bft\bfu{q_R})_{tcm})\calf^{Uh\psi^4}_{145}(h), \notag\\
\calo^{Uh\psi^4}_{146} &= \caly[{\tiny\yt{s,t}}]\epsilon^{abc}\epsilon^{km}\epsilon^{ln}((\bft{{l_L}^T})_{pm}C\gamma_\mu C(\bfu{q_R}_{sb})_k)((\bft{q_L}^T)_{ran}C\gamma^\mu C(\bfu{q_R}_{tc})_l)\calf^{Uh\psi^4}_{146}(h), \notag\\
\calo^{Uh\psi^4}_{147} &= \caly[{\tiny\yt{st}}]\epsilon^{abc}\epsilon^{ik}\epsilon^{jl}({{q_L}^T}_{pai}C\gamma_\mu C(\bfu{l_R}_{s})_k)({{q_L}^T}_{rbj}C\gamma^\mu C(\bfu{q_R}_{tc})_l)\calf^{Uh\psi^4}_{147}(h), 
\end{align}
\begin{align}
{\color{red}\calo^{Uh\psi^4}_{148}} &= \caly[{\tiny\yt{s,t}}]\epsilon^{abc}\epsilon^{ik}\epsilon^{jl}({{q_L}^T}_{pai}C\gamma_\mu C(\bfu{l_R}_{s})_k)({{q_L}^T}_{rbj}C\gamma^\mu C(\bfu{q_R}_{tc})_l)\calf^{Uh\psi^4}_{148}(h), \notag\\
\calo^{Uh\psi^4}_{149} &= {\color{gray}\caly[{\tiny\yt{st}}]\epsilon^{abc}\epsilon^{jk}\epsilon^{lm}((\bft{{q_L}^T})_{pam}C\gamma_\mu C(\bfu{l_R}_{s})_k)({{q_L}^T}_{rbj}C\gamma^\mu C(\bfu{q_R}_{tc})_l)\calf^{Uh\psi^4}_{149}(h),} \notag\\
\calo^{Uh\psi^4}_{150} &= \caly[{\tiny\yt{s,t}}]\epsilon^{abc}\epsilon^{jl}\epsilon^{km}((\bft{{q_L}^T})_{pam}C\gamma_\mu C(\bfu{l_R}_{s})_k)({{q_L}^T}_{rbj}C\gamma^\mu C(\bfu{q_R}_{tc})_l)\calf^{Uh\psi^4}_{150}(h),
\end{align}
\begin{align}
{\color{red}\calo^{Uh\psi^4}_{151}} &= {\color{gray}\caly[{\tiny\yt{s,t}}]\epsilon^{abc}\epsilon^{jk}\epsilon^{lm}((\bft{{q_L}^T})_{pam}C\gamma_\mu C (\bfu {l_R}_{s})_k)({{q_L}^T}_{rbj}C\gamma^\mu C(\bfu{q_R}_{tc})_l)\calf^{Uh\psi^4}_{151}(h),} \notag\\
\calo^{Uh\psi^4}_{152} &= {\color{gray}\caly[{\tiny\yt{s,t}}]\epsilon^{abc}\epsilon^{km}\epsilon^{ln}((\bft{{q_L}^T}_{pam})C\gamma_\mu C(\bfu{L_R}_{s})_k)((\bft{{q_L}^T})_{rbn}C\gamma^\mu C(\bfu{q_R}_{tc})_l)\calf^{Uh\psi^4}_{152}(h),} \notag\\
\calo^{Uh\psi^4}_{153} &= \caly[{\tiny\yt{rst}}]\epsilon^{abc}\epsilon^{ik}\epsilon^{jl}((\bfu{{l_R}^T}_{p})_iC(\bfu{q_R}_{ra})_j)((\bfu{{q_R}^T}_{sb})_kC(\bfu{q_R}_{tc})_l)\calf^{Uh\psi^4}_{153}(h), \notag\\
{\color{red}\calo^{Uh\psi^4}_{154}} &= \caly[{\tiny\yt{rs,t}}]\epsilon^{abc}\epsilon^{ik}\epsilon^{jl}((\bfu{{l_R}^T}_{p})_iC(\bfu{q_R}_{ra})_j)((\bfu{{q_R}^T}_{sb})_kC(\bfu{q_R}_{tc})_l)\calf^{Uh\psi^4}_{154}(h), 
\end{align}
\begin{align}
{\color{red}\calo^{Uh\psi^4}_{155}} &= \caly[{\tiny\yt{r,s,t}}]\epsilon^{abc}\epsilon^{ik}\epsilon^{jl}((\bfu{{l_R}^T}_{p})_iC(\bfu{q_R}_{ra})_j)((\bfu{{q_R}^T}_{sb})_kC(\bfu{q_R}_{tc})_l)\calf^{Uh\psi^4}_{155}(h), \notag\\
\calo^{Uh\psi^4}_{156} &= {\color{gray}\caly[{\tiny\yt{rst}}]\epsilon^{abc}\epsilon^{ij}\epsilon^{km}((\bfu{{l_R}^T}_{p})_iC\sigma_{\mu\nu}(\bfu{q_R}_{ra})_j)((\bfu{{q_R}^T}_{sb})_kC\sigma^{\mu\nu}(\bft\bfu{q_R})_{tcm})\calf^{Uh\psi^4}_{156}(h)}, \notag\\
{\color{red}\calo^{Uh\psi^4}_{157}} &= {\color{gray}\caly[{\tiny\yt{rs,t}}]\epsilon^{abc}\epsilon^{ij}\epsilon^{km}((\bfu{{l_R}^T}_{p})_iC\sigma_{\mu\nu}(\bfu{q_R}_{ra})_j)((\bfu{{q_R}^T}_{sb})_kC\sigma^{\mu\nu}(\bft\bfu{q_R})_{tcm})\calf^{Uh\psi^4}_{157}(h)},
\end{align}
\begin{align}
{\color{red}\calo^{Uh\psi^4}_{158}} &= \caly[{\tiny\yt{rs,t}}]\epsilon^{abc}\epsilon^{ij}\epsilon^{km}((\bfu{{l_R}^T}_{p})_iC(\bfu{q_R}_{ra})_j)((\bfu{{q_R}^T}_{sb})_kC(\bft\bfu{q_R})_{tcm})\calf^{Uh\psi^4}_{158}(h), \notag\\
{\color{red}\calo^{Uh\psi^4}_{159}} &={\color{gray} \caly[{\tiny\yt{r,s,t}}]\epsilon^{abc}\epsilon^{ij}\epsilon^{km}((\bfu{{l_R}^T}_{p})_iC(\bfu{q_R}_{ra})_j)((\bfu{{q_R}^T}_{sb})_kC(\bft\bfu{q_R})_{tcm})\calf^{Uh\psi^4}_{159}(h)}, \notag\\
{\color{red}\calo^{Uh\psi^4}_{160}} &= {\color{gray}\caly[{\tiny\yt{rs,t}}]\epsilon^{abc}\epsilon^{km}\epsilon^{ln}((\bft\bfu{{l_R}^T})_{pm}    C(\bft\bfu{q_R})_{ran})((\bfu{{q_R}^T}_{sb})_kC(\bfu{q_R}_{tc})_l)\calf^{Uh\psi^4}_{160}(h),}
\end{align}
}
$\calf$ functions in this type takes the form that
\beq
\calf^{Uh\psi^4}_i=1+\sum_{n=1}c^{Uh\psi^4}_n(\frac{h}{v})^n,\quad(i=133,\dots,160).
\eeq

\section{Conclusion}
\label{sec:conclu}

In this work, we present the complete and independent sets of the HEFT operators at the next-to-leading order, enumerating all the 224 (7704) operators for one (three) generation fermions without right-handed neutrino, and the 295 (11307) operators for one (three) generation fermions with right-handed neutrino, for the first time. Compared with the results in literature~\cite{Buchalla:2013rka,Brivio:2016fzo,Merlo:2016prs}, we find that there were 6 (9) terms of operators missing, corresponding to 420(663) operators for three generation fermions, without (with) right-handed sterile neutrinos, and there were many redundant operators. Furthermore, the numbers of the HEFT operators are consistent with the counting result via the Hilbert series obtained in Ref.~\cite{Graf:2022rco,Sun:2022aag}. Comparison with the literature is presented in the Appendix \ref{app:operatorlist} in detail. 


Although the Young tensor method has been developed to obtain the operator basis for generic EFTs up to any mass dimension \cite{Li:2022tec}, the operators involving in the Goldstone bosons and the spurion fields, appears in the HEFT, need further treatments. According to the operator-amplitude correspondence, for a type of operators, the on-shell amplitudes form a complete and independent basis in the Lorentz space, and if such operators involve in the Goldstone bosons, applying the soft theorem on the amplitude would pick up the Lorentz subspace satisfying the Adler zero condition in the soft momentum limit. All the HEFT operators have been rewritten with the on-shell contact amplitudes, as shown in Appendix \ref{app:amplitude}.

In the HEFT, the spurion field is also quite special because it is not the dynamical degree of freedom but it transforms as the adjoint representation under the $SU(2)$ symmetry. The spurion field is introduced to parametrizes the custodial symmetry breaking through freezing the dynamical field degree of freedom, and thus it could only form the $SU(2)$ invariant together other dynamical fields in the operator. Thus we remove the spurion from the Lorentz tensor construction, but keep the spurion to be involved in the gauge and flavor tensor construction.
 

The complete sets of operator basis, investigated in this work, would benefit various phenomenological studies below the electroweak symmetry breaking scale, especially phenomenologies involving in the three generation fermions. On the other hand, when performing the matching between the UV non-decoupling physics and the HEFT operators, the complete HEFT basis is necessarily needed during the matching procedure. Furthermore, based on the power counting rules, the one-loop renormalization of the HEFT~\cite{Guo:2015isa,Alonso:2017tdy,Buchalla:2017jlu,Buchalla:2020kdh}, including the renormalization group equations, can be organized systematically using the complete operator basis.    
With the one-loop renormalization on the NLO operators, the next-to-next-to-leading-order operators would be quite relevant, and we leave the discussion in the coming work.


	
\section*{Acknowledgments} 
We thank Xiaochuan Lu and other authors of Ref.~\cite{Graf:2022rco} for valuable correspondence on the total numbers of the 4-fermion operators in the HEFT without right-handed neutrinos. J.-H.Y. and H. S. are supported by the National Science Foundation of China under Grants No. 12022514, No. 11875003 and No. 12047503, and National Key Research and Development Program of China Grant No. 2020YFC2201501, No. 2021YFA0718304, and CAS Project for Young Scientists in Basic Research YSBR-006, the Key Research Program of the CAS Grant No. XDPB15. 
M.-L.X. is supported in part by the U.S. Department of Energy under contracts No. DE-AC02-06CH11357 at Argonne and No.DE-SC0010143 at Northwestern.


\appendix
	
\section{Conversion between 2- and 4-Spinors}
\label{app:converting}
We use $\Psi$ to denote the 4-component spinors, whose conjugate is denoted by $\bar{\Psi} = \Psi^\dagger\gamma^0$. Denoting the 2-component left-handed spinors by $\chi,\xi$, and the 2-component right-handed spinors are denoted by $\chi^\dagger,\xi^\dagger$. Thus the 4-component spinors can be expressed by the 2-component spinors,
\be
\psi=\left(\begin{array}{c}\xi_\alpha\\{\chi^\dagger}^{\dot{\alpha}}\end{array}\right),\quad \bar{\Psi}=\left(\begin{array}{cc}\chi^\alpha,&{\xi^\dagger}_{\dot{\alpha}}\end{array}\right)\,.
\ee
Then Dirac bilinears can be expressed by 2-components spinors,
\bea
\bar{\Psi}_1\Psi_2 &=& {\chi_1}^\alpha{\xi_2}_\alpha + {\xi^\dagger_1}_{\dot{\alpha}}{\chi^\dagger_2}^{\dot{\alpha}}\,,\notag \\
\bar{\Psi}_1\gamma^\mu\Psi_2 &=& {\chi_1}^\alpha\sigma^{\mu}_{\alpha\dot{\alpha}}{\chi^\dagger_2}^{\dot{\alpha}}+{\xi^\dagger_1}_{\dot{\alpha}}\bar{\sigma}^{\mu\dot{\alpha}\alpha}{\xi_2}_\alpha \,,\notag \\
\bar{\Psi}_1\sigma^{\mu\nu}\psi_2 &=& {\chi_1}^\alpha{(\sigma^{\mu\nu})_\alpha}^\beta{\xi_2}_\beta+{\xi^\dagger_1}_{\dot{\alpha}}{(\bar{\sigma}^{\mu\nu})^{\dot{\alpha}}}_{\dot{\beta}}{\chi_2}^{\dot{\beta}}\,,\notag \\
\Psi_1^T C\Psi_2 &=& {\xi_1}^\alpha{\xi_2}_\alpha + {\chi^\dagger_1}_{\dot{\alpha}}{\chi^\dagger_2}^{\dot{\alpha}} \,,\notag \\
\Psi_1^T C\gamma^\mu\Psi_2 &=& {\xi_1}^\alpha\sigma^\mu_{\alpha\dot{\alpha}}{\chi^\dagger_2}^{\dot{\alpha}}+{\chi^\dagger_1}_{\dot{\alpha}}\bar{\sigma}^{\mu\dot{\alpha}\alpha}{\xi_2}_\alpha\,, \notag\\
\Psi_1^T C\sigma^{\mu\nu}\Psi_2 &=& {\xi_1}^\alpha{(\sigma^{\mu\nu})_\alpha}^\beta{\xi_2}_\beta+{\chi\dagger_1}_{\dot{\alpha}}{(\bar{\sigma}^{\mu\nu})^{\dot{\alpha}}}_{\dot{\beta}}{\chi^\dagger_2}^{\dot{\beta}}\,,\notag \\
\bar{\Psi}_1C\bar{\Psi}_2^T &=& {\xi^\dagger_1}_{\dot{\alpha}}{\xi^\dagger_2}^{\dot{\alpha}}+{\chi_1}^\alpha{\chi_2}_\alpha\,,\notag \\
\bar{\Psi}_1\gamma^\mu C\bar{\Psi}_2^T &=& {\chi_1}^\alpha{\sigma^\mu}_{\alpha\dot{\alpha}}{\xi^\dagger_2}^{\dot{\alpha}} + {\xi^\dagger_1}_{\dot{\alpha}}\bar{\sigma}^{\mu\dot{\alpha}\alpha}{\chi_2}_\alpha\,,\notag \\
\bar{\Psi}_1\sigma^{\mu\nu}C\bar{\Psi}_2^T &=& {\xi^\dagger_1}_{\dot{\alpha}}{(\bar{\sigma}^{\mu\nu})^{\dot{\alpha}}}_{\dot{\beta}}{\xi^\dagger_2}^{\dot{\beta}}+{\chi_1}^\alpha{(\sigma^{\mu\nu})_\alpha}^\beta{\chi_2}_\beta\,,
\eea
where the gamma matrices in the equations above take the form  
\bea
\gamma^\mu &=& \left(\begin{array}{cc}0&\sigma^\mu_{\alpha\dot{\beta}}\\\bar{\sigma}^{\mu\dot{\alpha}\beta}\end{array}\right)\,,\notag \\
C &=& i\gamma^0\gamma^2 = \left(\begin{array}{cc}\epsilon_{\alpha\beta}&0\\0&\epsilon^{\dot{\alpha}\dot{\beta}}\end{array}\right) = \left(\begin{array}{cc}-\epsilon^{\alpha\beta}&0\\0&-\epsilon_{\dot{\alpha}\dot{\beta}}\end{array}\right)\,,\notag \\
\sigma^{\mu\nu} &=& \frac{i}{2}[\gamma^\mu,\gamma^\nu] = \left(\begin{array}{cc}{(\sigma^{\mu\nu})_\alpha}^{\beta}&0\\0&{(\bar{\sigma}^{\mu\nu})^{\dot{\alpha}}}_{\dot{\beta}}\end{array}\right)\,.
\eea
In this work, the left- and right-handed fermions are defined as
\be
q_L = \left(\begin{array}{c}Q_L\\0\end{array}\right),\quad l_L=\left(\begin{array}{c}L_L\\0\end{array}\right),\quad q_R = \left(\begin{array}{c}0\\Q_R\end{array}\right),\quad l_R=\left(\begin{array}{c}0\\L_R\end{array}\right),
\ee
and their conjugates are
\be
\bar{q}_L = \left(\begin{array}{cc}0,&Q_L^\dagger\end{array}\right),\quad \bar{l}_L = \left(\begin{array}{cc}0,&L_L^\dagger\end{array}\right),\quad \bar{q}_R = \left(\begin{array}{cc}Q^\dagger_R,&0\end{array}\right),\quad \bar{l}_R = \left(\begin{array}{cc}L^\dagger_R,&0\end{array}\right) \,.
\ee
According to these correspondences, we can convert the 2-component notations adopted in this paper to 4-component notations, for example, the operator $\calo^{Uh\psi^4}_1$ can by expressed in 4-component notation that
\be
\calo^{Uh\psi^4}_1 = \caly[{\tiny\yt{pr},\yt{st}}]({\bar{q_L}}_p\bfu{q_R}_s)({\bar{q_L}}_r\bfu{q_R}_t)\calf^{Uh\psi^4}_1(h)\,.
\ee



\section{Comparison with Literatures on NLO Operators}
\label{app:operatorlist}

		In this appendix we present the comparision with the NLO operators obtained in this work and previous literature such as \cite{Buchalla:2013rka}, \cite{Brivio:2016fzo} and \cite{Merlo:2016prs}~\footnote{Ref.~\cite{Pich:2016lew,Krause:2018cwe} also present the NLO operator basis, but a different power counting on the spurion is adopted. So the LO + NLO operators in total in Ref.~\cite{Pich:2016lew,Krause:2018cwe} should be a subset of the operator basis listed here. }. The operators in this section follow the conventions that: first, $D_\mu\Pi = D_\mu\phi^I\tau^I$ are used to represent the NGBs building block $\bfv_\mu$, as discussed in sec.~\ref{subsec:adler}. 
		The operator involving in the $D_\mu\phi^I$ would be equivalent to the one with the $\bfv_\mu$, which can be seen from the soft-recursion relation for the NGBs.   
		Second, the physical Higgs $h$ and its dimensionless function $\calf$ are omitted in all the operators here. For comparision, all the operators that were missing are marked by red in the following.

		\subsection{Bosonic Operators}
		
		The following notations and conversions among this work and previous ones are utilized for the comparison.
		\bit
		\item The field degrees of freedom of gauge bosons in this work are chosen to be
		\beq
		{X_L}_{\mu\nu} = \frac{1}{2}(X_{\mu\nu}+i\tilde{X}_{\mu\nu}),\quad {X_R}_{\mu\nu} = \frac{1}{2}(X_{\mu\nu}-i\tilde{X}_{\mu\nu})\,,
		\eeq
		where $\tilde{X}_{\mu\nu}=\epsilon_{\mu\nu\rho\sigma}X^{\rho\sigma}$. 
		Thus for the operators involving in the gauge bosons, we follow the convention that $X_L$ corresponds $X$ and $X_R$ corresponds $\tilde{X}$.

		\item If there are many derivatives in a type of operators, the Lorentz structure could be complicate, and thus there are different choices of the independent Lorentz basis. For the type $D^4h^4$, the Lorentz structure chosen in this paper is $(hD_\mu D_\nu h)(hD^\mu D^\nu h)$, while in Ref.~\cite{Buchalla:2013rka} the form $D_\mu hD_\nu h D^\mu h D^\nu h$ was chosen. They are different but can be converted to each other by integration-by-part(IBP):
				\begin{align}
						(D_\mu h D_\nu h)(D^\mu h D^\nu h) &= -(hD_\mu D_\nu h)(D^\mu h D^\nu h)-(hD_\nu h)(D^2hD^\nu h) -(hD_\nu h)(D^\mu h D_\mu D^\nu h) \notag\\
														   &=(hD_\mu D_\nu h)(h D^\mu D^\nu h)+(hD_\nu h)(D^\mu h D_\mu D^\nu h)-(hD_\nu h)(D^\mu h D_\mu D^\nu h) \notag\\
														   &=(hD_\mu D_\nu h)(h D^\mu D^\nu h) + \cdots\,,
				\end{align}
		where terms involving d'Alembert operator $D^2$ are eliminated, this procedure can be completed  by the ABC4EFT code automatically.
		
		\item Because the choice of the invariant tensors in the gauge group could be arbitrary, there are several operators which are not exactly the same with their correspondences in previous literature. For example, the operator $\mathcal{O}^{XUhD^2}_8$ corresponds to the $\mathcal{O}_{XU12}$ in the Ref.~\cite{Buchalla:2013rka} via
		\begin{align}
				\mathcal{O}_{XU12}&\leftrightarrow ig\epsilon_{\mu\nu\lambda\rho}\langle W^{\mu\nu}\tau_L\rangle\langle\tau_L[L^\lambda,L^\rho]\rangle F_{XU12}(h) \notag\\
		&\leftrightarrow \epsilon_{JLM}\bft^I\bft^J W^{I\mu\nu}_R D_\mu\phi^L D_\nu\phi^M\,,
		\end{align}
		but is of a different form. This equivalence can be checked via group tensor identities.
		For the $SU(2)$ group, there is the identity:
		\begin{equation}
		    \delta_{IJ}\epsilon_{KLM}-\delta_{IK}\epsilon_{JLM}+\delta_{IL}\epsilon_{JKM}-\delta_{IM}\epsilon_{JKL}=0\,.
		\end{equation}
		Contracting with building blocks $\bft^I\bft^JX^K_{\mu\nu}D_\mu\phi^L D_\nu\phi^M$, the first one is spurion self-contraction, which should be eliminated, the second one is of the form of $\mathcal{O}_{XU12}$, and the third one is of the form of $\mathcal{O}^{X^2Uh}_7$. The last one can be converted to the third one by
		\begin{align}
				&\delta_{IM}\epsilon_{JKL}\bft^I\bft^JX^K_{\mu\nu}D_\mu\phi^L D_\nu\phi^M\notag\\
				=& \delta_{IM}\epsilon_{JKL}\bft^I\bft^JX^K_{\mu\nu} D_\nu\phi^MD_\mu\phi^L  \quad([D_\mu\phi^L,D_\nu\phi^M]\rightarrow 0)\notag\\
				=&-\delta_{IL}\epsilon_{JKM}\bft^I\bft^JX^K_{\mu\nu} D_\mu\phi^L D_\nu\phi^M \quad (L\leftrightarrow M,X_{\mu\nu}=-X_{\nu\mu})\,.
		\end{align}
		Thus $\mathcal{O}_{XU12}\propto \mathcal{O}^{XUhD^2}_8$, as expected.
		\eit
Most of the identifications of the operators in this paper and the previous literature in this appendix are realised by this way. Next we list the complete operator list. 
		
		\paragraph{Type $UhD^4$:}
\begin{center}
\begin{longtable}{cccc}
  \hline
  $\mathcal{O}^{UhD^4}_i$&Operators&\cite{Buchalla:2013rka}&\cite{Brivio:2016fzo}\\
  \hline
  1&$\left(D_{\mu } \phi _{\text{}}^I\right) \left(D_{\nu } \phi _{\text{}}^J\right) \left(D^{\mu } \phi _{\text{}}^I\right) \left(D^{\nu } \phi_{\text{}}^J\right)$ &$ \mathcal{O}_{D1}$& $ \mathcal{P}_6$ \\
  2&$\left(D_{\mu } \phi _{\text{}}^I\right) \left(D_{\nu } \phi _{\text{}}^I\right) \left(D^{\mu } \phi _{\text{}}^J\right) \left(D^{\nu } \phi _{\text{}}^J\right)$ & $ \mathcal{O}_{D2}$ & $ \mathcal{P}_{11}$ \\
  3&$\bft_{\text{}}^J\bft_{\text{}}^K \left(D_{\mu } \phi _{\text{}}^I\right) \left(D_{\nu } \phi _{\text{}}^K\right) \left(D^{\nu } \phi _{\text{}}^I\right) \left(D^{\mu } \phi _{\text{}}^J\right)$ & $ \mathcal{O}_{D5}$ & $ \mathcal{P}_{24}$ \\
  4&$\bft_{\text{}}^J\bft_{\text{}}^K \left(D_{\mu } \phi _{\text{}}^I\right) \left(D_{\nu } \phi _{\text{}}^J\right) \left(D^{\mu } \phi _{\text{}}^I\right) \left(D^{\nu } \phi _{\text{}}^K\right)$ & $ \mathcal{O}_{D4}$ & $ \mathcal{P}_{23}$ \\
  5&$\bft_{\text{}}^I\bft_{\text{}}^J\bft_{\text{}}^K\bft_{\text{}}^M \left(D_{\mu } \phi _{\text{}}^I\right) \left(D_{\nu } \phi _{\text{}}^K\right) \left(D^{\mu } \phi _{\text{}}^J\right) \left(D^{\nu } \phi _{\text{}}^M\right)$ & $ \mathcal{O}_{D3}$ & $ \mathcal{P}_{26}$ \\
  6&$\bft_{\text{}}^J \left(D_{\mu } \phi_{\text{}}^I\right) \left(D_{\nu } \phi _{\text{}}^J\right) \left(D^{\nu } h_{\text{}}^{\text{}}\right) \left(D^{\mu } \phi _{\text{}}^I\right)$ & $ \mathcal{O}_{D12}$ & $ \mathcal{S}_6$ \\
  7&$\bft_{\text{}}^J \left(D_{\mu } \phi_{\text{}}^I\right) \left(D_{\nu } \phi _{\text{}}^I\right) \left(D^{\nu } h_{\text{}}^{\text{}}\right) \left(D^{\mu } \phi _{\text{}}^J\right)$ & $ \mathcal{O}_{D13}$& $ \mathcal{S}_5$ \\
  8&$\epsilon _{\text{}}^{\text{JKM}}\bft_{\text{}}^I\bft_{\text{}}^K \left(D_{\nu } \phi _{\text{}}^J\right) \left(D_{\mu } \phi _{\text{}}^M\right) \left(D^{\nu } h_{\text{}}^{\text{}}\right) \left(D^{\mu } \phi _{\text{}}^I\right)$ & $ \mathcal{O}_{D6}$ &$ \mathcal{P}_{18}$ \\
  9&$\bft_{\text{}}^I\bft_{\text{}}^J\bft_{\text{}}^K \left(D_{\mu } \phi _{\text{}}^I\right) \left(D_{\nu } \phi _{\text{}}^K\right) \left(D^{\nu } h_{\text{}}^{\text{}}\right) \left(D^{\mu } \phi _{\text{}}^J\right)$ & $ \mathcal{O}_{D14}$ & $ \mathcal{S}_{15}$ \\
  10&$h_{\text{}}^{\text{}} \left(D_{\mu } \phi _{\text{}}^I\right) \left(D_{\nu } \phi _{\text{}}^I\right) \left(D^{\mu } D^{\nu } h_{\text{}}^{\text{}}\right)$ & $ \mathcal{O}_{D8}$ & $ \mathcal{P}_8$ \\
  11&$\left(D_{\nu } h_{\text{}}^{\text{}}\right) \left(D_{\mu } \phi _{\text{}}^I\right) \left(D^{\nu } h_{\text{}}^{\text{}}\right) \left(D^{\mu } \phi _{\text{}}^I\right)$ & $ \mathcal{O}_{D7}$ & $ \mathcal{P}_{20}$ \\
  12&$h_{\text{}}^{\text{}}\bft_{\text{}}^I\bft_{\text{}}^J \left(D_{\mu } \phi _{\text{}}^I\right) \left(D_{\nu } \phi _{\text{}}^J\right) \left(D^{\mu } D^{\nu } h_{\text{}}^{\text{}}\right)$ & $ \mathcal{O}_{D10}$ & $ \mathcal{P}_{22}$ \\
  13&$\bft_{\text{}}^I\bft_{\text{}}^J \left(D_{\nu } h_{\text{}}^{\text{}}\right) \left(D_{\mu } \phi _{\text{}}^I\right) \left(D^{\nu } h_{\text{}}^{\text{}}\right) \left(D^{\mu } \phi _{\text{}}^J\right)$ & $ \mathcal{O}_{D9}$ & $ \mathcal{P}_{21}$ \\
  {\color{red}14}&$h_{\text{}}^{\text{}}\bft_{\text{}}^I \left(D_{\mu } \phi _{\text{}}^I\right) \left(D^{\nu } h_{\text{}}^{\text{}}\right) \left(D_{\nu } D^{\mu } h_{\text{}}^{\text{}}\right)$ &$ \mathcal{O}_{D15}$ & \\
  15&$h_{\text{}}^{\text{}}{}^2 \left(D_{\mu } D_{\nu } h_{\text{}}^{\text{}}\right) \left(D^{\mu } D^{\nu } h_{\text{}}^{\text{}}\right)$& $ \mathcal{O}_{D11}$ & $ \mathcal{P}_{DH}$ \\
  \hline
\end{longtable}
\end{center}

\paragraph{Type $X^2Uh$:}

\begin{center}		
\begin{longtable}{cccccc}
  \hline
  $\mathcal{O}^{X^2Uh}_i$&Operators&\cite{Buchalla:2013rka}&\cite{Brivio:2016fzo}\\
  \hline
  1&$B_L^{\mu\nu}B_{L\mu\nu}$ & $\mathcal{O}_{Xh1}$ & $\mathcal{P}_B$  \\
  2&$B_R^{\mu\nu}B_{R\mu\nu}$ & $\mathcal{O}_{Xh4}$ & $\mathcal{S}_{\tilde{B}}$  \\
  3&${G_L}^{A\mu\nu}{G_L}^A_{L\mu\nu}$ & $\mathcal{O}_{Xh3}$ & $\mathcal{P}_G$ \\
  4&$G_R^{A\mu\nu}G^A_{R\mu\nu}$ & $\mathcal{O}_{Xh6}$ & $\mathcal{S}_{\tilde{G}}$ \\
  5&$W_L^{I\mu\nu}W^I_{L\mu\nu}$ & $\mathcal{O}_{Xh2}$ & $\mathcal{P}_W$ \\
  6&$W_R^{I\mu\nu}W^I_{R\mu\nu}$ & $\mathcal{O}_{Xh5}$ & $\mathcal{S}_{\tilde{W}}$ \\
  7&$\bft_{\text{}}^IB_{L\mu\nu}W_L^{I{\mu \nu }}$ & $\mathcal{O}_{XU1}$ & $\mathcal{P}_1$  \\
  8&$\bft_{\text{}}^IB_{R\mu\nu}W_R^{I{\mu \nu }}$ & $\mathcal{O}_{XU4}$ & $\mathcal{S}_1$ \\
  9&$\bft_{\text{}}^I\bft_{\text{}}^JW_{L\mu\nu}^IW_L^{J\mu\nu}$ & $\mathcal{O}_{XU2}$  \\
  10&$\bft_{\text{}}^I\bft_{\text{}}^JW_{R\mu\nu}^IW_R^{J\mu\nu}$ & $\mathcal{O}_{XU5}$  \\
  \hline
\end{longtable}
\end{center}
\label{t2}

\paragraph{Type $XUhD^2$:}
\begin{center}
\begin{longtable}{cccc}
  \hline
  $\mathcal{O}^{XUhD^2}_i$&Operators&\cite{Buchalla:2013rka}&\cite{Brivio:2016fzo}\\
  \hline
  1&$\epsilon^{IJK} h W_L^{I\nu\mu}(D_\mu\phi^J)(D_\nu\phi^K)$ & $O_{XU8}$ & $P_3$ \\
  {\color{red}2}&$\epsilon^{IJK} h W_R^{K\nu\mu}(D_\mu\phi^I)(D_\nu\phi^J)$ & $O_{XU11}$ & \\
  3&$\epsilon^{IJK} h\bft^K B_L^{\nu\mu}(D_\mu\phi^I)(D_\nu\phi^J)$ & $O_{XU7}$ & $P_2$ \\
  {\color{red}4}&$\epsilon^{IJK} h\bft^K B_R^{\nu\mu}(D_\mu\phi^I)(D_\nu\phi^J)$ & $O_{XU10}$ & \\
  5&$h\bft^JW_L^{I\nu\mu}(D_\nu\phi^I)(D_\mu\phi^J)$ & $O_{XU6}$ & $S_4$  \\
  6&$h\bft^JW_R^{I\nu\mu}(D_\nu\phi^I)(D_\mu\phi^J)$ & $O_{XU3}$ & $P_{14}$ \\
  7&$\epsilon^{JKM}h\bft^I\bft^K W_L^{I\nu\mu}(D_\nu\phi^J)(D_\mu\phi^M)$ & $O_{XU9}$ &$P_{13}$ \\
  {\color{red}8}&$\epsilon^{JKM}h\bft^I\bft^M W_R^{K\nu\mu}(D_\nu\phi^I)(D_\mu\phi^J)$ & $O_{XU12}$ & \\
  \hline
\end{longtable}
\end{center}
\label{t3}

\paragraph{Type$ X^3$:}
\begin{center}
\begin{longtable}{ccc}
		\hline
  $\mathcal{O}^{X^3}_i$&Operators&\cite{Buchalla:2013rka}\\
  \hline
  1&$f^{ABC}{G}^A_{L\mu\nu}{G}^{C\mu\lambda}_LG_{L\lambda}^{B\nu}$&$\mathcal{P}_{GGG}$ \\
  2&$f^{ABC}\mathcal{D}^A_{R\mu\nu}\mathcal{D}^{C\mu\lambda}_RG_{R\lambda}^{B\nu}$&$\mathcal{S}_{\tilde{GGG}}$\\
  3&$\epsilon^{IJK}W^I_{L\mu\nu}W^{K\mu\lambda}_LW_{L\lambda}^{J\nu}$&$\mathcal{P}_{WWW}$ \\
  4&$\epsilon^{IJK}W^I_{R\mu\nu}W^{K\mu\lambda}_RW_{R\lambda}^{J\nu}$&$\mathcal{P}_{\tilde{WWW}}$ \\
  5&$\epsilon^{IJK}\bft^K{B_L}_{\mu\nu}{W^J_L}^{\lambda\mu}{{W^I_L}_\lambda}^\nu$ \\
  6&$\epsilon^{IJK}\bft^K{B_R}_{\mu\nu}{W^J_R}^{\lambda\mu}{{W^I_R}_\lambda}^\nu$ \\
  \hline
\end{longtable}
\end{center}
\label{t4}

\subsection{Fermionic Operators}
For the operators involving in the fermions, the treatment of the spurion fields would be complicate.	\bit
	\item
	There are 3 useful identities have been proven to be useful for the comparison:
	\begin{align}
			[\bft,\bfv_\mu]&=\frac{i}{2}\epsilon^{IJK}\langle \bft\,, \sigma^I\rangle\langle\bfv_\mu\sigma^J\rangle\sigma^K \,,\\
			\{\bft,\bfv_\mu\}&=\langle \bft\bfv_\mu\rangle \mathbf{I}\,, \\
			\bft\bfv_\mu\bft &=\bft\langle \bft\bfv_\mu\rangle -\bfv_\mu\,,
	\end{align}
	where $\langle...\rangle$ represents matrix trace.
	For example, the correspondence of $\mathcal{O}^{\psi^2UhD^2}_{10}$ and $\mathcal{N}^Q_{18}$ can be verified as following:
	\begin{align}
			\mathcal{N}^Q_{18}:\bft\bfv_\mu\bft\bfv^\mu &= (\bft\bfv_\mu\bft)\bfv^\mu\bft \rightarrow \bft\langle\bft\bfv_\mu\rangle\bfv^\mu\bft \notag\\
		&=\langle \bft\bfv_\mu\rangle\bft\bfv^\mu\bft \rightarrow \langle\bft\bfv_\mu\rangle\langle\bft\bfv^\mu\rangle \bft \notag\\
		&\rightarrow \bft^I\bft^J\bft^K(D_\mu\phi^I)(D^\mu\phi^J)\,.
	\end{align}
	\item
	In type $\psi^2 UhD^2$, operators $\mathcal{O}^{\psi^2UhD^2}_{9},\mathcal{O}^{\psi^2UhD^2}_{8}$ are quite different from their correspondences $\mathcal{N}^{Q}_{17},\mathcal{N}^Q_{20}$ in \cite{Buchalla:2013rka}. To verify this relation, recombine the $\mathcal{N}^Q_{17},\mathcal{N}^Q_{20}$ as
	\begin{align}
			&\mathcal{N}^Q_{17}+\mathcal{N}^Q_{20}\rightarrow \langle \bft\bfv_\mu\rangle\langle\bft\bfv^\mu\rangle \,,\\
			&\mathcal{N}^Q_{17}-\mathcal{N}^Q_{20}\rightarrow \langle \bft\bfv\rangle\epsilon^{IJK}\langle\bft\sigma^I\rangle\langle\bfv^\mu\sigma^J\rangle\sigma^K\,.
	\end{align}
	It is straightforward to figure out that $\mathcal{O}^{\psi^2 UhD^2}_{8}$ corresponds to $\mathcal{N}^Q_{17}-\mathcal{N}^Q_{20}$. As for the first combination, it actually can be converted to $\mathcal{O}^{\psi^2UhD^2}_{9}$, since there is the equation that
	\beq
	\epsilon^{IJK}{\tau^L}^j_i = -\epsilon^{ILM}\epsilon^{MKJ}\delta^j_i-\epsilon^{IKM}\epsilon^{MLJ}\delta^j_i-\epsilon^{JIM}\epsilon^{MLK}\delta^j_i\,.
	\eeq
	The right hand has the expression that $-2(\delta^{IK}\delta^{LJ}+\delta^{IJ}\delta^{KL})\delta_i^j$, one of them corresponds to spurion self-contraction, and the other corresponds to operator $\mathcal{O}^{\psi^2UhD^2}_{9}$.
	\item
	In the case of 4-fermion operators, Fierz identities in Appendix \ref{app:converting} are used for correspondence,	
\begin{align}
	& \delta_{il}\delta_{kj}=\frac{1}{2}(\sigma_I)_{ij}(\sigma_I)_{kl}+\frac{1}{2}\delta_{ij}\delta_{jl}\,, \notag \\
	& \delta_{ik}\delta_{jl}=\frac{1}{2}(\sigma_I)_{ij}(\sigma_I)_{kl}+\frac{1}{2}\delta_{ij}\delta_{jl}\,, \notag \\
	& \delta_{il}\delta_{kj}=\frac{1}{2}(\lambda_A)_{ij}(\lambda_A)_{kl}+\frac{1}{3}\delta_{ij}\delta_{jl}\,, \notag \\
	& \delta_{ik}\delta_{jl}=\frac{1}{2}(\lambda_A)_{ij}(\lambda_A)_{kl}+\frac{1}{3}\delta_{ij}\delta_{jl}\,, \notag \\
	& \delta_{il}\delta_{kj}=\frac{1}{6}(\bar{\sigma}^{\mu\nu})_{ij}(\bar{\sigma}_{\mu\nu})_{kl}+\frac{1}{2}\delta_{ij}\delta_{kl}\,, \notag\\
	& \delta_{ik}\delta_{jl}=\frac{1}{6}(\bar{\sigma}^{\mu\nu})_{ij}(\bar{\sigma}_{\mu\nu})_{kl}+\frac{1}{2}\delta_{ij}\delta_{kl}\,.
\end{align}
For example,
	\begin{align}
	& R^Q_1 = \frac{1}{2}\calo^{\psi^4Uh}_3 + \frac{1}{2}\calo^{\psi^4Uh}_4 \notag\\
	& R^Q_2 = \frac{1}{2}\calo^{\psi^4Uh}_1+\frac{1}{6}\calo^{\psi^4Uh}_2-\frac{1}{2}\calo^{\psi^4Uh}_3-\frac{1}{6}\calo^{\psi^4Uh}_4 \notag\\
	& R^Q_5 = \calo^{\psi^4Uh}_1+\frac{1}{3}\calo^{\psi^4Uh}_2-\frac{1}{3}\calo^{\psi^4Uh}_3-\calo^{\psi^4Uh}_4 \notag \\
	& R^Q_6 = \frac{1}{2}\calo^{\psi^4Uh}_1 + \frac{1}{2}\calo^{\psi^4Uh}_2
	\end{align}
	\item
	 As fermions' number increases, there is a general case not considered in previous literature, which gives extra independent operators. Because the product of $\delta$ symbol $\delta^i_j\delta^k_l$ is invariant tensor under $SU(2)$ group, it satisfies the relation that
	\beq
	{\tau^I}^i_j\delta^k_l + {\tau^I}^k_l\delta^i_j+{\tau^I}^k_j\delta^i_l+{\tau^I}^i_l\delta^k_j =0
	\eeq
	Thus there are 3 independent invariant tensors of this kind. This fact implies that in many types involving spurions such as ${Q^\dagger_L}{Q_L}{Q^\dagger_R}{Q_R}$, ${Q^\dagger_L}{Q^L}{L^\dagger_L}{L_L}$ and so on, operators considered before are not complete, in which only 2 independent tensors were considered before.
	\eit

\subsubsection{Fermion-current Operators}
\paragraph{Type $\psi^2UhD$:}
\begin{center}
\begin{longtable}{ccccc}
  \hline
  $\mathcal{O}^{\psi^2 UhD}_i$&Operators&\cite{Buchalla:2013rka}&\cite{Brivio:2016fzo}\\
  \hline
  1&${\tau^{Ii}}_j h (Q_L^{\dagger aj}\sigma^\mu Q_{Lai}) (D_\mu\phi^I)$ & $O_{\psi V3}\&O_{\psi V3}^\dagger$ \\
  2&$h\bft^I(Q_L^{\dagger ai}\sigma^\mu Q_{Lai})(D_\mu\phi^I)$& $O_{\psi V1}$ & $N^Q_5$ \\
  3&$\epsilon^{IJK}{\tau^{Ki}}_j h \bft^J (Q_L^{\dagger aj}\sigma^\mu Q_{Lai}) (D_\mu\phi^I)$ &  \\
  4&${\tau^{Ji}}_j h \bft^I\bft^J(Q_L^{\dagger aj}\sigma^\mu Q_{Lai}) (D_\mu\phi^I)$ & $O_{\psi V2}$ \\
  5&${\tau^{Ii}}_j h (Q_R^{\dagger aj}\sigma^\mu Q_{Rai}) (D_\mu\phi^I)$ & $O_{\psi V6}\&O_{\psi V6}^\dagger$ & $N^Q_2$ \\
  6&$h\bft^I(Q_R^{\dagger ai}\sigma^\mu Q_{Rai})(D_\mu\phi^I)$ & $O_{\psi V4}$ & $N^Q_6$ \\
  7&$\epsilon^{IJK}{\tau^{Ki}}_j h \bft^J (Q_R^{\dagger aj}\sigma^\mu Q_{Rai}) (D_\mu\phi^I)$ & & $N^Q_4$ \\
  8&${\tau^{Ji}}_j h \bft^I\bft^J(Q_R^{\dagger aj}\sigma^\mu Q_{Rai}) (D_\mu\phi^I)$ & $O_{\psi V5}$ & $N^Q_8$ \\
  \hline
\end{longtable}
\end{center}

\paragraph{Type $\psi^2UhD^2$:}
\begin{center}
\begin{longtable}{ccccc}
  \hline
  $\mathcal{O}^{\psi^2UhD^2}_i$&Operators&\cite{Buchalla:2013rka}&\cite{Brivio:2016fzo}\\
  \hline
  1&$(Q_L^{\dagger ai}Q_{Rai})(D_\mu\phi^I)(D^\mu\phi^I)$ &$O_{\psi S1}$ &$N_{15}^Q$ & \\
  2&${\tau^{Kj}}_i\epsilon^{IJK}(Q_L^{\dagger ai}\bar{\sigma}^{\mu\nu}Q_{Raj})(D_\nu\phi^I)(D_\mu\phi^J)$ & &$N_{27}^Q$ & \\
  3&$\epsilon^{IJK}\bft^K(Q_L^{\dagger ai}\bar{\sigma}^{\mu\nu}Q_{Raj})(D_\nu\phi^I)(D_\mu\phi^J)$ &$O_{\psi T1}$ &$N_{28}^Q$ & \\
  4&${\tau^{Jj}}_i\bft^I(Q_L^{\dagger ai}Q_{Raj})(D_\mu\phi^I)(D^\mu\phi^J)$ &$O_{\psi S5}\& O_{\psi S6}$ &$N_{19}^Q$ & \\
  5&${\tau^{Jj}}_i\bft^I(Q_L^{\dagger ai}\bar{\sigma}^{\mu\nu}Q_{Raj})(D_\nu\phi^I)(D_\mu\phi^J)$ &$O_{\psi T3}\& O_{\psi T4}$ &$N_{25}^Q$ & \\
  6&${\tau^{Jj}}_i\bft^J(Q_L^{\dagger ai}Q_{Raj})(D_\mu\phi^I)(D^\mu\phi^I)$ &$O_{\psi S2}$ &$N_{16}^Q$ & \\
  7&${\tau^{Mj}}_i\epsilon^{JKM}\bft^I\bft^K(Q_L^{\dagger ai}\bar{\sigma}^{\mu\nu}Q_{Raj})(D_\nu\phi^I)(D_\mu\phi^J)$ & &$N_{26}^Q$ & \\
\hline
\end{longtable}
\end{center}
\paragraph{Type $\psi^2XUh$:}

\begin{center}
\begin{longtable}{ccccc}
\hline
  $\mathcal{O}^{\psi^2 UhX}_i$&Operators&\cite{Buchalla:2013rka}&\cite{Brivio:2016fzo}\\
  \hline
  1&$B_{R\mu\nu}(Q_L^{\dagger ai}\bar{\sigma}^{\mu\nu}Q_{Rai})$ &$O_{\psi X1}$ &$N_{29}^Q$  \\
  2&${\tau^{Ij}}_i\bft^IB_{R\mu\nu}(Q_L^{\dagger ai}\bar{\sigma}^{\mu\nu}Q_{Raj})$ &$O_{\psi X2}$ &$N_{30}^Q$  \\
  3&${\lambda^{Ab}}_a G^A_{R\mu\nu}(Q_L^{\dagger ai}\bar{\sigma}^{\mu\nu}Q_{Rai})$&$O_{\psi X7}$ &$N_{31}^Q$ \\
  4&${\lambda^{Ab}}_a{\tau^{Ij}}_i\bft^I G^A_{R\mu\nu}(Q_L^{\dagger ai}\bar{\sigma}^{\mu\nu}Q_{Raj})$ &$O_{\psi X8}$ &$N_{32}^Q$ \\
  5&$\bft^I W^I_{R\mu\nu}(Q_L^{\dagger ai}\bar{\sigma}^{\mu\nu}Q_{Rai})$ &$O_{\psi X3}$ &$N_{34}^Q$ \\
  6&${\tau^{Kj}}_i\epsilon^{IJK}\bft^J W^I_{R\mu\nu}(Q_L^{\dagger ai}\bar{\sigma}^{\mu\nu}Q_{Raj})$ & & $N_{35}^Q$ \\
  7&${\tau^{Ij}}_iW^I_{R\mu\nu}(Q_L^{\dagger ai}\bar{\sigma}^{\mu\nu}Q_{Raj})$ &$O_{\psi X5}\& O_{\psi X6}$ &$N_{33}^Q$ \\
  8&${\tau^{Jj}}_i\bft^I\bft^J W^I_{R\mu\nu}(Q_L^{\dagger ai}\bar{\sigma}^{\mu\nu}Q_{Raj})$ &$O_{\psi X4}$ &$N_{36}^Q$ \\
  \hline
\end{longtable}
\end{center}

\subsubsection{Four-Fermion Operators}
\paragraph{Type ${Q_L^\dagger}^2{Q_R}^2$:}
\begin{center}
\begin{longtable}{ccccc}
\hline
$\mathcal{O}^{Uh\psi^4}_{1-24}$&Operator&\cite{Brivio:2016fzo}\\
\hline
1&$\mathcal{Y}[\tiny{\yt{pr},\yt{st}}](Q_{Lp}^{\dagger ai}Q_{Lr}^{\dagger bj})(Q_{Rsai}Q_{Rtbj})$&$R_1^Q$\\
2&$\mathcal{Y}[\tiny{\ytableaushort{pr},\ytableaushort{st}}](Q_{Lp}^{\dagger ai}\bar{\sigma}^{\mu\nu}Q_{Lr}^{\dagger bj})(Q_{Rsai}\bar{\sigma}_{\mu\nu}Q_{Rtbj})$ \\
3&$\mathcal{Y}[\tiny{\ytableaushort{pr},\ytableaushort{st}}](Q_{Lp}^{\dagger ai}Q_{Lr}^{\dagger bj})(Q_{Rsaj}Q_{Rtbi})$&$R_2^Q$\\
4&$\mathcal{Y}[\tiny{\ytableaushort{pr},\ytableaushort{st}}](Q_{Lp}^{\dagger ai}\bar{\sigma}^{\mu\nu}Q_{Lr}^{\dagger bj})(Q_{Rsaj}\bar{\sigma}_{\mu\nu}Q_{Rtbi})$ & $R_6^Q$\\
{\color{red}5}&$\mathcal{Y}[\tiny{\yt{p,r},\yt{s,t}}](Q_{Lp}^{\dagger ai}Q_{Lr}^{\dagger bj})(Q_{Rsai}Q_{Rtbj})$ \\
{\color{red}6}&$\mathcal{Y}[\tiny{\ytableaushort{p,r},\ytableaushort{s,t}}](Q_{Lp}^{\dagger ai}\bar{\sigma}^{\mu\nu}Q_{Lr}^{\dagger bj})(Q_{Rsai}\bar{\sigma}_{\mu\nu}Q_{Rtbj})$ \\
{\color{red}7}&$\mathcal{Y}[\tiny{\ytableaushort{p,r},\ytableaushort{s,t}}](Q_{Lp}^{\dagger ai}Q_{Lr}^{\dagger bj})(Q_{Rsaj}Q_{Rtbi})$ \\
\hline
\end{longtable}
\end{center}

\paragraph{Type ${Q^\dagger_L}{Q_L}{Q^\dagger_R}{Q_R}$}
\begin{center}
\begin{longtable}{ccccc}
\hline
$\mathcal{O}^{Uh\psi^4}_{25-36}$&Operator&\cite{Brivio:2016fzo}\\
\hline
25&$(Q_{Lpai}\sigma_\mu Q_{Ls}^{\dagger ai})(Q_{Rr}^{\dagger bj}\sigma^\mu Q_{Rtbj})$& $R^Q_{17}$ \\
26&$(Q_{Lpai}\sigma_\mu Q_{Ls}^{\dagger ak})(Q_{Rr}^{\dagger bi}\sigma^\mu Q_{Rtbk})$&$R^Q_{21}$\\
27&$(Q_{Lpai}\sigma_\mu Q_{Ls}^{\dagger ci})(Q_{Rr}^{\dagger aj}\sigma^\mu Q_{Rtcj})$& $R^Q_{22}$  \\
28&$(Q_{Lpai}\sigma_\mu Q_{Ls}^{\dagger ck})(Q_{Rr}^{\dagger ai}\sigma^\mu Q_{Rtck})$& $R^Q_{26}$ \\
29&${\tau^{Ii}}_k\bft^I(Q_{Lpai}\sigma_\mu Q_{Ls}^{\dagger ak})(Q_{Rr}^{\dagger bj}\sigma^\mu Q_{Rtbj})$& $R_{19}^Q$ \\
30&${\tau^{Ii}}_k\bft^I(Q_{Lpai}\sigma_\mu Q_{Ls}^{\dagger ck})(Q_{Rr}^{\dagger aj}\sigma^\mu Q_{Rtcj})$& $R_{24}^Q$\\
31&${\tau^{Il}}_k\bft^I(Q_{Lpai}\sigma_\mu Q_{Ls}^{\dagger ak})(Q_{Rr}^{\dagger bi}\sigma^\mu Q_{Rtbl})$&$R_{23}^Q$ \\
32&${\tau^{Il}}_k\bft^I(Q_{Lpai}\sigma_\mu Q_{Ls}^{\dagger ck})(Q_{Rr}^{\dagger aj}\sigma^\mu Q_{Rtcl})$& $R_{18}^Q$ \\
{\color{red}33}&${\tau^{Ii}}_j\bft^I(Q_{Lpai}\sigma_\mu Q_{Ls}^{\dagger ak})(Q_{Rr}^{\dagger bj}\sigma^\mu Q_{Rtbk})$& \\
{\color{red}34}&${\tau^{Ii}}_j\bft^I(Q_{Lpai}\sigma_\mu Q_{Ls}^{\dagger ck})(Q_{Rr}^{\dagger aj}\sigma^\mu Q_{Rtck})$& \\
35&${\tau^{Ii}}_j{\tau^{Jl}}_k\bft^I\bft^J(Q_{Lpai}\sigma_\mu Q_{Ls}^{\dagger ck})(Q_{Rr}^{\dagger aj}\sigma^\mu Q_{Rtcl})$& $R^Q_{20}$\\
36&${\tau^{Il}}_j{\tau^{Ji}}_k\bft^I\bft^J(Q_{Lpai}\sigma_\mu Q_{Ls}^{\dagger ck})(Q_{Rr}^{\dagger aj}\sigma^\mu Q_{Rtcl})$&$R^Q_{25}$\\
\hline
\end{longtable}
\end{center}

\paragraph{Type ${Q^\dagger_L}^2{Q_L}^2\&{Q^\dagger_R}^2{Q_R}^2$: }
\begin{center}
\begin{longtable}{ccccc}
\hline
$\mathcal{O}^{Uh\psi^4}_{37-60}$&Operator&\cite{Brivio:2016fzo}\\
\hline
37& $\mathcal{Y}[\tiny{\yt{pr},\yt{st}}](Q_{Lpai}\sigma_\mu Q_{Ls}^{\dagger ai})(Q_{Lrbj}\sigma^\mu Q_{Lt}^{\dagger bj})$ &$R_9^Q$ \\
38& $\mathcal{Y}[\tiny{\yt{pr},\yt{st}}](Q_{Lpai}\sigma_\mu Q_{Ls}^{\dagger aj})(Q_{Lrbj}\sigma^\mu Q_{Lt}^{\dagger bi})$ &$R_{12}^Q$ \\
{\color{red}39}& $\mathcal{Y}[\tiny{\yt{p,r},\yt{s,t}}](Q_{Lpai}\sigma_\mu Q_{Ls}^{\dagger ai})(Q_{Lrbj}\sigma^\mu Q_{Lt}^{\dagger bj})$ \\
{\color{red}40}& $\mathcal{Y}[\tiny{\yt{p,r},\yt{s,t}}](Q_{Lpai}\sigma_\mu Q_{Ls}^{\dagger aj})(Q_{Lrbj}\sigma^\mu Q_{Lt}^{\dagger bi})$ \\
41& $\mathcal{Y}[\tiny{\yt{pr},\yt{st}}]{\tau^{Ij}}_l\bft^I(Q_{Lpai}\sigma^\mu Q^{\dagger ai}_{Ls})(Q_{Lrbj}\sigma_\mu Q^{\dagger bl}_{Lt})$ &$R_{10}^Q$ & \\
{\color{red}42}& $\mathcal{Y}[\tiny{\yt{pr},\yt{s,t}}]{\tau^{Ij}}_l\bft^I(Q_{Lpai}\sigma^\mu Q^{\dagger ai}_{Ls})(Q_{Lrbj}\sigma_\mu Q^{\dagger bl}_{Lt})$ \\
{\color{red}43}& $\mathcal{Y}[\tiny{\yt{pr},\yt{s,t}}]{\tau^{Ii}}_l\bft^I(Q_{Lpai}\sigma^\mu Q^{\dagger aj}_{Ls})(Q_{Lrbj}\sigma_\mu Q^{\dagger bl}_{Lt})$ \\
\hline
\end{longtable}
\end{center}

\paragraph{Pure Lepton operators:} Including types ${L_L^\dagger}^2{L_L}^2,{L_L}^2{L_L^\dagger}^2,{L_R}^2{L_R^\dagger}^2,{L_L}{L_R^\dagger}{L_R}{L_L^\dagger}$.
\begin{center}
\begin{longtable}{ccccc}
\hline
$\mathcal{O}^{Uh\psi^4}_{61-90}$&Operator&\cite{Brivio:2016fzo}\\
\hline
61&$\mathcal{Y}[\tiny{\yt{pr},\yt{st}}](L^{\dagger i}_{Lp}L^{\dagger j}_{Lr})(L_{Rsi}L_{Rtj})$ & $R_1^l$ & \\
62&{\color{gray}$\mathcal{Y}[\tiny{\yt{pr},\yt{st}}](L^{\dagger i}_{Lp}\bar{\sigma}^{\mu\nu}L^{\dagger j}_{Lr})(L_{Rsi}\bar{\sigma}_{\mu\nu}L_{Rtj})$} & $R^l_8$\\
{\color{red}63}&$\mathcal{Y}[\tiny{\yt{p,r},\yt{s,t}}](L^{\dagger i}_{Lp}L^{\dagger j}_{Lr})(L_{Rsi}L_{Rtj})$  \\
{\color{red}64}&{\color{gray}$\mathcal{Y}[\tiny{\yt{p,r},\yt{s,t}}](L^{\dagger i}_{Lp}\bar{\sigma}^{\mu\nu}L^{\dagger j}_{Lr})(L_{Rsi}\bar{\sigma}_{\mu\nu}L_{Rtj})$} \\
65&{\color{gray}$\mathcal{Y}[\tiny{\yt{pr},\yt{st}}]{\tau^{Ik}}_j\bft^I(L^{\dagger i}_{Lp}L^{\dagger j}_{Lr})(L_{Rsk}L_{Rti})$} & $R^l_9$& \\
{\color{red}66}&{\color{gray}$\mathcal{Y}[\tiny{\yt{pr},\yt{s,t}}]{\tau^{Ik}}_j\bft^I(L^{\dagger i}_{Lp}L^{\dagger j}_{Lr})(L_{Rsk}L_{Rti})$} \\
{\color{red}67}&{\color{gray}$\mathcal{Y}[\tiny{\yt{pr},\yt{s,t}}]{\tau^{Ik}}_j\bft^I(L^{\dagger i}_{Lp}\bar{\sigma}_{\mu\nu}L^{\dagger j}_{Lr})(L_{Rsk}\bar{\sigma}^{\mu\nu}L_{Rti})$} \\
\hline
\end{longtable}
\end{center}

\paragraph{Type ${Q^\dagger_L}{Q_R}{L^\dagger_L}{L_R}$:}
\begin{center}
\begin{longtable}{ccccc}
\hline
$\mathcal{O}^{Uh\psi^4}_{91-102}$&Operator&\cite{Brivio:2016fzo}\\
\hline
91&$(L^{\dagger i}_{Lp}L_{Rri})(Q^{\dagger ak}_{Ls}Q_{Rtak})$ & $R^{Ql}_1$  \\
92&$(L^{\dagger i}_{Lp}\bar{\sigma}_{\mu\nu}L_{Rri})(Q^{\dagger ak}_{Ls}\bar{\sigma}^{\mu\nu}Q_{Rtak})$ & $R^{Ql}_6$ \\
93&$(L^{\dagger i}_{Lp}L_{Rrj})(Q^{\dagger aj}_{Ls}Q_{Rtai})$ & $R^{Ql}_2$ \\
94&$(L^{\dagger i}_{Lp}\bar{\sigma}_{\mu\nu}L_{Rrj})(Q^{\dagger aj}_{Ls}\bar{\sigma}^{\mu\nu}Q_{Rtai})$ & $R^{Ql}_5$ \\
95&${\tau^{Il}}_k\bft^I(L^{\dagger i}_{Lp}L_{Rri})(Q^{\dagger ak}_{Ls}Q_{Rtal})$ & $R^{Ql}_3$  \\
96&${\tau^{Il}}_k\bft^I(L^{\dagger i}_{Lp}\bar{\sigma}_{\mu\nu}L_{Rri})(Q^{\dagger ak}_{Ls}\bar{\sigma}^{\mu\nu}Q_{Rtal})$ &$R^{Ql}_4$ \\
97&{\color{gray}${\tau^{Ij}}_i\bft^I(L^{\dagger i}_{Lp}L_{Rrj})(Q^{\dagger ak}_{Ls}Q_{Rtak})$} & $R^{Ql}_{25}$ \\
98&{\color{gray}${\tau^{Ij}}_i\bft^I(L^{\dagger i}_{Lp}\bar{\sigma}_{\mu\nu}L_{Rrj})(Q^{\dagger ak}_{Ls}\bar{\sigma}^{\mu\nu}Q_{Rtak})$} & $R^{Ql}_{28}$ \\
99&{\color{gray}${\tau^{Ij}}_k\bft^I(L^{\dagger i}_{Lp}L_{Rrj})(Q^{\dagger ak}_{Ls}Q_{Rtai})$} & $R^{Ql}_{24}$  \\
100&{\color{gray}${\tau^{Ij}}_k\bft^I(L^{\dagger i}_{Lp}\bar{\sigma}_{\mu\nu}L_{Rrj})(Q^{\dagger ak}_{Ls}\bar{\sigma}^{\mu\nu}Q_{Rtai})$} & $R^{Ql}_{29}$ \\
101&{\color{gray}${\tau^{Il}}_i{\tau^{Jj}}_k\bft^I\bft^J(L^{\dagger i}_{Lp}L_{Rrj})(Q^{\dagger ak}_{Ls}Q_{Rtal})$} & $R^{Ql}_{26}$ \\
102&{\color{gray}${\tau^{Il}}_i{\tau^{Jj}}_k\bft^I\bft^J(L^{\dagger i}_{Lp}\bar{\sigma}_{\mu\nu}L_{Rrj})(Q^{\dagger ak}_{Ls}\bar{\sigma}^{\mu\nu}Q_{Rtal})$} & $R^{Ql}_{27}$ \\
\hline
\end{longtable}
\end{center}

\paragraph{Type ${Q^\dagger_L}{Q_L}{L^\dagger_L}{L_L}\& {Q^\dagger_R}{Q_R}{L^\dagger_R}{L_R} $:}
\begin{center}
\begin{longtable}{ccccc}
\hline
$\mathcal{O}^{Uh\psi^4}_{103-114}$&Operator&\cite{Brivio:2016fzo}\\
\hline
103&$(L_{Lpi}\sigma_\mu L^{\dagger i}_{Ls})(Q_{Lraj}\sigma^\mu Q^{\dagger aj}_{Lt})$ & $R^{Ql}_7$ \\
104&$(L_{Lpi}\sigma_\mu L^{\dagger j}_{Ls})(Q_{Lraj}\sigma^\mu Q^{\dagger ai}_{Lt})$ & $R^{Ql}_{13}$  \\
105&${\tau^{Ii}}_k\bft^I(L_{Lpi}\sigma_\mu L^{\dagger k}_{Ls})(Q_{Lraj}\sigma^\mu Q^{\dagger aj}_{Lt})$ & $R^{Ql}_{11}$ & \\
106&${\tau^{Ij}}_l\bft^I(L_{Lpi}\sigma_\mu L^{\dagger i}_{Ls})(Q_{Lraj}\sigma^\mu Q^{\dagger al}_{Lt})$ & $R^{Ql}_9$ & \\
{\color{red}107}&${\tau^{Ii}}_l\bft^I(L_{Lpi}\sigma_\mu L^{\dagger j}_{Ls})(Q_{Lraj}\sigma^\mu Q^{\dagger al}_{Lt})$ & & \\
108&${\tau^{Ii}}_k{\tau^{Jj}}_l\bft^I\bft^J (L_{Lpi}\sigma_\mu L^{\dagger k}_{Ls})(Q_{Lraj}\sigma^\mu Q^{\dagger al}_{Lt})$ &$R^{Ql}_{12}$& \\
109&$(L_{Rpi}\sigma_\mu L^{\dagger i}_{Rs})(Q_{Rraj}\sigma^\mu Q^{\dagger aj}_{Rt})$ & $R^{Ql}_8$\\
110&{$(L_{Rpi}\sigma_\mu L^{\dagger j}_{Rs})(Q_{Rraj}\sigma^\mu Q^{\dagger ai}_{Rt})$} & $R^{Ql}_{32}$\\
111&{\color{gray}${\tau^{Ii}}_k\bft^I(L_{Rpi}\sigma_\mu L^{\dagger k}_{Rs})(Q_{Rraj}\sigma^\mu Q^{\dagger aj}_{Rt})$} & $R^{QL}_{30}$ \\
112&{$\color{gray}{\tau^{Ij}}_l\bft^I(L_{Rpi}\sigma_\mu L^{\dagger i}_{Rs})(Q_{Rraj}\sigma^\mu Q^{\dagger al}_{Rt})$} & $R^{Ql}_{10}$ \\
{\color{red}113}&{\color{gray}${\tau^{Ii}}_l\bft^I(L_{Rpi}\sigma_\mu L^{\dagger j}_{Rs})(Q_{Rraj}\sigma^\mu Q^{\dagger al}_{Rt})$} & \\
114&{\color{gray}${\tau^{Ii}}_k{\tau^{Jj}}_l\bft^I\bft^J (L_{Rpi}\sigma_\mu L^{\dagger k}_{Rs})(Q_{Rraj}\sigma^\mu Q^{\dagger al}_{Rt})$} & $R^{Ql}_{31}$ \\
\hline
\end{longtable}
\end{center}

\paragraph{Type ${Q^\dagger_L}{Q_L}{L^\dagger_R}{L_R}\& {Q^\dagger_R}{Q_R}{L^\dagger_L}{L_L} $:}
\begin{center}
\begin{longtable}{ccccc}
\hline
$\mathcal{O}^{Uh\psi^4}_{115-126}$&Operator&\cite{Brivio:2016fzo}\\
\hline
115&$(L_{Rpi}\sigma_\mu L^{\dagger i}_{Rs})(Q_{Lraj}\sigma^\mu Q^{\dagger aj}_{Lt})$ & $R^{Ql}_{15}$ \\
116&{$(L_{Rpi}\sigma_\mu L^{\dagger k}_{Rs})(Q_{Lrai}\sigma^\mu Q^{\dagger ak}_{Lt})$} & $R^{Ql}_{35}$\\
117&{\color{gray}${\tau^{Ij}}_l\bft^I(L_{Rpi}\sigma_\mu L^{\dagger i}_{Rs})(Q_{Lraj}\sigma^\mu Q^{\dagger al}_{Lt})$} & $R^{Ql}_{17}$ \\
118&{\color{gray}${\tau^{Ij}}_i\bft^I(L_{Rpi}\sigma_\mu L^{\dagger k}_{Rs})(Q_{Lraj}\sigma^\mu Q^{\dagger al}_{Lk})$} & $R^{Ql}_{33}$ \\
{\color{red}119}&{\color{gray}${\tau^{Ik}}_l\bft^I(L_{Rpi}\sigma_\mu L^{\dagger k}_{Rs})(Q_{Lrai}\sigma^\mu Q^{\dagger al}_{Ll})$} & \\
120&{\color{gray}${\tau^{Ik}}_i{\tau^{Jj}}_l\bft^I\bft^J (L_{Rpi}\sigma_\mu L^{\dagger k}_{Rs})(Q_{Lraj}\sigma^\mu Q^{\dagger al}_{Lt})$}& $R^{Ql}_{34}$ \\
121&$(L_{Lpi}\sigma_\mu L^{\dagger i}_{Ls})(Q_{Rraj}\sigma^\mu Q^{\dagger aj}_{Rt})$ & $R^{Ql}_{14}$ \\
122&$(L_{Lpi}\sigma_\mu L^{\dagger k}_{Ls})(Q_{Rrai}\sigma^\mu Q^{\dagger ak}_{Rt})$ & $R^{Ql}_{20}$ \\
123&${\tau^{Ii}}_k\bft^I(L_{Lpi}\sigma_\mu L^{\dagger k}_{Ls})(Q_{Rraj}\sigma^\mu Q^{\dagger aj}_{Rt})$ & $R^{Ql}_{16}$ \\
124&${\tau^{Ii}}_j\bft^I(L_{Lpi}\sigma_\mu L^{\dagger k}_{Ls})(Q_{Rraj}\sigma^\mu Q^{\dagger al}_{Rk})$ & $R^{Qj}_{18}$ \\
{\color{red}125}&${\tau^{Il}}_k\bft^I(L_{Lpi}\sigma_\mu L^{\dagger k}_{Ls})(Q_{Rrai}\sigma^\mu Q^{\dagger al}_{Rl})$ & \\
126&${\tau^{Ii}}_j{\tau^{Jl}}_k\bft^I\bft^J (L_{Lpi}\sigma_\mu L^{\dagger k}_{Ls})(Q_{Rraj}\sigma^\mu Q^{\dagger al}_{Rt})$ &$R^{Ql}_{19}$ \\
\hline
\end{longtable}
\end{center}

\paragraph{Type ${Q^\dagger_L}{Q_R}{L^\dagger_R}{L_L}$:}
\begin{center}
\begin{longtable}{ccccc}
\hline
$\mathcal{O}^{Uh\psi^4}_{127-132}$&Operator&\cite{Brivio:2016fzo}\\
\hline
127&$(L_{Lpi}\sigma_\mu Q^{\dagger ai}_{Ls})(L^{\dagger j}_{Rr}\sigma^\mu Q_{Rtaj})$ & $R^{Ql}_{21}$ \\
128&$(L_{Lpi}\sigma_\mu Q^{\dagger ak}_{Ls})(L^{\dagger j}_{Rr}\sigma^\mu Q_{Rtak})$ & $R^{Ql}_{23}$ \\
129&${\tau^{Ii}}_k\bft^I(L_{Lpi}\sigma_\mu Q^{\dagger ak}_{Ls})(L^{\dagger j}_{Rr}\sigma^\mu Q_{Rtaj})$ & $R^{Ql}_{22}$ \\
130&{\color{gray}${\tau^{Il}}_k\bft^I(L_{Lpi}\sigma_\mu Q^{\dagger ak}_{Ls})(L^{\dagger i}_{Rr}\sigma^\mu Q_{Rtal})$} &$R^{Ql}_{38}$ \\
131&{\color{gray}${\tau^{Ii}}_j\bft^I(L_{Lpi}\sigma_\mu Q^{\dagger ak}_{Ls})(L^{\dagger j}_{Rr}\sigma^\mu Q_{Rtak})$} & $R^{Ql}_{36}$ \\
132&{\color{gray}${\tau^{Ii}}_j{\tau^{Jl}}_k\bft^I\bft^J (L_{Lpi}\sigma_\mu Q^{\dagger ak}_{Ls})(L^{\dagger j}_{Rr}\sigma^\mu Q_{Rtal})$} & $R^{Ql}_{37}$ \\
\hline
\end{longtable}
\end{center}

\paragraph{Type $Q^3L$:}
Literature \cite{Merlo:2016prs} considering only 1 generation, so operators obtained here have a surjection relation with operators in \cite{Merlo:2016prs}.
\begin{center}
\begin{longtable}{ccccc}
\hline
$\mathcal{O}^{Uh\psi^4}_{133-160}$&Operator&\cite{Merlo:2016prs} \\
\hline
133&$\mathcal{Y}[\tiny{\yt{rst}}]\epsilon^{abc}\epsilon^{ik}\epsilon^{jl}(L_{Lpi}Q_{Lraj})(Q_{Lsbk}Q_{Ltcl})$ & $R_1$\\
{\color{red}134}&$\mathcal{Y}[\tiny{\yt{rs,t}}]\epsilon^{abc}\epsilon^{ik}\epsilon^{jl}(L_{Lpi}Q_{Lraj})(Q_{Lsbk}Q_{Ltcl})$ & \\
{\color{red}135}&$\mathcal{Y}[\tiny{\yt{r,s,t}}]\epsilon^{abc}\epsilon^{ik}\epsilon^{jl}(L_{Lpi}Q_{Lraj})(Q_{Lsbk}Q_{Ltcl})$ & \\
136&$\mathcal{Y}[\tiny{\yt{rst}}]{\tau^{Il}}_m\epsilon^{abc}\epsilon^{ij}\epsilon^{km}\bft^I(L_{Lpi}\sigma_{\mu\nu}Q_{Lraj})(Q_{Lsbk}\sigma^{\mu\nu}Q_{Ltcl})$ & $R_2$\\
{\color{red}137}&$\mathcal{Y}[\tiny{\yt{rs,t}}]{{\tau^{Il}}_m\epsilon^{abc}\epsilon^{ij}\epsilon^{km}\bft^I(L_{Lpi}Q_{Lraj})(Q_{Lsbk}Q_{Ltcl})}$ & \\
{\color{red}138}&$\mathcal{Y}[\tiny{\yt{rs,t}}]{{\tau^{Il}}_m\epsilon^{abc}\epsilon^{ij}\epsilon^{km}\bft^I(L_{Lpi}\sigma_{\mu\nu}Q_{Lraj})(Q_{Lsbk}\sigma^{\mu\nu}Q_{Ltcl})}$ & \\
{\color{red}139}&$\mathcal{Y}[\tiny{\yt{r,s,t}}]{{\tau^{Il}}_m\epsilon^{abc}\epsilon^{ij}\epsilon^{km}\bft^I(L_{Lpi}Q_{Lraj})(Q_{Lsbk}Q_{Ltcl})}$ & \\
\hline
\end{longtable}
\end{center}

\section{On-shell Amplitude Basis}
\label{app:amplitude}


In this section, we present operators' amplitude basis, in which spurions are treated as part of gauge factors and are written as $\bft_i = \bft^{I_i}$.
\paragraph{Type $UhD^4$:}
\renewcommand\arraystretch{1.4}
\begin{center}
\begin{longtable}{r|l}
\hline
\multirow{2}{*}{$\phi^4$}
& $(\delta^{I_1I_3}\delta^{I_2I_4})\lra{34}^2\lrs{34}^2$ \\
& $(\delta^{I_1I_3}\delta^{I_2I_4})\lra{34}^2\lrs{34}^2$ \\
\hline
\multirow{2}{*}{$\phi^4\bft^2$}
& $(\delta^{I_1I_4}\delta^{I_2I_5}\delta^{I_3I_6}\bft_5\bft_6)\lra{34}^2\lrs{34}^2$\\
& $(\delta^{I_1I_4}\delta^{I_2I_5}\delta^{I_3I_6}\bft_5\bft_6)\lra{24}^2\lrs{24}^2$\\
\hline
$\phi^4\bft^4$& $(\delta^{I_1I_5}\delta^{I_2I_6}\delta^{I_3I_7}\delta^{I_4I_8}\bft_5\bft_6\bft_7\bft_8)\lra{34}^2\lrs{34}^2$ \\
\hline
\multirow{2}{*}{$h\phi^3\bft$}
& $(\delta^{I_2I_4}\delta^{I_3I_5}\bft_5)\lra{34}^2\lrs{34}^2$ \\
& $(\delta^{I_2I_4}\delta^{I_3I_5}\bft_5)\lra{24}^2\lrs{24}^2$ \\
\hline
$h\phi^3\bft^2$ & $(\delta^{I_2I_5}\epsilon^{I_3I_4I_6}\bft_5\bft_6)\lra{24}^2\lrs{24}^2$ \\
$h\phi^3\bft^3$ & $(\delta^{I_2I_5}\delta^{I_3I_6}\delta^{I_4I_7}\bft_5\bft_6\bft_7)\lra{34}^2\lrs{34}^2$ \\
\hline
\multirow{2}{*}{$h^2\phi^2$}
& $(\delta^{I_3I_4})\lra{34}^2\lrs{34}^2$\\
& $(\delta^{I_3I_4})\lra{24}^2\lrs{24}^2$\\
\hline
\multirow{2}{*}{$h^2\phi^2\bft^2$}
& $(\delta^{I_3I_5}\delta^{I_4I_6}\bft_5\bft_6)\lra{34}^2\lrs{34}^2$\\
& $(\delta^{I_3I_5}\delta^{I_4I_6}\bft_5\bft_6)\lra{34}^2\lrs{34}^2$\\
\hline
$h^3\phi\bft$ & $(\delta^{I_4I_5}\bft_5)\lra{34}^2\lrs{34}^2$ \\
\hline
$h^4$ & $\lra{34}^2\lrs{34}^2$ \\
\hline
\end{longtable}
\end{center}

\paragraph{Type $X^2Uh$:
}
\begin{center}
\begin{longtable}{r|l}
\hline
${B_L}^2$& $\lra{12}^2 $\\
\hline
${B_R}^2$& $\lrs{12}^2$ \\
\hline
${G_L}^2$ & $ (\delta^{A_1A_2})\lra{12}^2$\\
\hline
${G_R}^2$ & $ (\delta^{A_1A_2})\lrs{12}^2$\\
\hline
${W_L}^2$ & $ (\delta^{I_1I_2})\lra{12}^2$ \\
\hline
${W_R}^2$ & $ (\delta^{I_1I_2})\lrs{12}^2 $\\
\hline
${B_L}{W_L}\bft$ & $ (\delta^{I_2I_3}\bft_3)\lra{12}^2$ \\
\hline
${B_R}{W_R}\bft$ & $ (\delta^{I_2I_3}\bft_3)\lrs{12}^2$ \\
\hline
${W_L}^2\bft^2$ & $ (\delta^{I_1I_3}\delta^{I_2I_4}\bft_3\bft_4)\lra{12}^2$ \\
\hline
${W_R}^2\bft^2$ & $(\delta^{I_1I_3}\delta^{I_2I_4}\bft_3\bft_4)\lrs{12}^2$ \\
\hline
\end{longtable}
\end{center}

\paragraph{Type $XUhD^2$:}
\begin{center}
\begin{longtable}{r|l}
\hline
${W_L}h\phi^2$ & $ (\epsilon^{I_1I_3I_4})\lrs{34}\lra{13}\lra{14}$ \\
\hline
$h\phi^2{W_R}$ & $ (\epsilon^{I_2I_3I_4})\lrs{24}\lrs{34}\lra{23}$ \\
\hline
${B_L}h\phi^2\bft$ & $ (\epsilon^{I_3I_4I_5}\bft_5)\lrs{34}\lra{13}\lra{14}$ \\
\hline
$h\phi^2{B_R}\bft$ & $ (\epsilon^{I_2I_3I_5}\bft_5)\lrs{24}\lrs{34}\lra{23}$ \\
\hline
${W_L}h\phi^2\bft$ & $ (\delta^{I_1I_4}\delta^{I_3I_5}\bft_5)\lrs{34}\lra{13}\lra{14}$ \\
\hline
$h\phi^2{W_R}\bft$ & $(\delta^{I_2I_4}\delta^{I_3I_5}\bft_5)\lrs{24}\lrs{34}\lra{23}$ \\
\hline
${W_L}h\phi^2\bft^2$ & $(\delta^{I_1I_5}\epsilon^{I_3I_4I_6}\bft_5\bft_6)\lrs{34}\lra{13}\lra{14}$ \\
\hline
$h\phi^2{W_R}\bft^2$ & $(\delta^{I_2I_5}\epsilon^{I_3I_4I_6}\bft_5\bft_6)\lrs{24}\lrs{34}\lra{23}$ \\
\hline
\end{longtable}
\end{center}

\paragraph{Type $X^3$:}

\begin{center}
\begin{longtable}{r|l}
\hline
\multirow{4}{*}{$X^3$}
& $(f^{A_1A_2A_3})\lra{12}\lra{13}\lra{23}$ \\
& $(f^{A_1A_2A_3})\lrs{12}\lrs{13}\lrs{23}$ \\
& $(\epsilon^{I_1I_2I_3})\lra{12}\lra{13}\lra{23}$ \\
& $(\epsilon^{I_1I_2I_3})\lrs{12}\lrs{13}\lrs{23}$ \\
& $(\epsilon^{I_2I_3I_4}\bft_4)\lra{12}\lra{13}\lra{23}$ \\
& $(\epsilon^{I_2I_3I_4}\bft_4)\lrs{12}\lrs{13}\lrs{23}$ \\
\hline
\end{longtable}
\end{center}

\paragraph{Type $\psi^2UhD$:}

\begin{center}
\begin{longtable}{r|l}
\hline
${Q_L}h\phi{Q^\dagger_L}$ & $(\delta^{a_1}_{a_4}{(\tau^{I_3})}_{i_4}^{i_1})\lrs{34}\lra{13} $ \\
\hline
\multirow{2}{*}{${Q_L}h\phi{Q^\dagger_L}\bft $}
& $(\delta^{i_1}_{i_4}\delta^{a_1}_{a_4}\delta^{I_5I_3}\bft_5)\lrs{34}\lra{13}$ \\
& $(\delta^{a_1}_{a_4}\epsilon^{I_3I_5K}{(\tau^K)}_{i_4}^{i_1}\bft_5)\lrs{34}\lra{13}$ \\
\hline
${Q_L}h\phi{Q^\dagger_L}\bft^2$ & $(\delta^{a_1}_{a_4}\delta^{I_3I_6}\delta^{I_5M}{(\tau^M)}_{i_4}^{i_1}\bft_5\bft_6)\lrs{34}\lra{13}$ \\
\hline
${Q^\dagger_R}h\phi{Q_R}$ & $(\delta_{a_1}^{a_4}{(\tau^{I_3})}_{i_1}^{i_4})\lrs{34}\lra{13} $ \\
\hline
\multirow{2}{*}{${Q^\dagger_R}h\phi{Q_R}\bft $}
& $(\delta^{i_4}_{i_1}\delta^{a_4}_{a_1}\delta^{I_5I_3}\bft_5)\lrs{34}\lra{13}$ \\
& $(\delta^{a_4}_{a_1}\epsilon^{I_3I_5K}{(\tau^K)}^{i_4}_{i_1}\bft_5)\lrs{34}\lra{13}$ \\
\hline
${Q^\dagger_R}h\phi{Q_R}\bft^2$ & $(\delta^{a_4}_{a_1}\delta^{I_3I_6}\delta^{I_5M}{(\tau^M)}^{i_4}_{i_1}\bft_5\bft_6)\lrs{34}\lra{13}$ \\
\hline
\end{longtable}
\end{center}

\paragraph{Type $\psi^2UhD^2$:}

\begin{center}
\begin{longtable}{r|l}
\hline
\multirow{2}{*}{$\phi^2{Q^\dagger_L}{Q_R}$}
& $(\delta_{i_3}^{i_4}\delta_{a_3}^{a_4}\delta^{I_1I_2})\lra{34}\lrs{34}^2$\\
& $(\delta_{a_3}^{a_4}\epsilon^{I_1I_2K}{(\tau^K)}^{i_4}_{i_3})\lrs{34}\lra{24}\lrs{24}$ \\
\hline
\multirow{4}{*}{$\phi^2{Q^\dagger_L}{Q_R}\bft$}
& $(\delta_{i_3}^{i_4}\delta_{a_3}^{a_4}\delta^{I_5M}\epsilon^{I_1I_2M}\bft_5)\lrs{34}\lra{24}\lrs{24}$ \\
& $(\delta_{a_3}^{a_4}\delta^{I_1M}\delta^{I_2I_5}{(\tau^{M})}^{i_4}_{i_3}\bft_5)\lra{34}\lrs{34}^2$ \\
& $(\delta_{a_3}^{a_4}\delta^{I_1M}\delta^{I_2I_5}{(\tau^{M})}^{i_4}_{i_3}\bft_5)\lrs{34}\lra{24}\lrs{24}$ \\
& $(\delta_{a_3}^{a_4}\delta^{I_1I_2}{(\tau^{I_5})}^{i_4}_{i_3}\bft_5)\lrs{34}\lra{3,4}\lrs{34}$ \\
\hline
\multirow{3}{*}{$\phi^2{Q^\dagger_L}{Q_R}\bft^2$}
& $(\delta_{a_3}^{a_4}\delta^{I_2N}\epsilon^{I_1I_5N}{(\tau^{I_6})}^{i_4}_{i_3}\bft_5\bft_6)\lrs{34}\lra{24}\lrs{24}$ \\
& $(\delta_{a_3}^{a_4}\delta^{I_1I_5}\epsilon^{I_2I_6N}{(\tau^N)\bft_5\bft_6}^{i_4}_{i_3})\lra{34}\lrs{34}^2$ \\
\hline
\end{longtable}
\end{center}

\paragraph{Type $\psi^2UhX$:}

\begin{center}
\begin{longtable}{r|l}
\hline
${Q^\dagger_L}{Q_R}{B_R}$ & $(-\delta^{i_2}_{i_1}\delta^{a_2}_{a_1})\lrs{13}\lrs{23}$ \\
\hline
${Q^\dagger_L}{Q_R}{B_R}\bft$ & $(-\delta^{a_2}_{a_1}{(\tau^{I_4})}^{i_2}_{i_1}\bft_4)\lrs{13}\lrs{23}$ \\
\hline
${Q^\dagger_L}{Q_R}{G_R}$ & $(-\delta^{i_2}_{i_1}{(\lambda^{A_3})}^{a_1}_{a_2})\lrs{13}\lrs{23}$ \\
\hline
${Q^\dagger_L}{Q_R}{G_R}\bft$ & $(-{(\lambda^{A_3})}^{a_1}_{a_2}{(\tau^{I_4})}^{i_2}_{i_1}\bft_4)\lrs{13}\lrs{23}$ \\
\hline
${Q^\dagger_L}{Q_R}{W_R}$ & $(-\delta^{a_2}_{a_1}{(\tau^{I_3})}^{i_2}_{i_1})\lrs{13}\lrs{23}$ \\
\hline
\multirow{2}{*}{${Q^\dagger_L}{Q_R}{W_R}\bft$}
& $(-\delta^{a_2}_{a_1}\epsilon^{I_3I_4K}{(\tau^K)}^{i_2}_{i_1}\bft_4)\lrs{13}\lrs{23}$ \\
& $(-\delta^{i_2}_{i_1}\delta^{a_2}_{a_1}\delta^{I_3I_4}\bft_4)\lrs{13}\lrs{23}$ \\
\hline
${Q^\dagger_L}{Q_R}{W_R}\bft^2$ & $(\delta^{a_2}_{a_1}\delta^{I_3I_4}\delta^{I_5M}{(\tau^M)}^{i_2}_{i_1}\bft_4\bft_5)\lrs{13}\lrs{23}$ \\
\hline
\end{longtable}
\end{center}

\paragraph{Type ${Q^\dagger_L}^2{Q_R}^2$:}

\begin{center}
\begin{longtable}{r|l}
\hline
\multirow{8}{*}{${Q^\dagger_L}^2{Q_R}^2$} & $C^{[2],[2]}_{f_1f_2,f_3f_4}(\delta^{i_3}_{i_1}\delta^{i_4}_{i_2}\delta^{a_3}_{a_1}\delta^{a_4}_{a_2})\lrs{12}\lrs{34}$ \\
 & $C^{[2],[2]}_{f_1f_2,f_3f_4}(\delta^{i_3}_{i_1}\delta^{i_4}_{i_2}\delta^{a_3}_{a_1}\delta^{a_4}_{a_2})\lrs{13}\lrs{24}$ \\
 & $C^{[2],[2]}_{f_1f_2,f_3f_4}(\delta^{i_3}_{i_2}\delta^{i_4}_{i_1}\delta^{a_3}_{a_1}\delta^{a_4}_{a_2})\lrs{12}\lrs{34}$ \\
 & $C^{[2],[2]}_{f_1f_2,f_3f_4}(\delta^{i_3}_{i_2}\delta^{i_4}_{i_1}\delta^{a_3}_{a_1}\delta^{a_4}_{a_2})\lrs{13}\lrs{24}$ \\
 & $C^{[1,1],[1,1]}_{f_1f_2,f_3f_4}(\delta^{i_3}_{i_1}\delta^{i_4}_{i_2}\delta^{a_3}_{a_1}\delta^{a_4}_{a_2})\lrs{12}\lrs{34}$ \\
 & $C^{[1,1],[1,1]}_{f_1f_2,f_3f_4}(\delta^{i_3}_{i_1}\delta^{i_4}_{i_2}\delta^{a_3}_{a_1}\delta^{a_4}_{a_2})\lrs{13}\lrs{24}$ \\
 & $C^{[1,1],[1,1]}_{f_1f_2,f_3f_4}(\delta^{i_3}_{i_2}\delta^{i_4}_{i_1}\delta^{a_3}_{a_1}\delta^{a_4}_{a_2})\lrs{12}\lrs{34}$ \\
 & $C^{[1,1],[1,1]}_{f_1f_2,f_3f_4}(\delta^{i_3}_{i_2}\delta^{i_4}_{i_1}\delta^{a_3}_{a_1}\delta^{a_4}_{a_2})\lrs{13}\lrs{24}$ \\
\hline
\end{longtable}
\end{center}

\paragraph{Type: ${Q^\dagger_L}{Q_L}{Q^\dagger_R}{Q_R}$}
\begin{center}
\begin{longtable}{r|l}
\hline
\multirow{4}{*}{$Q_LQ^\dagger_R Q_L^\dagger Q_R$}
& $(\delta^{i_1}_{i_3}\delta^{i_4}_{i_2}\delta^{a_1}_{a_3}\delta^{a_4}_{a_2})\lrs{34}\lra{12}$ \\
& $(^{i_1}_{i_2}\delta^{i_4}_{i_3}\delta^{a_1}_{a_3}\delta^{a_4}_{a_2})\lrs{34}\lra{12}$ \\
& $(\delta^{i_1}_{i_3}\delta^{i_4}_{i_2}\delta^{a_1}_{a_2}\delta^{a_4}_{a_3})\lrs{34}\lra{12}$ \\
& $(^{i_1}_{i_2}\delta^{i_4}_{i_3}\delta^{a_1}_{a_2}\delta^{a_4}_{a_3})\lrs{34}\lra{12}$ \\
\hline
\multirow{6}{*}{$Q_LQ^\dagger_R Q_L^\dagger Q_R\bft $}
& $(\delta^{i_4}_{i_2}\delta^{a_1}_{a_3}\delta^{a_4}_{a_2}(\bft_5\tau)^{i_1}_{i_3})\lrs{34}\lra{12}$ \\
& $(\delta^{i_4}_{i_2}\delta^{a_1}_{a_2}\delta^{a_4}_{a_3}(\bft_5\tau)^{i_1}_{i_3})\lrs{34}\lra{12}$ \\
& $(^{i_1}_{i_2}\delta^{a_1}_{a_3}\delta^{a_4}_{a_2}(\bft_5\tau)^{i_4}_{i_3})\lrs{34}\lra{12}$ \\
& $(^{i_1}_{i_2}\delta^{a_1}_{a_2}\delta^{a_4}_{a_3}(\bft_5\tau)^{i_4}_{i_3})\lrs{34}\lra{12}$ \\
& $(\delta^{i_4}_{i_3}\delta^{a_1}_{a_3}\delta^{a_4}_{a_2}(\bft_5\tau)^{i_1}_{i_2})\lrs{34}\lra{12}$ \\
& $(\delta^{i_4}_{i_3}\delta^{a_1}_{a_2}\delta^{a_4}_{a_3}(\bft_5\tau)^{i_1}_{i_2})\lrs{34}\lra{12}$ \\
\hline
\multirow{2}{*}{$Q_LQ^\dagger_R Q_L^\dagger Q_R\bft^2 $}
& $(\delta^{a_1}_{a_2}\delta^{a_4}_{a_3}(\bft_5\tau)^{i_1}_{i_2}(\bft_6\tau)^{i_4}_{i_3})\lrs{34}\lra{12}$ \\
& $(\delta^{a_1}_{a_2}\delta^{a_4}_{a_3}(\bft_5\tau)^{i_2}_{i_4}(\bft_6\tau)^{i_1}_{i_3})\lrs{34}\lra{12}$ \\
\hline
\end{longtable}
\end{center}

\paragraph{Type ${Q^\dagger_L}^2{Q_L}^2 \& {Q^\dagger_R}^2{Q_R}^2$}
\begin{center}
\begin{longtable}{r|l}
\hline
\multirow{4}{*}{${Q_L}^2{Q_L^\dagger}^2$}
& $C^{[2],[2]}_{f_1f_2,f_3f_4}(\delta^{i_1}_{i_3}\delta^{i_2}_{i_4}\delta^{a_1}_{a_3}\delta^{a_2}_{a_4})\lrs{34}\lra{12}$ \\
& $C^{[2],[2]}_{f_1f_2,f_3f_4}(\delta^{i_1}_{i_4}\delta^{i_2}_{i_3}\delta^{a_1}_{a_3}\delta^{a_2}_{a_4})\lrs{34}\lra{12}$ \\
& $C^{[1,1],[1,1]}_{f_1f_2,f_3f_4}(\delta^{i_1}_{i_3}\delta^{i_2}_{i_4}\delta^{a_1}_{a_3}\delta^{a_2}_{a_4})\lrs{34}\lra{12}$ \\
& $C^{[1,1],[1,1]}_{f_1f_2,f_3f_4}(\delta^{i_1}_{i_4}\delta^{i_2}_{i_3}\delta^{a_1}_{a_3}\delta^{a_2}_{a_4})\lrs{34}\lra{12}$ \\
\hline
\multirow{6}{*}{${Q_L}^2{Q_L^\dagger}^2\bft $}
& $C^{[2],[2]}_{f_1f_2,f_3f_4}(\delta^{i_1}_{i_3}\delta^{a_1}_{a_3}\delta^{a_2}_{a_4}(\bft_5\tau)^{i_2}_{i_4})\lrs{34}\lra{12}$ \\
& $C^{[1,1],[2]}_{f_1f_2,f_3f_4}(\delta^{i_1}_{i_3}\delta^{a_1}_{a_3}\delta^{a_2}_{a_4}(\bft_5\tau)^{i_2}_{i_4})\lrs{34}\lra{12}$ \\
& $C^{[1,1],[2]}_{f_1f_2,f_3f_4}(\delta^{i_2}_{i_3}\delta^{a_1}_{a_3}\delta^{a_2}_{a_4}(\bft_5\tau)^{i_1}_{i_4})\lrs{34}\lra{12}$ \\
& $C^{[2],[1,1]}_{f_1f_2,f_3f_4}(\delta^{i_1}_{i_3}\delta^{a_1}_{a_3}\delta^{a_2}_{a_4}(\bft_5\tau)^{i_2}_{i_4})\lrs{34}\lra{12}$ \\
\hline
\end{longtable}
\end{center}

\paragraph{Pure lepton}

\begin{center}
\begin{longtable}{r|l}
\hline
\multirow{4}{*}{${L_L^\dagger}^2{L_R}^2$}
& $C^{[2],[2]}_{f_1f_2,f_3f_4}(\delta^{i_3}_{i_1}\delta^{i_4}_{i_2})\lrs{12}\lrs{34}$ \\
& $C^{[2],[2]}_{f_1f_2,f_3f_4}(\delta^{i_3}_{i_1}\delta^{i_4}_{i_2})\lrs{13}\lrs{24}$ \\
& $C^{[1,1],[1,1]}_{f_1f_2,f_3f_4}(\delta^{i_3}_{i_1}\delta^{i_4}_{i_2})\lrs{12}\lrs{34}$ \\
& $C^{[1,1],[1,1]}_{f_1f_2,f_3f_4}(\delta^{i_3}_{i_1}\delta^{i_4}_{i_2})\lrs{13}\lrs{24}$ \\
\hline
\multirow{6}{*}{${L_L^\dagger}^2{L_R}^2\bft$}
& $C^{[2],[2]}_{f_1f_2,f_3f_4}(\delta^{i_4}_{i_1}(\bft_5\tau)^{i_3}_{i_2})\lrs{12}\lrs{34}$ \\
& $C^{[2],[1,1]}_{f_1f_2,f_3f_4}(\delta^{i_4}_{i_1}(\bft_5\tau)^{i_3}_{i_2})\lrs{12}\lrs{34}$ \\
& $C^{[2],[1,1]}_{f_1f_2,f_3f_4}(\delta^{i_4}_{i_1}(\bft_5\tau)^{i_3}_{i_2})\lrs{13}\lrs{24}$ \\
& $C^{[1,1],[2]}_{f_1f_2,f_3f_4}(\delta^{i_4}_{i_1}(\bft_5\tau)^{i_3}_{i_2})\lrs{12}\lrs{34}$ \\
\hline
\end{longtable}
\end{center}

\paragraph{Type ${Q_L^\dagger}Q_RL^\dagger_L L_R$:}

\begin{center}
\begin{longtable}{r|l}
\hline
\multirow{4}{*}{$L^\dagger_LL_R Q^\dagger_L Q_R$}
& $(\delta^{i_2}_{i_1}\delta^{i_4}_{i_3}\delta^{a_4}_{a_3})\lrs{12}\lrs{34}$ \\
& $(\delta^{i_2}_{i_1}\delta^{i_4}_{i_3}\delta^{a_4}_{a_3})\lrs{13}\lrs{24}$ \\
& $(\delta^{i_2}_{i_3}\delta^{i_4}_{i_1}\delta^{a_4}_{a_3})\lrs{12}\lrs{34}$ \\
& $(\delta^{i_2}_{i_3}\delta^{i_4}_{i_1}\delta^{a_4}_{a_3})\lrs{13}\lrs{24}$ \\
\hline
\multirow{6}{*}{$L^\dagger_LL_R Q^\dagger_L Q_R\bft$}
& $(\delta^{i_2}_{i_1}\delta^{a_4}_{a_3}(\bft_5\tau)^{i_4}_{i_3})\lrs{12}\lrs{34}$ \\
& $(\delta^{i_2}_{i_1}\delta^{a_4}_{a_3}(\bft_5\tau)^{i_4}_{i_3})\lrs{13}\lrs{24}$ \\
& $(\delta^{i_4}_{i_3}\delta^{a_4}_{a_3}(\bft_5\tau)^{i_2}_{i_1})\lrs{12}\lrs{34}$ \\
& $(\delta^{i_4}_{i_3}\delta^{a_4}_{a_3}(\bft_5\tau)^{i_2}_{i_1})\lrs{13}\lrs{24}$ \\
& $(\delta^{i_4}_{i_1}\delta^{a_4}_{a_3}(\bft_5\tau)^{i_2}_{i_3})\lrs{12}\lrs{34}$ \\
& $(\delta^{i_4}_{i_1}\delta^{a_4}_{a_3}(\bft_5\tau)^{i_2}_{i_3})\lrs{13}\lrs{24}$ \\
\hline
\multirow{2}{*}{$L^\dagger_LL_R Q^\dagger_L Q_R\bft^2$}
& $C^{[2]}_{f_{56}}(\delta^{a_4}_{a_3}(\bft_5\tau)^{i_4}_{i_1}(\bft_6\tau)^{i_2}_{i_3})\lrs{12}\lrs{34}$ \\
& $C^{[2]}_{f_{56}}(\delta^{a_4}_{a_3}(\bft_5\tau)^{i_4}_{i_1}(\bft_6\tau)^{i_2}_{i_3})\lrs{13}\lrs{24}$ \\
\hline
\end{longtable}
\end{center}

\paragraph{Type ${Q^\dagger_L}{Q_L}{L^\dagger_L}{L_L}\& {Q^\dagger_R}{Q_R}{L^\dagger_R}{L_R}$:}

\begin{center}
\begin{longtable}{r|l}
\hline
\multirow{2}{*}{$L_L Q_L L^\dagger_L Q^\dagger_L$}
& $(\delta^{i_1}_{i_3}\delta^{i_2}_{i_4}\delta^{a_2}_{a_4})\lrs{34}\lra{12} $\\
& $(\delta^{i_1}_{i_4}\delta^{i_2}_{i_3}\delta^{a_2}_{a_4})\lrs{34}\lra{12} $\\
\hline
\multirow{3}{*}{$L_L Q_L L^\dagger_L Q^\dagger_L \bft$}
& $(\delta^{i_2}_{i_4}\delta^{a_2}_{a_4}(\bft_5\tau)^{i_1}_{i_3})\lrs{34}\lra{12}$ \\
& $(\delta^{i_1}_{i_3}\delta^{a_2}_{a_4}(\bft_5\tau)^{i_2}_{i_4})\lrs{34}\lra{12}$ \\
& $(\delta^{i_2}_{i_3}\delta^{a_2}_{a_4}(\bft_5\tau)^{i_1}_{i_4})\lrs{34}\lra{12}$ \\
\hline
$L_L Q_L L^\dagger_L Q^\dagger_L \bft^2$ & $(\delta^{a_2}_{a_4}(\bft_5\tau)^{i_1}_{i_3}(\bft_6\tau)^{i_2}_{i_4})\lrs{34}\lra{12}$ \\
\hline
\multirow{2}{*}{$L^\dagger_R Q^\dagger_R L_R Q_R$}
& $(\delta_{i_1}^{i_3}\delta_{i_2}^{i_4}\delta_{a_2}^{a_4})\lrs{34}\lra{12} $\\
& $(\delta_{i_1}^{i_4}\delta_{i_2}^{i_3}\delta_{a_2}^{a_4})\lrs{34}\lra{12} $\\
\hline
\multirow{3}{*}{$L^\dagger_R Q^\dagger_R L_R Q_R \bft$}
& $(\delta_{i_2}^{i_4}\delta_{a_2}^{a_4}(\bft_5\tau)^{i_3}_{i_1})\lrs{34}\lra{12}$ \\
& $(\delta^{i_3}_{i_1}\delta_{a_2}^{a_4}(\bft_5\tau)^{i_4}_{i_2})\lrs{34}\lra{12}$ \\
& $(\delta_{i_2}^{i_3}\delta_{a_2}^{a_4}(\bft_5\tau)^{i_4}_{i_1})\lrs{34}\lra{12}$ \\
\hline
$L^\dagger_R Q^\dagger_R L_R Q_R \bft^2$ & $(\delta_{a_2}^{a_4}(\bft_5\tau)^{i_3}_{i_1}(\bft_6\tau)^{i_4}_{i_2})\lrs{34}\lra{12}$ \\
\hline
\end{longtable}
\end{center}

\paragraph{Type ${L^\dagger_L}{L_L}{Q^\dagger_R}{Q_R}\& {L^\dagger_R}{L_R}{Q^\dagger_L}{Q_L}$:}

\begin{center}
\begin{longtable}{r|l}
\hline
\multirow{2}{*}{$L^\dagger_R Q_L L_R Q^\dagger_L$}
& $(\delta^{i_2}_{i_4}\delta^{i_3}_{i_1}\delta^{a_2}_{a_4})\lrs{34}\lra{12}$ \\
& $(\delta^{i_2}_{i_1}\delta^{i_3}_{i_4}\delta^{a_2}_{a_4})\lrs{34}\lra{12}$ \\
\hline
\multirow{3}{*}{$L^\dagger_R Q_L L_R Q^\dagger_L\bft $}
& $(\delta^{i_3}_{i_1}\delta^{a_2}_{a_4}(\bft_5\tau)^{i_2}_{i_4})\lrs{34}\lra{12}$ \\
& $(\delta^{i_3}_{i_4}\delta^{a_2}_{a_4}(\bft_5\tau)^{i_2}_{i_1})\lrs{34}\lra{12}$ \\
& $(\delta^{i_2}_{i_1}\delta^{a_2}_{a_4}(\bft_5\tau)^{i_3}_{i_4})\lrs{34}\lra{12}$ \\
\hline
$L^\dagger_R Q_L L_R Q^\dagger_L\bft^2 $ & $(\delta^{a_2}_{a_4}(\bft_5\tau)^{i_3}_{i_1}(\bft_6\tau)^{i_2}_{i_4})\lrs{34}\lra{12}$ \\
\hline
\multirow{2}{*}{$L_L Q^\dagger_R L^\dagger_L Q_R $}
& $(\delta^{i_4}_{i_2}\delta^{i_1}_{i_3}\delta^{a_4}_{a_2})\lrs{34}\lra{12}$ \\
& $(\delta^{i_1}_{i_2}\delta^{i_4}_{i_3}\delta^{a_4}_{a_2})\lrs{34}\lra{12}$ \\
\hline
\multirow{3}{*}{$L_L Q^\dagger_R L^\dagger_L Q_R \bft $}
& $(\delta_{i_3}^{i_1}\delta_{a_2}^{a_4}(\bft_5\tau)^{i_4}_{i_2})\lrs{34}\lra{12}$ \\
& $(\delta_{i_3}^{i_4}\delta_{a_2}^{a_4}(\bft_5\tau)^{i_1}_{i_2})\lrs{34}\lra{12}$ \\
& $(\delta_{i_2}^{i_1}\delta_{a_2}^{a_4}(\bft_5\tau)^{i_4}_{i_3})\lrs{34}\lra{12}$ \\
\hline
$L_L Q^\dagger_R L^\dagger_L Q_R \bft^2 $ & $(\delta_{a_2}^{a_4}(\bft_5\tau)^{i_1}_{i_2}(\bft_6\tau)^{i_4}_{i_3})\lrs{34}\lra{12}$ \\
\hline
\end{longtable}
\end{center}

\paragraph{Type ${Q^\dagger_L}{Q_R}{L^\dagger_R}{L_L}$:}

\begin{center}
\begin{longtable}{r|l}
\hline
\multirow{2}{*}{${L_L}{L^\dagger_R}{Q^\dagger_L}{Q_R}$}
& $(\delta^{i_1}_{i_3}\delta^{i_4}_{i_2}\delta^{a_4}_{a_3})\lrs{34}\lra{12}$ \\
& $(\delta^{i_1}_{i_2}\delta^{i_4}_{i_3}\delta^{a_4}_{a_3})\lrs{34}\lra{12}$ \\
\hline
\multirow{3}{*}{${L_L}{L^\dagger_R}{Q^\dagger_L}{Q_R}\bft $}
& $(\delta^{i_4}_{i_2}\delta^{a_4}_{a_3}(\bft_5\tau)^{i_1}_{i_3})\lrs{34}\lra{12}$ \\
& $(\delta^{i_1}_{i_2}\delta^{a_4}_{a_3}(\bft_5\tau)^{i_4}_{i_3})\lrs{34}\lra{12}$ \\
& $(\delta^{i_4}_{i_3}\delta^{a_4}_{a_3}(\bft_5\tau)^{i_1}_{i_2})\lrs{34}\lra{12}$ \\
\hline
${L_L}{L^\dagger_R}{Q^\dagger_L}{Q_R}\bft^2 $ & $(\delta^{a_4}_{a_3}(\bft_5\tau)^{i_1}_{i_2}(\bft_6\tau)^{i_4}_{i_3})$\\
\hline
\end{longtable}
\end{center}

\paragraph{Type $Q^3L$:}
\begin{center}
\begin{longtable}{r|l}
\hline
\multirow{3}{*}{${L_L}{Q_L}^3$}
& $C^{[3]}_{f_2f_3f_4}(\epsilon^{i_1i_3}\epsilon^{i_2i_4}\epsilon^{a_2a_3a_4})\lra{12}\lra{34}$ \\
& $C^{[2,1]}_{f_2f_3f_4}(\epsilon^{i_1i_3}\epsilon^{i_2i_4}\epsilon^{a_2a_3a_4})\lra{12}\lra{34}$ \\
& $C^{[1,1,1]}_{f_2f_3f_4}(\epsilon^{i_1i_3}\epsilon^{i_2i_4}\epsilon^{a_2a_3a_4})\lra{12}\lra{34}$ \\
\hline
\multirow{3}{*}{${L_L}{Q_L}^3\bft$}
& $C^{[3]}_{f_2f_3f_4}(\epsilon^{i_1i_2}\epsilon^{i_3m}\epsilon^{a_2a_3a_4}(\bft_5\tau)_{m}^{i_4})\lra{13}\lra{24}$ \\
& $C^{[2,1]}_{f_2f_3f_4}(\epsilon^{i_1i_2}\epsilon^{i_3m}\epsilon^{a_2a_3a_4}(\bft_5\tau)_{m}^{i_4})\lra{12}\lra{34}$ \\
& $C^{[2,1]}_{f_2f_3f_4}(\epsilon^{i_2i_4}\epsilon^{i_3m}\epsilon^{a_2a_3a_4}(\bft_5\tau)_{m}^{i_1})\lra{12}\lra{34}$ \\
& $C^{[1,1,1]}_{f_2f_3f_4}(\epsilon^{i_1i_2}\epsilon^{i_3m}\epsilon^{a_2a_3a_4}(\bft_5\tau)_{m}^{i_4})\lra{12}\lra{34}$ \\
\hline
${L_L}{Q_L}^3\bft^2 $ & $C^{[2,1]}_{f_2f_3f_4}(\epsilon^{i_3m}\epsilon^{i_4n}\epsilon^{a_2a_3a_4}(\bft_5\tau)_m^{i_1}(\bft_6\tau)_n^{i_2})\lra{12}\lra{34}$ \\
\hline
\end{longtable}
\end{center}



\bibliographystyle{JHEP}
\bibliography{ref}

\end{document}